\documentclass{aa}  

\usepackage{graphicx}
\usepackage{txfonts}
\usepackage{float}
\usepackage[colorlinks=true,citecolor=blue,linkcolor=blue]{hyperref}
\usepackage[caption=false]{subfig}
\begin{document}

   \title{Self-generated ultraviolet radiation in molecular shock waves}
   \titlerunning{Self-generated UV in molecular shocks II}
   \subtitle{II. CH$^+$ and the interpretation of emission from shock ensembles}

   
   \author{A. Lehmann\inst{1}, B. Godard\inst{1,2}, G. Pineau des For\^{e}ts\inst{2,3}, A. Vidal-Garc\'{i}a\inst{1}, \and E. Falgarone\inst{1}}
   \authorrunning{Lehmann, Godard, Pineau des For\^{e}ts, Vidal-Garc\'{i}a\inst{1}, \& Falgarone}

   \institute{Laboratoire de Physique de l'ENS, ENS, Universit\'{e} PSL, CNRS, Sorbonne Universit\'{e}, Universit\'{e} Paris-Diderot, Paris, France
   		\and
   Observatoire de Paris, PSL University, Sorbonne Universit\'{e}, LERMA, 75014 Paris, France
   		\and
   Universit\'{e} Paris-Saclay, CNRS, Institut d’Astrophysique Spatiale, 91405 Orsay, France\\
   \email{benjamin.godard@obspm.fr}
   }
   \date{Submitted Month Day, Year}

 
  \abstract
  
\abstract{The energetics and physical conditions of the interstellar medium and feedback processes remain challenging to probe.}
{Shocks, modelled over a broad range of parameters, are used to construct a new tool to deduce the mechanical energy and physical conditions from observed atomic or molecular emission lines.}
{We compute magnetised, molecular shock models with velocities $V_s=5$--$80$~km~s$^{-1}$, preshock proton densities $n_{\rm H}=10^2$--$10^6$~cm$^{-3}$, weak or moderate magnetic field strengths, and in the absence or presence of an external UV radiation field. These parameters represent the broadest published range of physical conditions for molecular shocks. As a key shock tracer we focus on the production of CH$^+$, and post-process the radiative transfer of its rovibrational lines. We develop a simple emission model of an ensemble of shocks for connecting any observed emission lines to the mechanical energy and physical conditions of the system.}
{For this range of parameters we find the full diversity (C-, C$^*$-, CJ-, and J-type) of magnetohydrodynamic shocks. H$_2$ and H are dominant coolants, with up to 30\% of the shock kinetic flux escaping in Ly$\alpha$ photons. The reformation of molecules in the cooling tail means H$_2$ is even a good tracer of dissociative shocks and shocks that were initially fully atomic. The known shock tracer CH$^+$ can also be a significant coolant, reprocessing up to one percent of the kinetic flux. Its production and excitation is intimately linked to the presence of H$_2$ and C$^+$. For each shock model we provide  integrated intensities of rovibrational lines of H$_2$, CO, and CH$^+$, atomic H lines, and atomic fine-structure and metastable lines. 
We demonstrate how to use these shock models to deduce the mechanical energy and physical conditions of extragalactic environments. As a template example, we interpret the CH$^+$(1-0) emission from the Eyelash starburst galaxy. A mechanical energy injection rate of at least $10^{11}$~$L_\odot$ into molecular shocks is required to reproduce the observed line. We find that shocks with velocities as low as 5~km~s$^{-1}$ irradiated by a strong UV field are compatible with the available energy budget. The low-velocity, externally irradiated shocks are at least an order magnitude more efficient than the most efficient shocks with no external irradiation, in terms of the total mechanical energy required. We predict differences of more than 2 orders of magnitude in intensities of the pure rotational lines of CO, Ly$\alpha$, metastable lines of O, S$^+$, and N, between representative models of low-velocity ($V_s \sim 10$~km~s$^{-1}$) externally irradiated shocks and higher velocity ($V_s \geq 50$~km~s$^{-1}$) shocks with no external irradiation.}
{Shock modeling over an extensive range of physical conditions allows for the interpretation of challenging observations of broad line emission from distant galaxies. Our new method opens up a promising avenue to quantitatively probe the physical conditions and mechanical energy of galaxy scale gas flows.}


   \maketitle
%

\section{Introduction}

At first sight, the problem of galaxy formation might appear simple: a reservoir of gas in a potential well of dark matter must collapse under self-gravity to form the stars that make up the galaxy. This process is complicated by stellar and black-hole feedback mechanisms that individually impinge on the small scale local environments, the dense molecular clouds in which stars ultimately form, and collectively produce a galaxy scale feedback mechanism that affects the total reservoir of infalling material. Observational probes of these processes promise to uncover the energetics of the systems.

The complex galactic lifecycle of baryonic matter always involves vast amounts of mechanical energy, often observed in the form of galaxy scale high velocity flows ($V > 500$~km~s$^{-1}$) that must result in shock waves. Such outflows have long been observed around starbursts and active galactic nuclei with atomic tracers \citep[see review by][]{veilleux_2005}. More recently, these kinematic signatures have been increasingly observed in molecular emission, including H$_2$ \citep{rupke_2013, petric_2018}, CO \citep{gowardhanl_2018,fluetsch_2019}, OH \citep{gonzalez-alfonso_2017}, HCN, HNC, and HCO$^+$ \citep{aalto_detection_2012,lindberg_2016}, and CH$^+$ \citep{falgarone_2017, vidal-garcia_2021}, revealing the multiphase nature of these flows. These molecular observations are challenging to explain, given that molecules cannot survive in the hot gas ($T > 10^6$~K) generated by shock waves at these velocities. An accurate modeling of the molecular component is important because it carries a significant fraction of the outflow mass and momentum, directly concerns the star forming activity of galaxies, and can dominate the cooling budget over the X-ray cooling of the hot gas.

A recent paradigm for interpreting such puzzling molecular emission has emerged wherein the mechanical energy of a large-scale high velocity shock dissipates in myriad low- and intermediate-velocity shocks ($V < 100$~km~s$^{-1}$) \citep{guillard_2009,lesaffre_low-velocity_2013,appleton_powerful_2017}. These lower velocity shocks are generated from the cascade of mechanical energy in the supersonic turbulence driven in the multiphase plasma overrun by the large-scale shock. Little work has been done to use this framework to quantitatively connect observations to the energetics and physical conditions of the observed system. Accurate models of molecular shocks over a broad range of velocities and gas conditions are required to capture the diversity of dissipative structures within this framework. We present in this paper both new shock models and make a first step in using grids of shock models to connect observed emission lines to physical information of the system.

The first of the two main goals of this work is to model molecular shocks in a broad range of environments. The microphysics and computational methods required to accurately model atomic and molecular shocks have been rigorously studied for decades \citep[see Introduction of][hereafter Paper 1]{lehmann_2020}. In Paper 1 we developed an iterative, post-processed radiative transfer algorithm to self-consistently compute the UV radiation from collisionaly excited atomic H in weakly magnetised (preshock transverse magnetic field $B_0 = 0.1 \, \mu {\rm G} \, (n_{\rm H} / {\rm cm^{-3}} )^{1/2}$ for preshock density $n_{\rm H}$), J-type molecular shocks at intermediate velocities (30~km~s$^{-1}~\leq V_s \leq$~60~km~s$^{-1}$) and a single preshock density $n_{\rm H}=10^4$~cm$^{-3}$. We extend that work by computing a grid of shock models over a range of densities ($10^2$~cm$^{-3}~\leq n_{\rm H} \leq 10^6$~cm$^{-3}$) and broaden the range of velocities (5~km~s$^{-1}~\leq V_s \leq$~80~km~s$^{-1}$). We also compute two more grids at moderate magnetic field strengths ($B_0 = 1 \, \mu {\rm G} \, (n_{\rm H} / {\rm cm^{-3}} )^{1/2}$) and with or without an external UV radiation field. This work therefore presents molecular shock models with the broadest published range of physical conditions. The shock code we use and the grids of models are presented in Sect.~\ref{sec:model}. While we focus on an extragalactic application, shocks in this parameter space have applications to numerous local interstellar molecular environments that are the subjects of ongoing research, for example protostellar outflows \citep{dopita_effects_2017,gusdorf_2017,fang_2018, van_dishoeck_2021}, infrared dark clouds \citep{pon3_2016}, stellar winds \citep{tram_2018}, supernovae remnants \citep{reach_supernova_2019}, or cloud-cloud collisions \citep{armijos-abendano_2020, enokiya_2021}.

Amongst the molecules observed in extragalactic environments, CH$^+$ is particularly interesting. Its dominant formation pathway is highly endothermic which yet seems to take place in cold molecular environments. This fact, in combination with a destruction timescale too short to allow for transport far from the site of formation, makes the presence of CH$^+$ in dense environments a strong indicator of bursts of turbulent dissipation. Recent observations of its rovibrational lines around high-redshift starburst galaxies \citep{falgarone_2017, vidal-garcia_2021}, in photodissociation regions in the Orion molecular cloud \citep{goicoechea_2019}, in the inner regions of NGC 1365 \citep{sandqvist_2021}, or in planetary nebulae \citep{neufeld_2021}, motivates a focused study on this molecule. In Sect.~\ref{sec:RT} we consider in detail the radiative transfer of CH$^+$ rovibrational lines in different kinds of shocks. In Sect.~\ref{sec:predictions} we also provide line intensities and column densities for species other than CH$^+$. Many of these lines are observable with state-of-the-art observational facilities such as the \textit{James Webb Space Telescope} (JWST) or the \textit{Atacama Large Millimeter/submillimeter Array} (ALMA).

The second main goal is to develop a simple framework to use shock model results to interpret observations of environments containing an ensemble of shocks, as in the cascade of supersonic turbulence. This toy model is presented in Sect.~\ref{sec:ensemble}, where we demonstrate how to use the results of the shock models to deduce the mechanical energy and physical conditions from an observed atomic or molecular line. As a template observation, we apply the ensemble model to the CH$^+$(1-0) broad ($\sim 1000$~km~s$^{-1}$) emission line from the high-$z$ starburst galaxy The Cosmic Eyelash \citep{falgarone_2017}. Finally, in Sect.~\ref{sec:predictions} we show how lines other than CH$^+$, such as rovibrational lines of H$_2$ and CO, and atomic fine structure and metastable lines, can be used to further constrain our knowledge of the Eyelash galaxy and similar systems.


\begin{table}
\caption{Shock parameters.}
\centering
\begin{tabular}{l c c c} \hline \hline
\multicolumn{4}{c}{\vspace{-0.3cm}} \\
Parameter & Symbol & Value(s) & Unit\\ \cline{1-4}
\multicolumn{4}{c}{\vspace{-0.3cm}} \\
Shock velocity & $V_s$ & 5-80 & km~s$^{-1}$ \\
Preshock proton density$^a$ & $n_{\rm{H}}$  & $10^2$-$10^6$ & cm$^{-3}$ \\
Magnetic parameter$^b$ & $b$ & 0.1 or 1 & \\
Cosmic-ray ionization rate & $\zeta$ & 10$^{-16}$ & s$^{-1}$ \\
External radiation field$^c$ & $G_0$ & 0 or 10$^3$ & \\
Viscous length & $l$ & $10^{13}/n_{\rm{H}}$ & cm \\
Number of H$_2$ levels & N$_{\rm{lev}}$ & 150 & \\ \hline \vspace{-0.2cm}
\end{tabular}
 
\tablefoot{(a) Defined as $n_{\rm H} = n({\rm H}) + 2n({\rm H}_2)$. (b) Sets the initial transverse magnetic field $B_0 = (1 \, \mu {\rm G}) \, b \times (n_{\rm H} / {\rm cm^{-3}} )^{1/2}$. (c) The scaling factor $G_0$ is applied to the standard ultraviolet radiation field of \cite{mathis_interstellar_1983}.}

\label{tab:shockparams}
\end{table}

\begin{figure*}
\centering
  \includegraphics[width=0.95\textwidth]{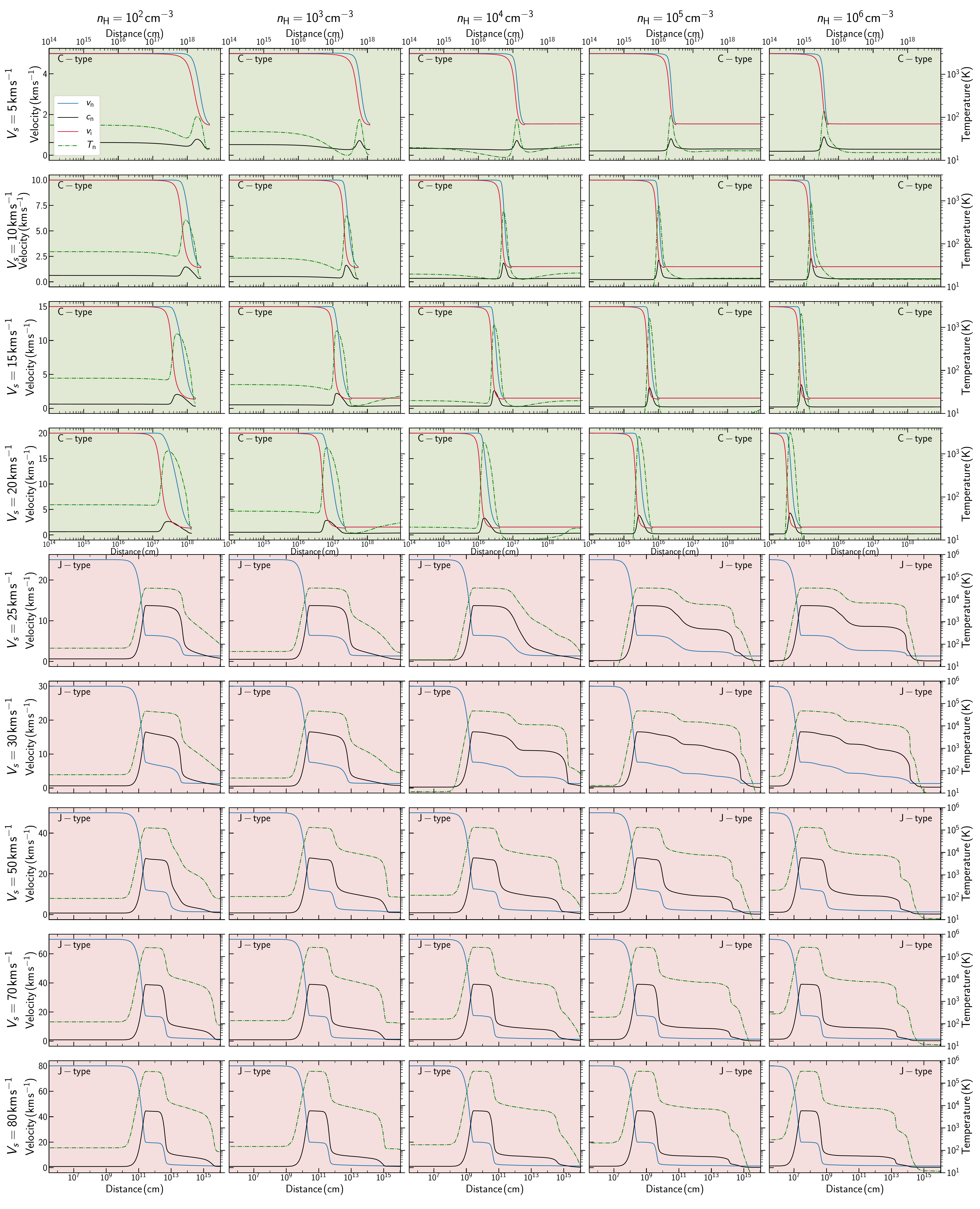}
  \caption{Shock profiles of ion (red solid) and neutral (blue solid) fluid velocities, local sound speed (black solid) and neutral fluid temperature (green dash-dotted). The shocks propagate from right to left, and the velocities are given in the reference frame in which the shock is stationary. These shocks were computed with magnetic parameter $b=1$ and no external radiation field $G_0=0$. Note that both the distance and temperature axes are different for the C- and J-type shocks, and that the logarithmic scaling can give the false impression that both types of shocks have similar sizes for a given preshock density. We have excluded profiles for shocks with velocities $V_s=40$ and 60~km~s$^{-1}$ for brevity.}
  \label{fig:temps_g0}
\end{figure*}

\section{Shock model and grid} \label{sec:model}
For this work we use the Paris-Durham shock code\footnote{available on the ISM platform https://ism.obspm.fr}, a publicly available state-of-the-art tool designed to study molecular chemistry in magnetised shocks. In this section we summarise some relevant theoretical aspects of magnetised shocks and introduce the grids of shocks that we will analyse in this work.

\subsection{Paris-Durham Shock Code}\label{sec:model_code}
The Paris-Durham shock code solves the equations of magnetohydrodynamics (MHD) coupled with cooling functions and an extensive chemical reaction network appropriate to the molecular phases of the interstellar medium. The code simultaneously solves ordinary differential equations for the conservation of number density, mass density, momentum and energy for both the ion and neutral fluid, for each of the 150 H$_2$ levels, and each of the 141 species in the chemical network. To do this, it uses the thoroughly tested \textsc{DVODE} forward integration algorithm \citep{hindmarsh_1983} to give steady-state, plane-parallel shock models. The version of the shock code used in this work is that of \cite{flower_influence_2003}, with updates described in \cite{lesaffre_low-velocity_2013}, \cite{godard_models_2019}, and Paper 1.

This shock code allows the computation of stationary MHD shocks which, depending on the physical conditions, may be continuous C- and C$^*$-type, and discontinuous J- and CJ-type shocks. It is important to capture the full diversity of shock types because differences in their thermal structure lead to wildly different observational predictions \citep[see review by][]{draine_1993}. Representative temperature and velocity profiles of all these types are shown in Figs.~\ref{fig:temps_g0} and \ref{fig:temps_g3}. The C-, C$^*$-, and CJ-type shocks can only be modelled with a multifluid code, as their internal dynamical, thermal, and chemical structure requires the inclusion of ion-neutral drift. The basic division between these types of shocks arises from how the shock velocity compares to the characteristic speeds in the medium and the interplay between heating and cooling. The three characteristic MHD speeds, the fast, intermediate and slow modes, allow for fast-, intermediate-, and slow-mode shocks \cite[see e.g.][]{kennel_mhd_1989,lehmann_signatures_2016}, but we focus only on fast-mode shocks in this work. In order to have a fast-mode MHD shock, the fluid velocity must be greater than the total fluid fast magnetosonic speed ($c_f = \sqrt{ c_s^2 + B^2/ 4 \pi \rho_n }$, for magnetic field strength $B$ and where the neutral mass density is much greater than the charged fluid mass density $\rho_n \gg \rho_c$ in a weakly ionized gas, and sound speed $c_s = \sqrt{k_{\rm B}T/\mu m_{\rm H}}$ for temperature $T$, Boltzman constant $k_{\rm B}$, mean particle mass $\mu$ and proton mass $m_{\rm H}$). However, if the shock velocity is below the magnetosonic speed of the charged fluid ($c_{\rm m} = \sqrt{ c_s^2 + B^2/ 4 \pi \rho_c }$), the charged fluid decouples from the neutrals and streams ahead to heat and deccelerate the gas via collisions in a \textit{magnetic precursor} \citep{mullan_structure_1971,draine_interstellar_1980}. In this case, if the cooling is efficient enough to keep the local sound speed below the local fluid velocity, then all fluid variables remain continuous throughout resulting in a C-type shock solution. If the cooling cannot keep up with the heating there will be a transition from the supersonic fluid to a subsonic fluid. In some conditions there exists a unique physical solution that crosses this sonic point smoothly giving a C$^*$-type shock. Otherwise a discontinuous transition occurs, mediated by viscosity over length scales on the order of the mean free path, resulting in CJ-type shock. These C$^*$ and CJ shock solutions require atypical numerical methods due to large gradients near the sonic point \citep[e.g.][]{chernoff_magnetohydrodynamic_1987,roberge_new_1990} and were recently implemented in the Paris-Durham code \citep{godard_models_2019}. Finally, if the shock velocity is greater than the ion magnetosonic speed, there can be no drift between the fluid types and the fluid transitions without a magnetic precursor discontinuously due to viscosity resulting in a J-type shock. Viscous heating in J-type shocks can generate temperatures large enough to produce UV photons that can heat and ionize the gas ahead of the shock in a \textit{radiative precursor}. As the self-generated UV photons affect the shock structure, and the shock structure is required to compute the UV radiative transfer, iterative methods are required to self-consistently model these shocks paying particular attention to non-equilibrium effects as gas is advected through the shock generated UV field \citep[e.g.][]{shull_theoretical_1979,sutherland_effects_2017}. This method has been implemented in the Paris-Durham code in Paper 1. 

The shock code assumes that the magnetic field is transverse to the direction of shock propagation and so we cannot compute oblique shocks, where there is an angle between the magnetic field and shock plane. However, these transverse shock models accurately describe oblique shocks for J-type shocks with the large Alfv\'{e}nic Mach numbers considered in this work \citep{hollenbach_molecule_1979}. For the C-type shocks, oblique shocks can be approximated by purely transverse models with a modified total field strength \citep{wardle_1987}.

\subsection{Grids of shock models}\label{sec:model_grids}

\begin{figure*}[!ht]
\centering
  \includegraphics[width=0.95\textwidth]{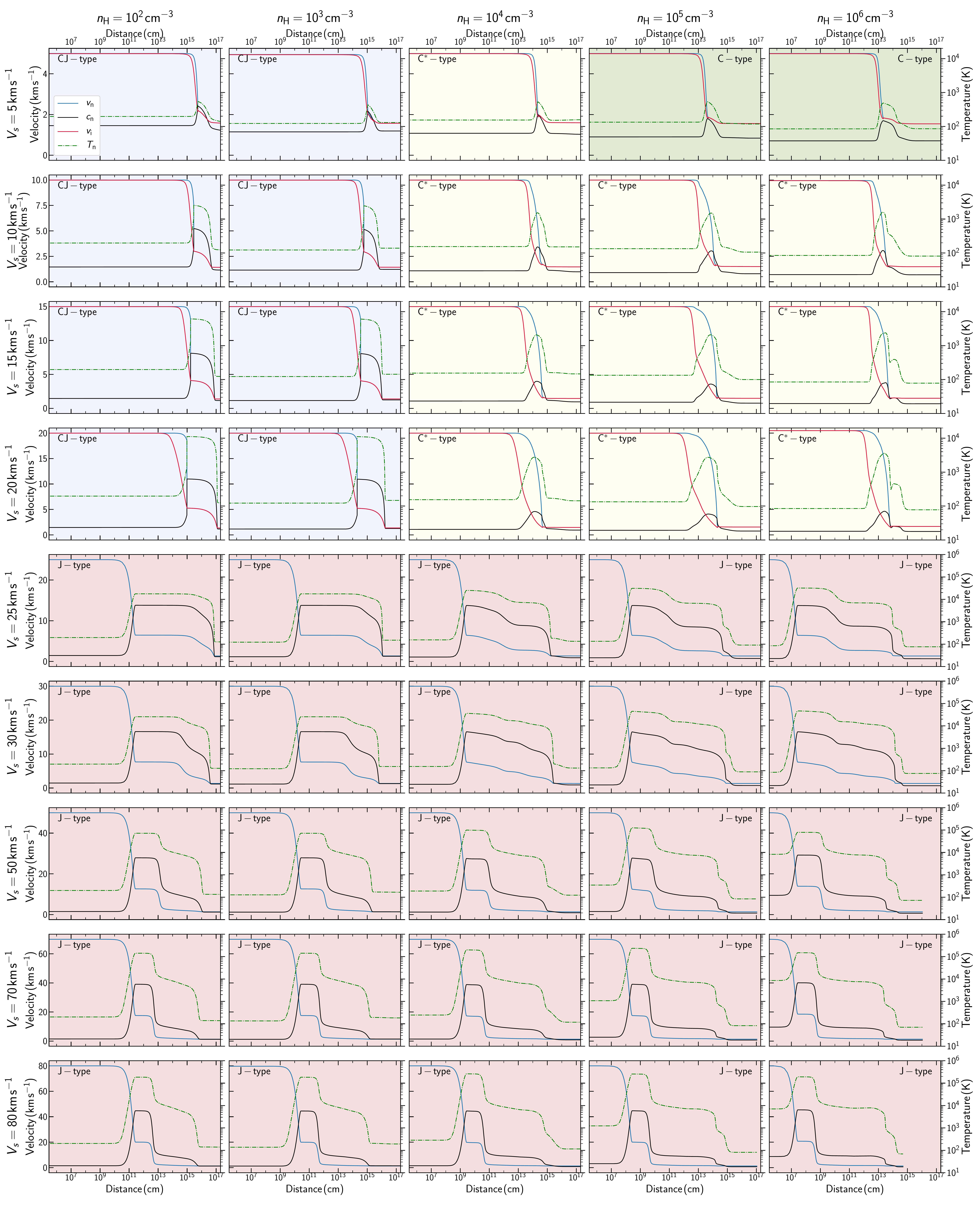}
  \caption{Same as Fig.~\ref{fig:temps_g0}, but for shocks with external radiation field $G_0=10^3$.}
  \label{fig:temps_g3}
\end{figure*}

The interstellar and circumgalactic medium displays an extraordinary diversity of physical conditions. In order to give a robust interpretation of observations we therefore model shocks over a broad range of conditions, including a range of densities, velocities, magnetic field strength, and external UV radiation fields. We consider shock velocities in the range 5-80~km~s$^{-1}$, with models every 5~km~s$^{-1}$ up to 30~km~s$^{-1}$ and then every 10~km~s$^{-1}$ up to 80~km~s$^{-1}$, with preshock densities ranging from $10^2$~cm$^{-3}$ every half dex to $10^6$~cm$^{-3}$. In this section we introduce three grids of shock models varying the magnetic parameter $b$ defining the preshock transverse magnetic field strength $B_0 = b \left(n_{\rm H}/{\rm cm^{-3}}\right)^{1/2}$, and the radiation parameter $G_0$, which is a scaling factor\footnote{The radiation parameter $G_0=1$ is equivalent to a photon flux of $1.55\times~10^8$~ph~s$^{-1}$~cm$^{-2}$ between 911~$\AA$ and 2400~$\AA$.} applied to the standard ultraviolet radiation field of \cite{mathis_interstellar_1983}. The three grids\footnote{The grids have been run on the computing cluster Totoro funded by the ERC Advanced Grant MIST.} have
\begin{itemize}
\item moderate magnetic field strength (magnetic parameter $b=1$) and no external radiation (radiation parameter $G_0=0$), with profiles shown in Fig.~\ref{fig:temps_g0},
\item weak magnetic field strength (magnetic parameter $b=0.1$) and no external radiation (radiation parameter $G_0=0$), with profiles shown in Appendix~\ref{app:g0b0}, 
\item moderate magnetic field strength (magnetic parameter $b=1$) and strong external radiation\footnote{The UV radiative energy flux is equal to the shock kinetic flux when $(n_{\rm H}/10^4 \, {\rm cm^{-3}} ) \left(V_s / 40 {\rm km \, s^{-1}} \right)^3=G_0/10^3$ (see Fig. 12 of \cite{godard_models_2019}).} (radiation parameter $G_0=10^3$), with profiles shown in Fig.~\ref{fig:temps_g3}.
\end{itemize}
In total, these grids comprise 297 shock models. The elemental abundances and chemical network for these models are the same as in Paper 1. The initial abundances of all chemical species are then computed by evolving a parcel of fixed density gas until thermal and chemical equilibrium for each of the 9 densities. For the models with external radiation, there is an additional computation where the parcel is slowly advected away from the source of radiation until the intervening material gives an attenuation of the UV field of $A_V=0.1$ magnitudes. The shock computation therefore begins with an initial state that depends on the density and radiation field considered. An analysis of the radiation field dependency for the initial temperature, electronic fraction, molecular fraction, and grain depletion can be found in Fig.~2 of \cite{godard_models_2019}.

Here we focus on the comparison of grids with or without external radiation, and leave the weak magnetic field models to Appendix~\ref{app:g0b0}. To prefigure the focus on CH$^+$ in this work, we present here its column densities in the shocks of these grids. The key shock parameters are summarised in Table~\ref{tab:shockparams}. 

\subsubsection{Shocks with no external UV irradiation: $G_0=0$}\label{sec:model_grids_g0}
Figure~\ref{fig:temps_g0} shows profiles of the neutral and ionized fluid velocities and neutral fluid temperature for the shocks with moderate magnetic field strength, $b=1$, and no external radiation field, $G_0=0$. For this grid the ion magnetosonic speed $c_{\rm ims}\sim 24$~km~s$^{-1}$, and so there is a separation of shock types into C- ($V_s\leq$20~km~s$^{-1}$) and J-type ($V_s\geq$25~km~s$^{-1}$). Following \cite{godard_models_2019}, both neutral and charged grains are assumed to be fully coupled with the charged fluid. The mass density of the charged fluid is dominated by dust grains rather than atomic or molecular ions. While the latter depends strongly on gas density and radiation field strength, the former does not, and so the ion magnetosonic speed is roughly constant and low compared to a dust-free gas \citep{guillet_shocks_2007}. The neutral and ion fluid temperatures and velocities are decoupled in the C-type shocks, whereas they act as a single fluid in J-type shocks. This decoupling has important chemical implications, for example the kinetic energy provided by the ion-neutral velocity difference boosts endothermic reactions. It should be noted that the logarithmic scale gives the impression that the J-type shocks are of comparable thickness to the C-type shocks. However, the lengthscales for the J-type shocks to cool down to 100~K range from $10^{14}$-$10^{16}$~cm, while for the C-type shocks they range from $10^{15}$-$10^{18}$~cm. The corresponding shock timescales are less dependent on shock type, with values ranging from 100~years at high density to 10$^5$~years for low density shocks.

The top panel of Fig~\ref{fig:chp_column} shows the CH$^+$ column densities through each shock in this grid, for the region of the shock before it cools down to a threshold temperature of 100~K. The column densities range from $3\times 10^3$ to $5\times 10^{13}$~cm$^{-2}$. For the J-type shocks ($V_s \geq 25$~km~s$^{-1}$), the column density increases with either density or velocity. There is a clear change in behaviour in the C-type shocks ($V_s \leq 20$~km~s$^{-1}$), with the column density being weakly dependent to velocity but inversely proportional to density.

\subsubsection{Shocks with external UV irradiation: $G_0=10^3$}\label{sec:model_grids_g3}
In the presence of an external UV field, we model the shocks as propagating perpendicularly to the source of UV. In practice this means that we take the external UV field to have been attenuated by a layer of gas with $A_V=0.1$ magnitudes, and the field is held constant at each point in the shock. Figure~\ref{fig:temps_g3} shows profiles of the neutral and ionized fluid velocities and neutral fluid temperature for the grid with $b=1$, $G_0=10^3$. The ion magnetosonic speed $c_{\rm ims}\sim 21-24$~km~s$^{-1}$ for these shocks, and so the shocks with velocity $V_s \geq 25$~km~s$^{-1}$ are all J-type as in the $G_0=0$ grid. Below this velocity we find the full variety of C-, C$^*$- and CJ-type shocks, in agreement with \cite{godard_models_2019}. Despite this variety of shock types, the lengthscales to cool down to 100~K are all similar for a given density, ranging from $10^{14}$-$10^{17}$~cm over the whole grid with no strong dependence on shock type. The corresponding shock timescales range from 100~years to 10$^5$~years.

\begin{figure}
\centering
  \includegraphics[width=\columnwidth]{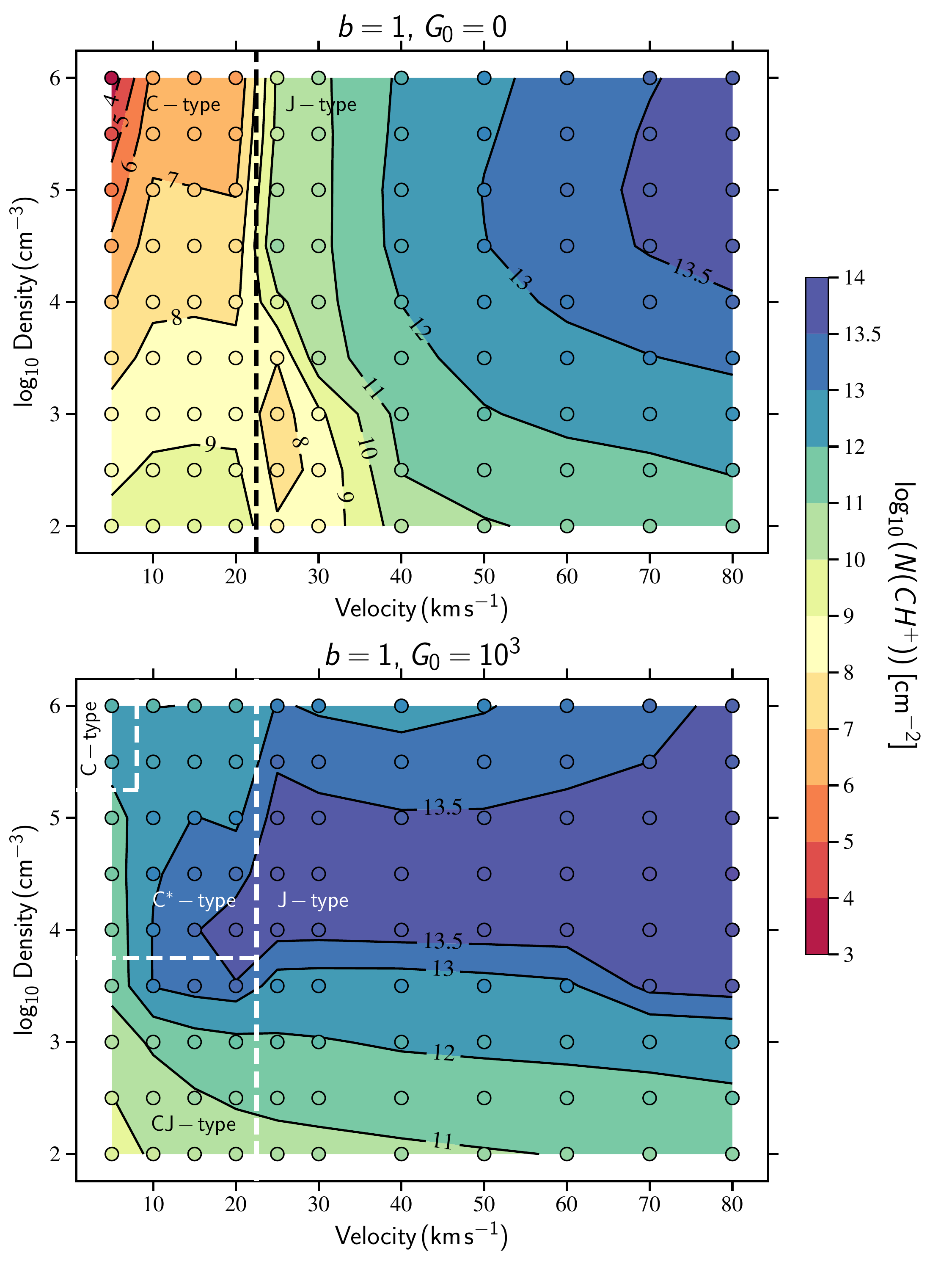}
  \caption{Column density of CH$^+$ for shocks with magnetic parameter $b=1$, and no external radiation field $G_0=0$ (top) or strong external radiation field $G_0=10^3$ (bottom). The dashed lines mark the boundaries between C-, C$^*$-, CJ-, and J-type shocks. The coloured circles show exact values for each model in the grids.}
  \label{fig:chp_column}
\end{figure}

The bottom panel of Fig~\ref{fig:chp_column} shows the CH$^+$ column densities through each shock in this grid, for the region of the shock before it cools down to a threshold temperature of 100~K. In shocks where the external radiation field is able to maintain a temperature above 100~K in the postshock, we instead take the temperature threshold to be 10\% larger than the final temperature. The column densities range from $6\times 10^9$ to $8\times 10^{13}$~cm$^{-2}$. The column density roughly increases with preshock density until $n_{\rm H}=10^4$~cm$^{-3}$ before dropping off at densities larger than $10^5$~cm$^{-3}$. There is a surprising continuity across the border between C$^*$- and J-type shocks.

\subsubsection{Comparison of shock grids}\label{sec:model_grids_comparison}
A consequence of the broad parameter space that we explore is the occurence of the full diversity of MHD shock types across the grids of models. All MHD shock types, J-, C-, C$^*$-, and CJ-type, arise in the grid with moderate magnetic field strength and a strong external radiation field. The UV field increases the ionisation fraction and hence the ion-neutral collisional heating rate, while strongly photodissociating molecules like H$_2$ and CO, reducing the cooling rate. Thus the cooling cannot keep the local sound speed below the fluid velocity in sections of the parameter space and a transition takes place across the sonic point giving C$^*$-, and CJ-type shock solutions. Taking away the radiation field removes this possibility, and the shock grid contains just C- and J-type shocks. Finally, taking a weak magnetic field (as in the grid in Appendix~\ref{app:g0b0}) reduces the total fluid ion-magnetosonic speed to below 5~km~s$^{-1}$, and so that shock grid contains just J-type shocks.

The peak temperature for the J-shocks is independent of preshock density and temperature. It is accurately estimated using the Rankine-Hugoniot jump conditions, assuming an adiabatic index of $5/3$ and shock velocity much larger than the Alfv\'en velocity ($v_{\rm A}=B/\sqrt{4\pi\rho}$ for fluid mass density $\rho$), as
\begin{align}\label{eq:rankineT}
T_{\rm{max}} \sim 53 \, \rm{K} \, \left(\frac{\mu}{\rm 2.33 \, a.m.u.} \right) \left( \frac{V_s}{\rm{km~s^{-1}}} \right)^2,
\end{align}
where $\mu$ is the mean particle mass in a.m.u. In a fully molecular medium, $\mu=2.33$~a.m.u. and we recover the peak temperatures in the J-type shocks in all the shocks with no external irradiation. In the shocks with external irradiation, the picture is more complex. At densities lower than $10^{3.5}$~cm$^{-3}$ the external UV field fully photodissociates H$_2$ and the atomic gas has mean mass $\mu \sim 1.3$~a.m.u. At densities larger than $10^5$~cm$^{-3}$ the H$_2$ is fully self-shielded and $\mu \sim 2.33$~a.m.u. However, at high velocities ($V_s \geq 50$~km~s$^{-1}$) the combined external and self-generated UV field is strong enough in the radiative precursor to fully dissociate H$_2$ and the preshock fluid enters the shock front as an atomic medium with mean particle mass $\mu \sim 1.3$~a.m.u. Hence the peak temperatures in these externally irradiated J-type shocks can be up to $\sim$2 times lower than the same shocks with no external irradiation. In addition, H$_2$ is a key coolant in a warm, fully molecular gas. Hence the preshock gas in the radiative precursors with UV fields that fully dissociate H$_2$ are heated to $10^4$~K, two orders of magnitude larger than any preshock temperatures in shocks with no external irradiation.

The chemistry controlling the production of CH$^+$ in shocks has been explained in detail in \cite{godard_models_2019}. In short, CH$^+$ formation requires the presence of H$_2$ and is linked to the available C$^+$, while its destruction is dominated by H or H$_2$. The C$^+$ abundance is enhanced by UV photoionisation of C. Hence the shocks with the external radiation field produce larger column densities of CH$^+$ than those shocks with no external radiation field at all densities and velocities. This is especially true at low velocities, e.g. $V_S \leq 25$~km~s$^{-1}$, where the column densities of CH$^+$ are boosted by up to 8 orders of magnitude with an external radiation field.

\section{Radiative Transfer of CH$^+$} \label{sec:RT}
The formation and destruction pathways of CH$^+$ are described in detail in \cite{godard_models_2019}. We note here that the shock code calculates the formation rate taking into account the excited rovibrational states of H$_2$. In this section we investigate the line emission from CH$^+$ generated in these shocks, which we calculate using an updated version of the Large Velocity Gradient (LVG) code of \cite{gusdorf_sio_2008}. We first summarise the details of the excitation and radiative transfer calculation. Then we show how the character of the emission is affected by, for example, the shock type, preshock density, shock velocities, and presence or absence of an external radiation field.

\subsection{Excitation processes} \label{sec:RT_LVG}
The full details of the included processes, extrapolations, and testing of the excitation and radiative transfer of CH$^+$ are given in \cite{neufeld_2021}, but we give a summary here for completeness. Updated Einstein $A$ coefficients are taken from \cite{changala_2021}. We include the excitation to ro-vibrational states by nonreactive collisions with e$^-$ \citep{hamilton_2016}, H \citep{faure_2017}, and He \citep{hammami_2009}. Rates for collisions with H$_2$ are deduced from the CH$^+$-He data within the framework of the rigid rotor approximation as done by \cite{godard_2013}. CH$^+$ can also form in an excited state in the reaction
\begin{align}
{\rm C}^+ + {\rm H}_2 \to {\rm CH}^+(v,j) + {\rm H},
\end{align}
which is the dominant formation pathway of CH$^+$. Since CH$^+$ is short lived it is important to include this formation pumping in the excitation model. Furthermore, since the destruction in the reverse reaction occurs on timescales longer than its radiative decay, chemical de-pumping is not necessary to include. Radiative pumping in the infrared rovibrational lines is modeled within the LVG approximation. Radiative pumping in the optical electronic lines followed by a fluorescent cascade is modeled taking into account self-shielding processes using the FGK approximation \citep{federman_1979}.

With all these excitation processes in place, the radiative transfer is computed from the outputs of the Paris-Durham shock code, e.g. the local temperature, ion fluid velocity, sound speed, CH$^+$ density, and densities of collisional partners. This calculation is an update on that in \cite{godard_models_2019}, with the main differences being the CH$^+$-e$^-$ collisional rates that are orders of magnitude different at low temperatures and the inclusion of chemical pumping and the impinging isotropic radiation field. In the end the only free parameters are the number of CH$^+$ rovibrational levels to include in the computation and the microturbulent velocity dispersion, $\sigma_{\rm turb}$, which we set to 100 levels and 2.5~km~s$^{-1}$, respectively, for the rest of this work.

\begin{table}
 \caption{A selection of CH$^+$ pure rotational ($v=0-0$) line data}
 
 \centering
 \begin{tabular}{l c c c c} \hline \hline
 \multicolumn{5}{c}{\vspace{-0.3cm}} \\
 Line & $A_{ij}$~(s$^{-1}$) & $\nu$~(GHz) & $E_u$~(K) & $\lambda$~($\mu$m) \\ \cline{1-5}
 \multicolumn{5}{c}{\vspace{-0.3cm}} \\
    $J=1-0$  & $6.40(-03)$  & 835.0997 & 40.078 & 359.0\\
    $J=2-1$  & $6.13(-02)$  & 1669.210 & 120.19 & 179.6\\
    $J=3-2$  & $2.21(-01)$  & 2501.344 & 240.23 & 119.9\\
    $J=4-3$  & $5.41(-01)$  & 3330.520 & 400.07 & 90.01\\
    $J=5-4$  & $1.08(+00)$  & 4155.763 & 599.52 & 72.14\\
    $J=6-5$  & $1.87(+00)$  & 4976.106 & 838.33 & 60.25\\
    $J=7-6$  & $2.99(+00)$  & 5790.597 & 1116.2 & 51.77\\
    $J=8-7$  & $4.46(+00)$  & 6598.296 & 1432.9 & 45.43\\
    $J=9-8$  & $6.32(+00)$  & 7398.279 & 1788.0 & 40.52\\
    $J=10-9$ & $8.62(+00)$  & 8189.644 & 2181.0 & 36.60\\
 \hline \vspace{-0.2cm}
 \end{tabular}
 \tablefoot{Columns give the transition label, Einstein coefficient, transition frequency, energy of the upper level, and transition wavelength.}
 \label{tab:chp_data}
\end{table}

\subsection{CH$^+$ emission line profiles} \label{sec:RT_chp_profiles}
\begin{figure}
\centering
  \includegraphics[width=\columnwidth]{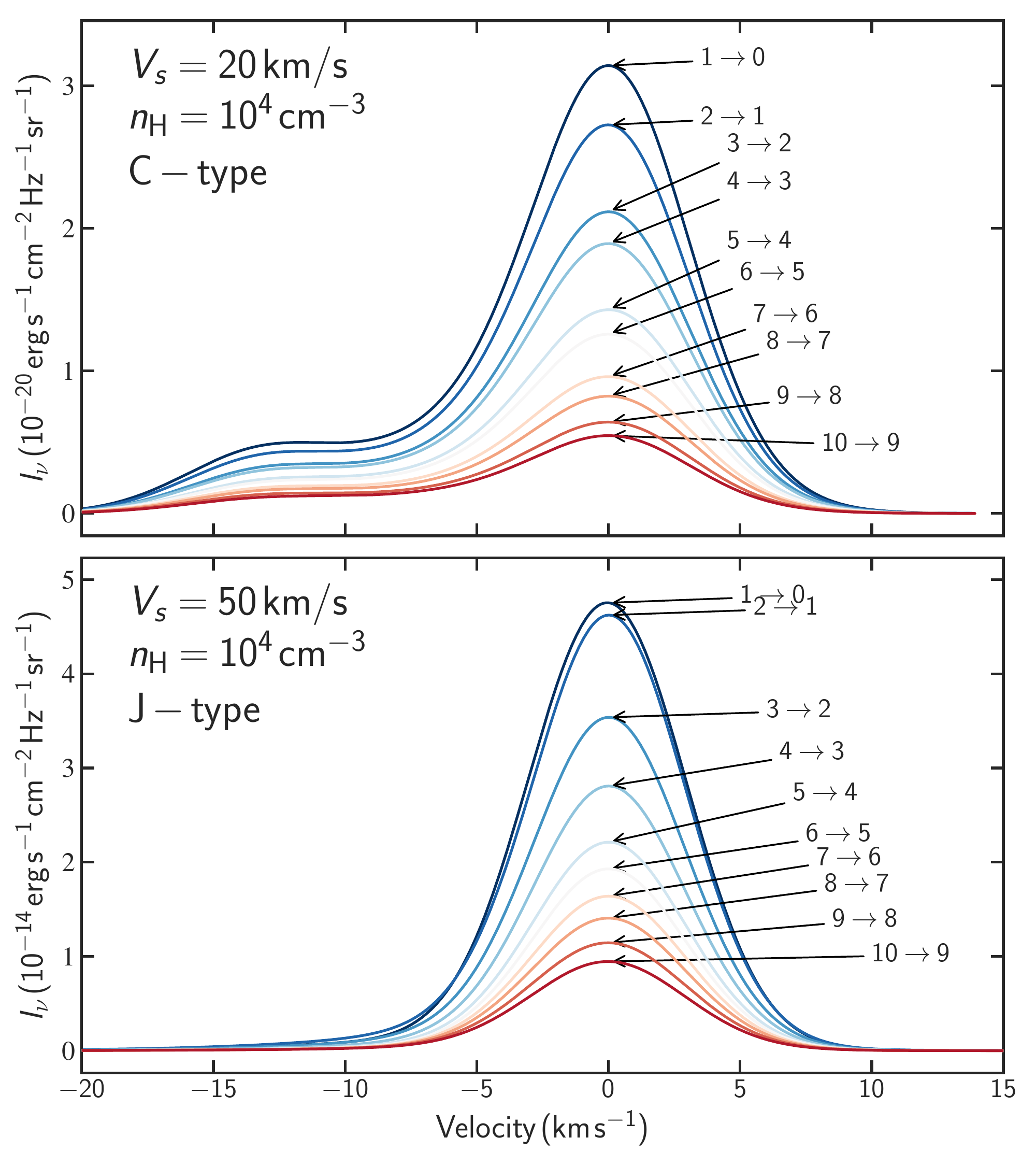}
  \caption{Specific intensities of CH$^+$ pure rotational lines $v=0-0$~$J=n\to n-1$ up to $J_{\rm up}=10$ emitted parallel to the propagation direction of a (top) C-type shock with velocity $V_s=20$~km~s$^{-1}$ or (bottom) J-type shock with velocity $V_s=50$~km~s$^{-1}$, both with preshock density $10^4$~cm$^{-3}$, magnetic parameter $b=1$, and no external radiation field $G_0=0$.}
  \label{fig:chp_intensities}
\end{figure}

In Fig~\ref{fig:chp_intensities} we show the specific intensity line profiles for the pure rotational lines of CH$^+$ $J=n\to n-1$ up to $J_{\rm up}=10$ emitted parallel to the shock propagation direction for a $20$~km/s C-type shock (upper panel) and a $50$~km/s J-type shock (lower panel) with no external radiation field. The Einstein $A$ coefficients, central line frequency and upper energy level of these lines are given in Table~\ref{tab:chp_data}. The emission from the C-type shock gives a skewed line profile, showing emission at velocities between the shock velocity and 0~km~s$^{-1}$. This reflects the character of C-type shocks in which the gas is heated and decelerated smoothly. The line shape near the peak is given by the microturbulent velocity dispersion, because the peak CH$^+$ production and excitation occurs where the thermal velocity dispersion is less than $\sigma_{\rm turb} = 2.5$~km~s$^{-1}$. In contrast, deceleration in the J-type shock occurs in the high temperature viscous jump that strongly dissociates all molecules. The CH$^+$ emission therefore uniquely arises from the region where the shock has cooled, the velocity is $\sim 0$~km~s$^{-1}$, and molecules have reformed. This region has cooled down to a temperature $\sim 10^3$~K and so the thermal velocity dispersion is also smaller than $\sigma_{\rm turb}$. 

\subsection{CH$^+$ integrated line intensities} \label{sec:RT_chp_fluxes}
\begin{figure}
\centering
  \includegraphics[width=\columnwidth]{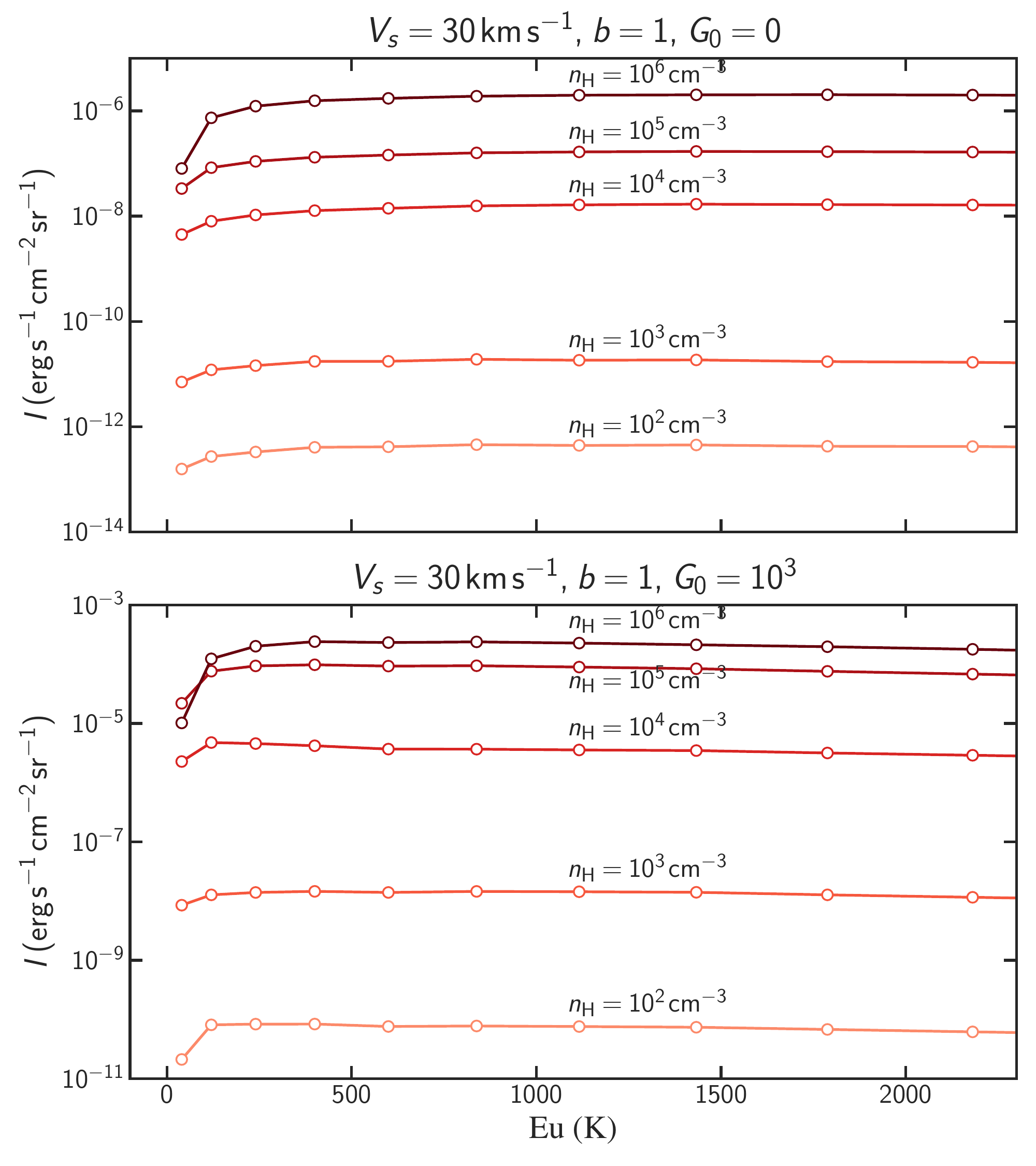}
  \caption{Line intensities of CH$^+$ $J=n\to n-1$ up to $J_{\rm up}=10$ emitted parallel to the propagation direction of shocks with velocity $V_s=30$~km/s, densities $n_{\rm H} = 10^2$-$10^6$~cm$^{-3}$, and radiation parameter $G_0=0$ (top) or $G_0=10^3$ (bottom). Note that each figure has a different $y$-axis scale.}
  \label{fig:chp_fluxes_vs_density}
\end{figure}

Figure~\ref{fig:chp_fluxes_vs_density} shows the integrated line intensities of the pure rotational lines of CH$^+$ up to $J=10\to 9$, emitted parallel to the shock propagation direction, for shocks with fixed velocity $V_s=30$~km~s$^{-1}$ over the density range, comparing the shocks with no external radiation field (upper panel) to those with an external radiation field $G_0=10^3$ (lower panel). Externally irradiated shocks produce CH$^+$ much more efficiently than non-irradiated shocks, and so their line intensities at a given density are always orders of magntitude stronger than shocks with no external UV field. The intensities for a given shock are remarkably similar over the rotational ladder, meaning an observation of multiple CH$^+$ pure rotational lines does not distinguish different preshock densities.

\begin{figure}
\centering
  \includegraphics[width=\columnwidth]{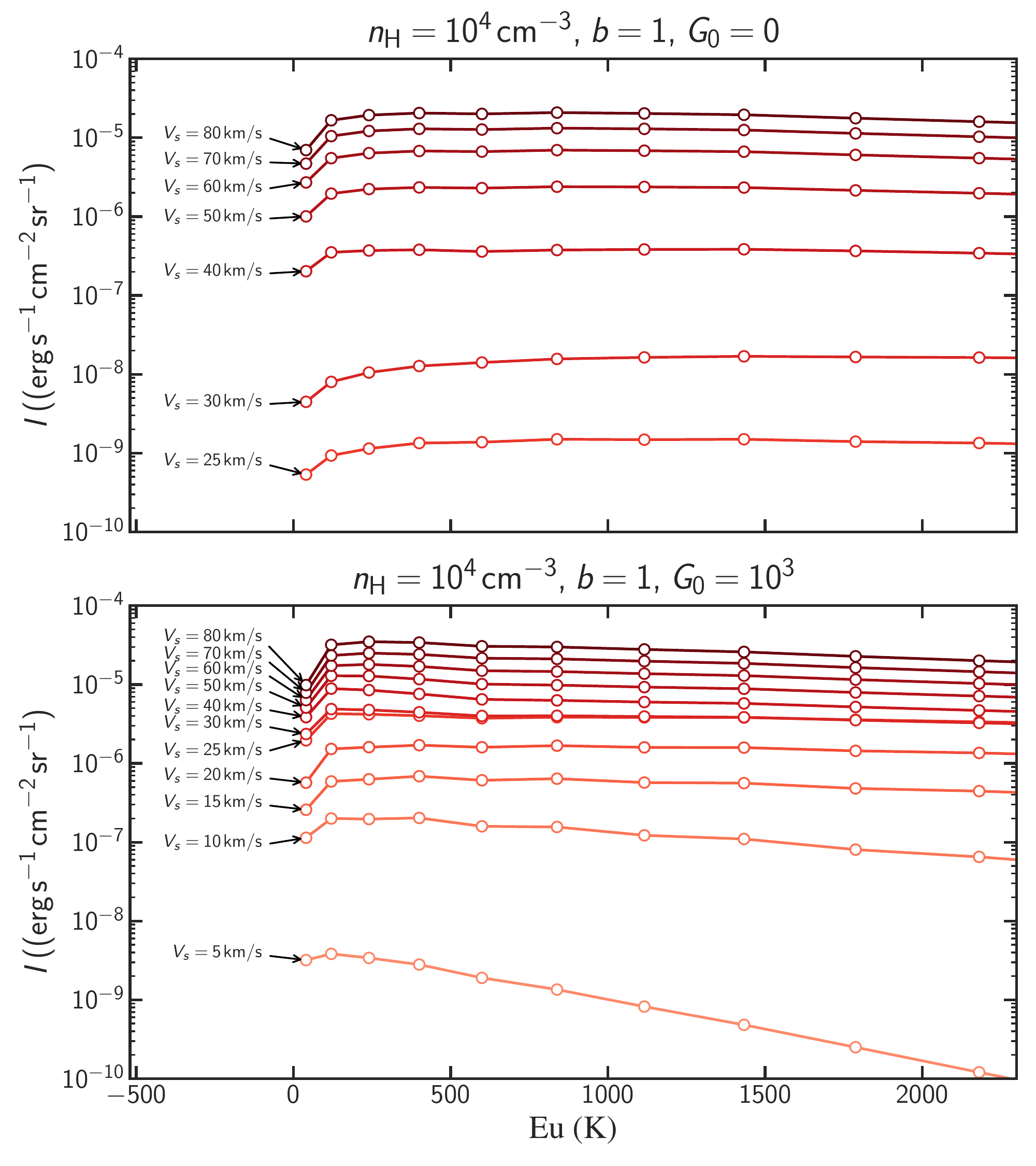}
  \caption{Line intensities of CH$^+$ $J=n\to n-1$ up to $J_{\rm up}=10$ emitted parallel to the propagation direction of shocks with velocities $V_s=5$-$80$~km/s, density $n_{\rm H} = 10^4$~cm$^{-3}$, and radiation parameter $G_0=0$ (top) or $G_0=10^3$ (bottom).}
  \label{fig:chp_fluxes_vs_velocity}
\end{figure}

Figure~\ref{fig:chp_fluxes_vs_velocity} shows the integrated line intensities of the pure rotational lines of CH$^+$ up to $J=10\to 9$ for shocks with fixed density $n_{\rm H}=10^4$~cm$^{-3}$ over the velocity range, comparing the shocks with no external radiation field (upper panel) to those with an external radiation field $G_0=10^3$ (lower panel). In the upper panel there is strong CH$^+$ emission for shocks with velocity $V_s \geq 40$~km~s$^{-1}$. At these velocities the shocks generate a strong Ly$\alpha$ flux, emphasising the need for a UV radiation field to initiate the CH$^+$ chemistry by ionising C. With such a UV field provided externally, the low velocity shocks in the lower panel produce just as much CH$^+$ emission at shock velocity $V_s =25$~km~s$^{-1}$ as the non-irradiated shocks with shock velocity $V_s=60$~km~s$^{-1}$. With the exception of the lowest velocity shocks that give negligible CH$^+$ emission, the intensities for a given shock are again remarkably similar. Hence the CH$^+$ rotational ladder does not strongly trace the shock velocity.

\begin{figure}
\centering
  \includegraphics[width=\columnwidth]{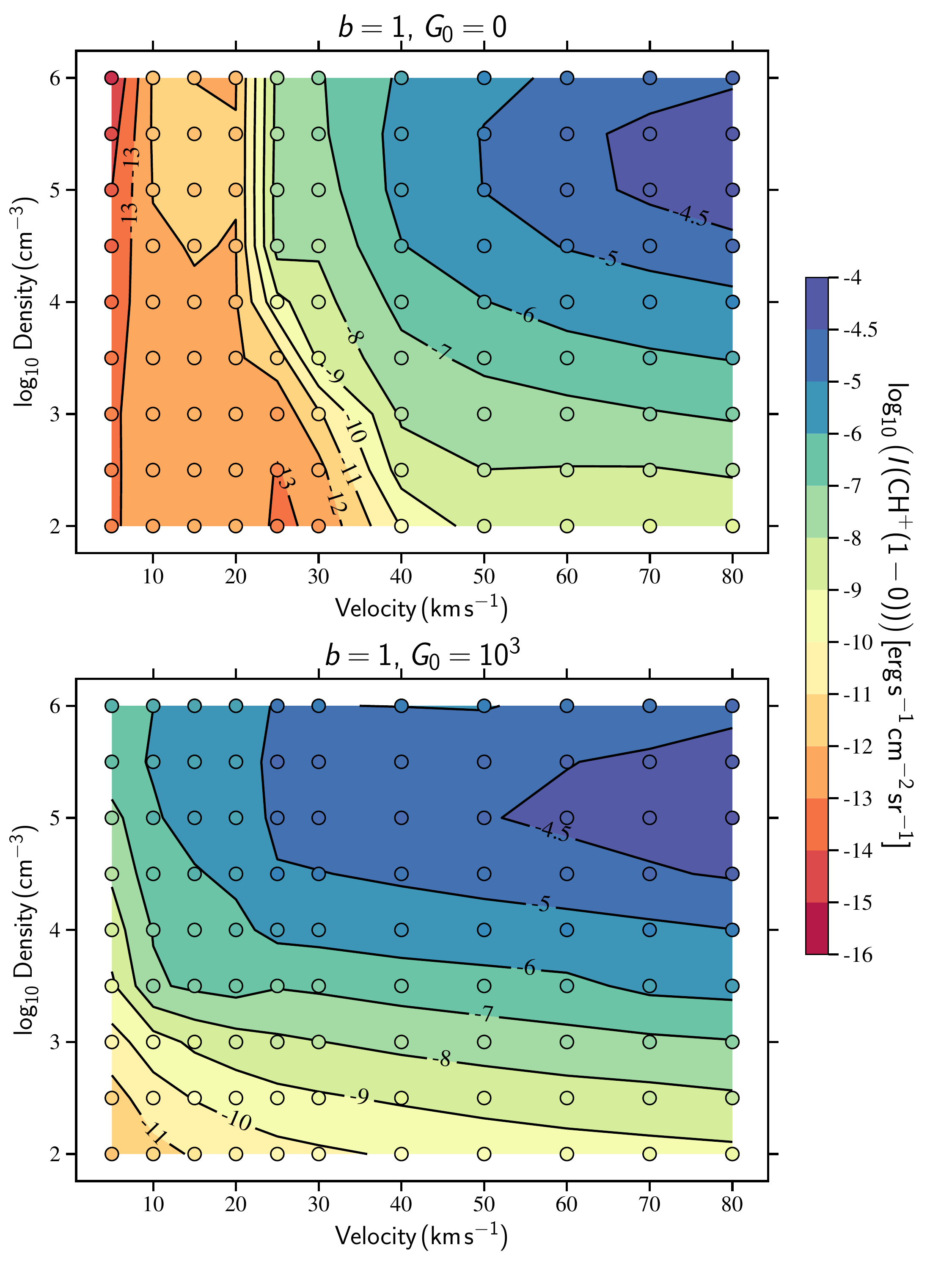}
  \caption{CH$^+$(1-0) line intensity for shocks with magnetic parameter $b=1$, and radiation parameter $G_0=0$ (top) or $G_0=10^3$ (bottom).}
  \label{fig:chp_1to0}
\end{figure}

The 1-0 line of CH$^+$ is the most commonly observed line, and so in Fig.~\ref{fig:chp_1to0} we show the line intensity over the entire grid of shocks with no external radiation field (upper panel) and for externally irradiated shocks (lower panel). This line is generally brighter in externally irradiated shocks over the whole parameter space, and is thus a good tracer for the presence of a strong UV field. In particular at low velocities ($V_s \leq 30$~km~s$^{-1}$) the intensity is up to 7 orders of magnitude stronger in externally irradiated shocks than shocks with no external UV field.

\begin{figure}
\centering
  \includegraphics[width=\columnwidth]{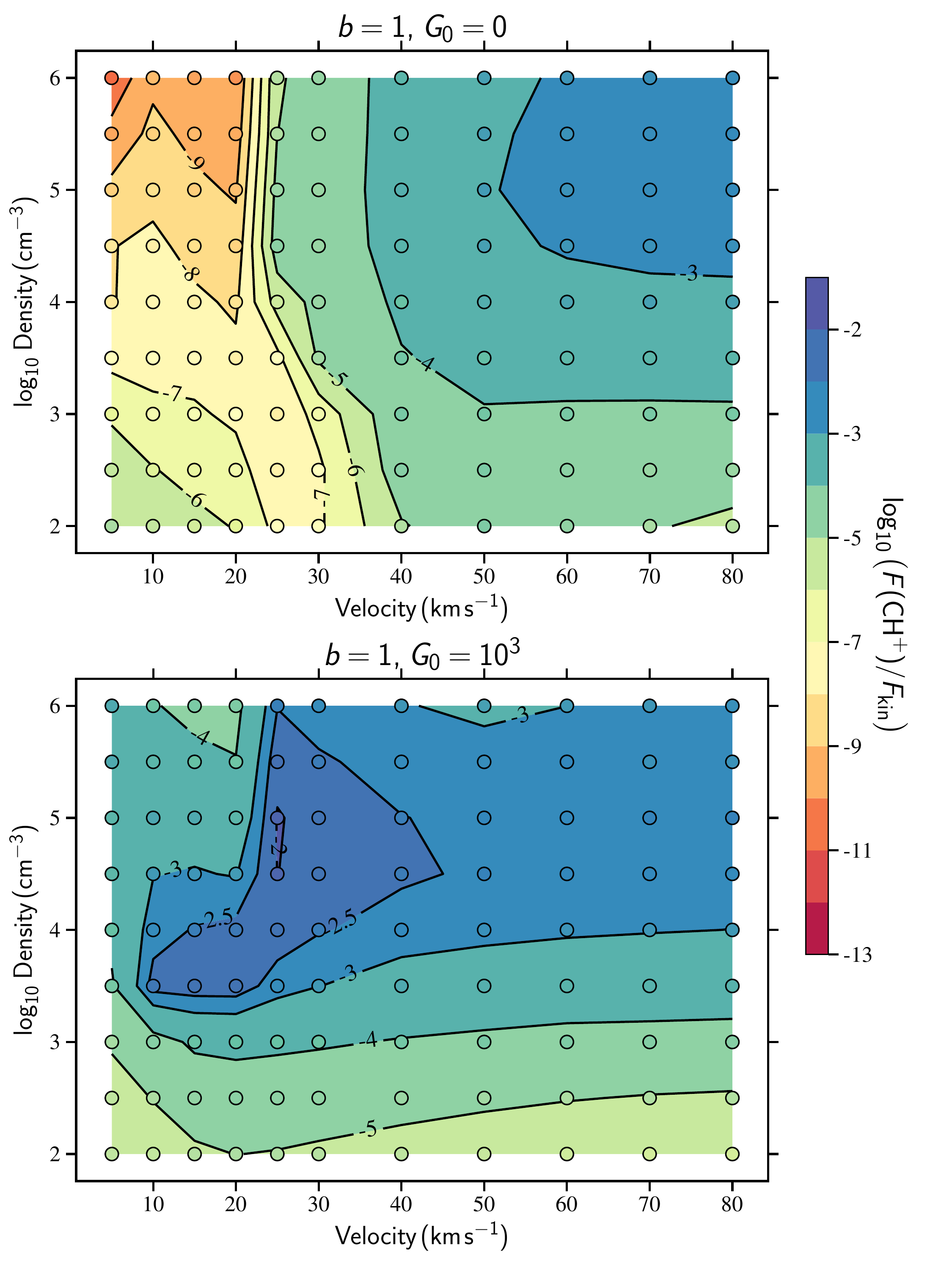}
  \caption{Total CH$^+$ flux as a fraction of shock kinetic flux for shocks with magnetic parameter $b=1$, and radiation parameter $G_0=0$ (top) or $G_0=10^3$ (bottom).}
  \label{fig:chp_energy_fraction}
\end{figure}

In Fig.~\ref{fig:chp_energy_fraction} we show the fraction of the shock kinetic flux reprocessed into all CH$^+$ rovibrational lines for shocks with no external radiation field (upper panel) and for externally irradiated shocks (lower panel). These panels show that CH$^+$ can be a significant pathway of the reprocessing of mechanical energy in shocks. In shocks with an external radiation field, 
up to 1 percent of the shock kinetic flux can escape in the rovibrational lines of CH$^+$.

\section{Ensembles of shocks}\label{sec:ensemble}

\begin{figure*}[h]
\centering
  \includegraphics[width=0.8\textwidth]{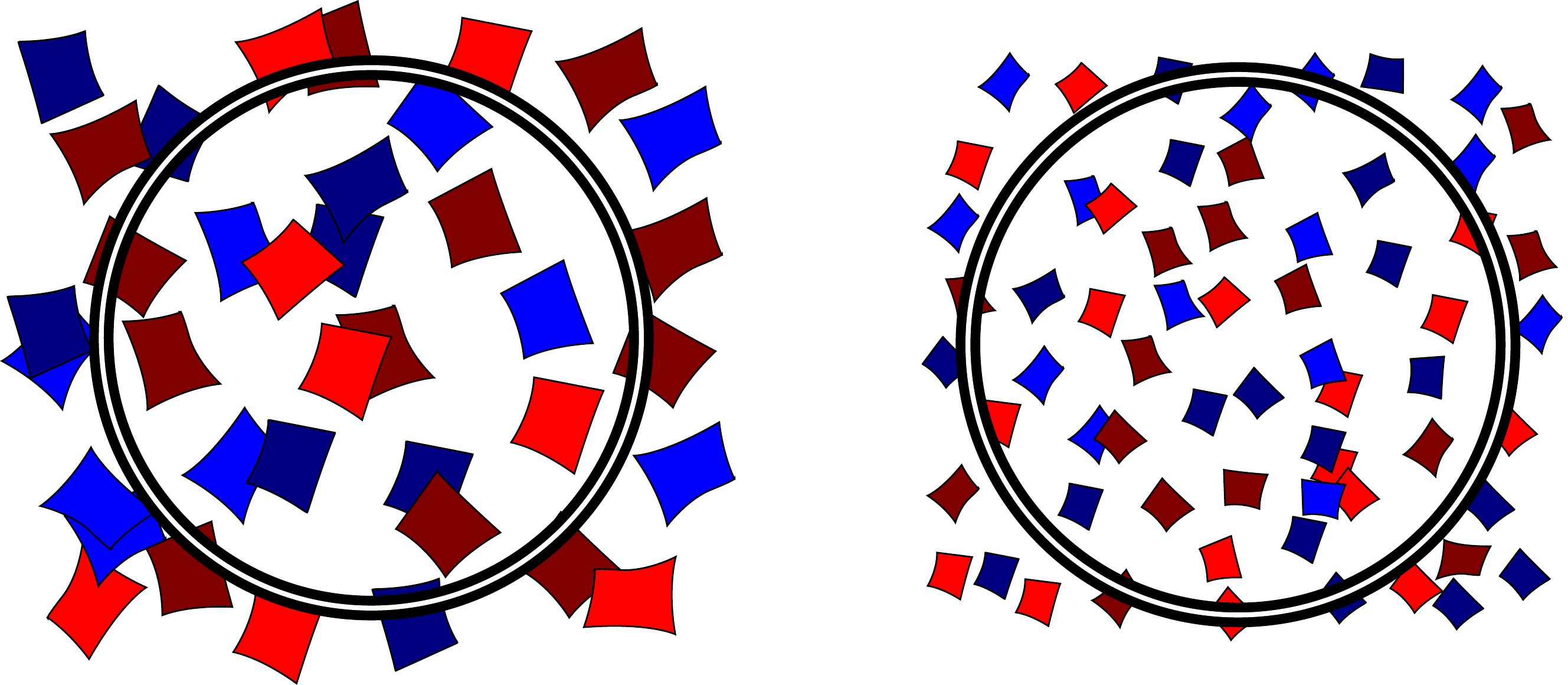}
  \caption{Cartoon of ensemble of blue and redshifted shocks within an observational solid angle (black and white circle). The left and right represent two scenarios with the same total shocked surface area, but different surface area per shock.}
  \label{fig:cartoon}
\end{figure*}

\begin{figure*}
\centering
  \includegraphics[width=\textwidth]{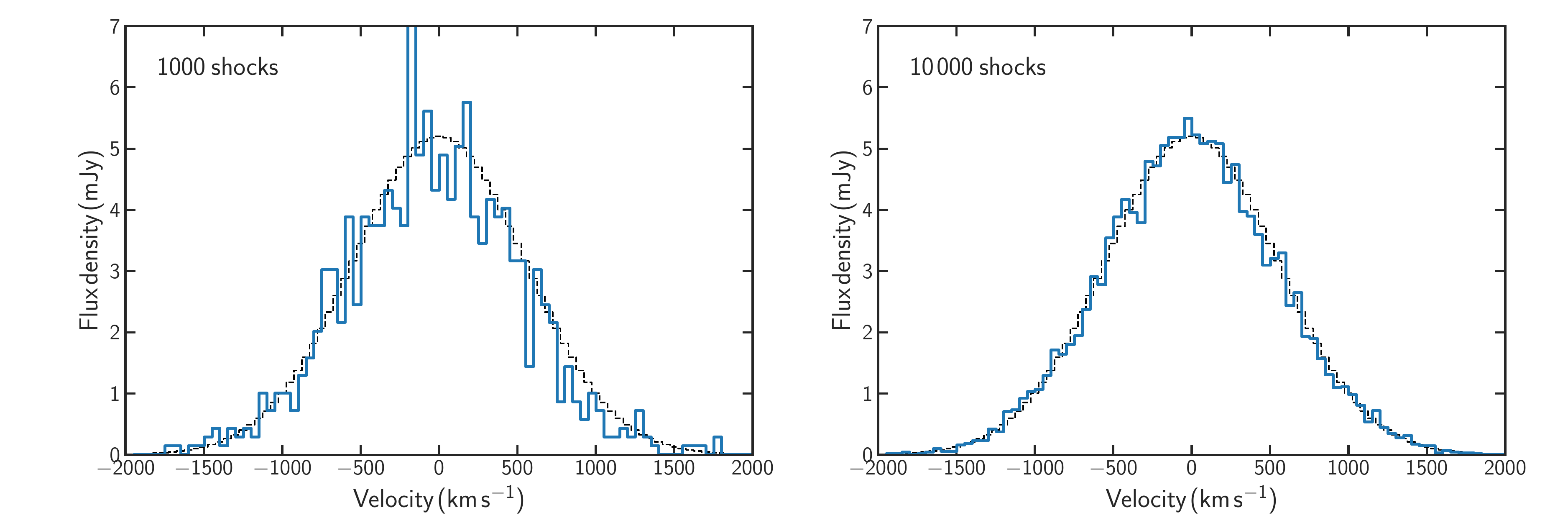}
  \caption{CH$^+$ (1-0) flux density for the Eyelash galaxy from ensembles of 1000 (left) or 10000 (right) shocks each with velocity $V_s$=50~km~s$^{-1}$ and preshock density $n_{\rm H}$=10$^4$~cm$^{-3}$. The peak of the average ensemble (black) has been fitted to $5.2 \times 10^{-26}$~erg~s$^{-1}$~cm$^{-2}$~Hz$^{-1}$, and the particular ensembles (blue) have resolution $\Delta W = 50$~km~s$^{-1}$.}
  \label{fig:distr_shocks}
\end{figure*}

Observations of extragalactic molecular line emission \citep[e.g.][]{guillard_2009, falgarone_2017} with extremely large linewidths ($\sim 1000$~km/s), to first approximation, imply temperatures at which such molecules cannot exist. These observations have inspired an explanatory framework in which energy injected at large scales enters a turbulent cascade that dissipates in structures at smaller scales and much lower velocities. Here we make a first step in using this framework to make quantitative predictions of the mechanical energy and physical conditions of an observed system, assuming these dissipative structures are MHD shocks.

We consider a toy model in which the emission is due to shocks with velocities much lower than the observed linewidth. In particular, the velocities are low enough to produce molecular emission lines, which are then spread over the observed velocity range by bulk flows. We stress that the emission model we develop can be used with any line whose radiative transfer can be calculated in these shocks. As an illustrative example we apply the model to a particular observation, the CH$^+$(1-0) line from the high-redshift starburst galaxy The Cosmic Eyelash. We show how to use the ensemble model with one shock output, giving a prediction of the total rate of kinetic energy dissipated in shocks and the minimum number of shocks needed in the observational solid angle. Finally we show how a grid of shock models can then be used to constrain physical conditions and mechanical energy of an observed system.

\subsection{Toy model}\label{sec:ensemble_model}

\begin{table*}
 \centering
 \caption{Symbols used in the ensemble model.}
 \begin{tabular}{l c c} \hline \hline
 \multicolumn{3}{c}{\vspace{-0.3cm}} \\
 Parameter & Symbol & Units \\ \cline{1-3}
 \multicolumn{3}{c}{\vspace{-0.3cm}} \\
 Total number of shocks                               & $\mathcal{N}$ & - \\
 Observational solid angle                            & $\Omega$ & sr \\ 
 Single shock solid angle                             & $\Omega_1$ & sr \\ 
 Single shock tophat specific intensity               & $I_{\nu,1}$ & erg~s$^{-1}$~cm$^{-2}$~Hz$^{-1}$~sr$^{-1}$ \\ 
 Single shock peak optical depth                      & $\tau_1$ & - \\
 Background continuum specific intensity              & $I_{\nu,c}$ & erg~s$^{-1}$~cm$^{-2}$~Hz$^{-1}$~sr$^{-1}$ \\ 
 Ensemble flux density                                & $F_\nu (V)$ & erg~s$^{-1}$~cm$^{-2}$~Hz$^{-1}$\\ 
 Gaussian distribution                                & $f(V,\sigma)$ & (km~s$^{-1}$)$^{-1}$ \\
 Emission line velocity dispersion                    & $\sigma$ & km~s$^{-1}$ \\
 Velocity bin size                                    & $\Delta V$ & km~s$^{-1}$ \\
 Number of shocks in velocity bin at $V$              & $N$(V) & - \\
 Average number of shocks in velocity bin             & $\bar{N}$(V) & - \\
 Average number of overlapping shocks in velocity bin & $\bar{N}_{\rm S}$(V) & - \\
 Standard deviation of shock number in velocity bin   & $\sigma_{\Delta V}$ & \\
 \hline \vspace{-0.2cm}
 \end{tabular}\label{tab:summary_ensemble}
\end{table*}

A cartoon of our toy model is shown in Fig.~\ref{fig:cartoon}, in which $\mathcal{N}$ shock surfaces are captured within an observational solid angle $\Omega$. We consider each of the shocks in the ensemble to have the same shock velocity $V_S$, preshock density, and to subtend the same solid angle, $\Omega_1$, on the plane of the sky. This model is a highly simplified representation of a large-scale injection of energy (for example the infall/outflow of material onto galaxies like the Eyelash, or the galaxy collision in Stephan's Quintet) exciting a dominant shock velocity in clumps dispersed according to a Gaussian distribution in velocity-space with linewidth equal to the observed linewidth.

Each shock is randomly placed in the observational solid angle. The specific intensity of any line emitted by one shock (e.g. see Fig.~\ref{fig:chp_intensities}) is centered at a velocity $V$ randomly drawn from a Gaussian distribution. For simplicity we replace the shock emission profile with a tophat profile with linewidth $\Delta \nu$, line intensity $I$ (e.g. see Fig.~\ref{fig:chp_1to0}), and optical depth  $\tau_1$. The amplitude of the tophat specific intensity of one shock is then $I_{\nu,1} = I/\Delta \nu$. $\Delta \nu$ is derived from the approximate velocity width $\Delta V$ of a shock line profile, found in Sect.~\ref{sec:RT_chp_profiles} to be dominated by the microturbulent velocity dispersion, and set to $\Delta V=2\sigma_{\rm turb}=5$ km~s$^{-1}$. Two shocks placed at velocities within this linewidth will both contribute to the intensity at $V$. In this bin, if there is no spatial overlap of shock surfaces then we model the ensemble flux density with
\begin{align} \label{eq:emission1}
F_\nu (V) &= \overbrace{I_{\nu,c} \left( \Omega - N(V) \Omega_1 \right)}^{\rm unattenuated \, background} + \overbrace{I_{\nu,c} \exp \left(-\tau_1\right) N(V) \Omega_1}^{\rm attenuated \, background} + \overbrace{I_{\nu,1} N(V) \Omega_1}^{\rm shock \, emission},
\end{align}
where $N(V)$ is the number of shocks in the velocity bin of size $\Delta V$ and $I_{\nu,c}$ is the background continuum specific intensity. The first term gives the background emission that passes through unattenuated by shock slabs at velocity $V$, the second term gives the background emission attenuated by a single shock layer scaled by the number of shocks in the bin, and the third term gives the emission generated in the shocks themselves. On the other hand, if there are many shocks overlapping in both velocity and space, the ensemble emission is given instead by
\begin{align} \label{eq:emission2}
F_\nu(V) = \overbrace{I_{\nu,c} \exp \left( - N(V) \tau_1\right) \Omega}^{\rm attenuated \, background} + \overbrace{I_{\nu,1} \Omega_1 \sum_{n=0}^{N(V)-1}\exp\left(-n \tau_1 \right) }^{\rm shock \, emission},
\end{align}
where now there is no unattenuated background. The first term gives the background attenuated by as many shock layers as there are in the velocity bin and the second term gives the emission generated in the shocks attenuated by the other shock layers between them and the observer in the same velocity bin. Equation~\eqref{eq:emission2} is given here for completeness, but we find for our application in the next section that we can always use Eq.~\eqref{eq:emission1}. This must be verified for each application, and can be decided with the following considerations.

The single shock peak emission is shifted to velocities randomly drawn according to a Gaussian distribution
\begin{align}
f(V,\sigma) = \frac{1}{\sigma\sqrt{2\pi}} \exp\left( - \frac{V^2}{2\sigma^2} \right).
\end{align}
After constructing many ensembles, the number of shocks in the velocity bin around $V$ will follow a Poisson distribution with average given by
\begin{align}\label{eq:overlap_vel}
\bar{N}\left( V\right) = \mathcal{N} \Delta V f(V,\sigma),
\end{align}
and standard deviation
\begin{align}
\sigma_{\Delta V} \left(V \right) = \sqrt{ \bar{N}\left( V \right) }.
\end{align}
Equation~\eqref{eq:emission1} can only be used if the shocks within a velocity bin are not overlapping in space. This can be verified by making sure the typical number of shocks overlapping in velocity and space,
\begin{align}\label{eq:overlap_vel_space}
\bar{N}_{\rm S}\left( V \right) = \bar{N}\left( V \right) \frac{\Omega_1}{\Omega} < 1,
\end{align}
otherwise we use Eq.~\eqref{eq:emission2}. It is worth noting that if the cumulative optical depth through the shock layers, $\bar{N}_{\rm S}\left( V \right)\tau_1$, is small, then Eq.~\eqref{eq:emission2} reduces to Eq.~\eqref{eq:emission1}. That is, both the background and shock emission are unattenuated by the shock layers in the ensemble. These conditions need to be checked when computing the shock ensemble emission to reproduce a particular observation.

\subsection{Example application: The Eyelash Starburst Galaxy}\label{sec:ensemble_app1}
In this section we show how the ensemble toy model can be used to extract information from an extragalactic observation of extremely broad CH$^+$(1-0) emission line towards the Eyelash galaxy. To apply the ensemble model, we need the peak value of the background subtracted flux density integrated over the galaxy $F_{\nu,p}^{\rm obs}$, the observational solid angle of the galaxy $\Omega_{\rm obs}$, the background continuum flux density $F_{\nu,c}^{\rm obs}$, and the line velocity dispersion $\sigma_{\rm obs}$. For further analysis we also use the observational velocity resolution $\Delta W$, gravitational lensing magnification factor $\mu$, and signal-to-noise ratio SNR, of the observed line. These values are all summarised in Table~\ref{tab:summary_eyelash}.

\begin{table*}
 \centering
 \caption{Summary of CH$^+$(1-0) observation of the Eyelash galaxy.}
 \begin{tabular}{l c c} \hline \hline
 \multicolumn{3}{c}{\vspace{-0.3cm}} \\
 Parameter & Symbol & Value \\ \cline{1-3}
 \multicolumn{3}{c}{\vspace{-0.3cm}} \\
 Line peak of the galaxy integrated flux density & $F_{\nu,p}^{\rm obs}$ & $5.2\times 10^{-26}$~erg~s$^{-1}$~cm$^{-2}$~Hz$^{-1}$ \\ 
 Galaxy integrated background flux density       & $F_{\nu,c}^{\rm obs}$ & $3.7\times 10^{-25}$~erg~s$^{-1}$~cm$^{-2}$~Hz$^{-1}$ \\
 Emission line velocity dispersion               & $\sigma_{\rm obs}$ & 552~km~s$^{-1}$ \\
 Observational solid angle                       & $\Omega_{\rm obs}$ & 1.42~arcsec$^2$ \\ 
 Resolution & $\Delta W$                & 50~km~s$^{-1}$ \\
 Magnification factor & $\mu$                    & 37.5 \\
 Redshift										 & $z$ & 2.3 \\
 Angular diameter distance                       & $D_A$ & 1734 Mpc \\
 Line signal-to-noise ratio                      & SNR & 6  \\ 
 Total bolometric luminosity                     & $L_{\rm bol}$ & $2.3 \times 10^{12}$~$L_\odot$ \\ 
 \hline \vspace{-0.2cm}
 \end{tabular}\label{tab:summary_eyelash}
 \tablefoot{These data come from Tables 1 and 2 of \cite{falgarone_2017}, except for the bolometric luminosity \citep{Ivison2010}. By conservation of energy, the line and continuum flux densities, $F_\nu$, have been derived from the observed flux densities, $F_\nu'$, as $F_\nu = F_\nu' \left( 1 + z \right)^3/\mu$. This flux density is what would be observed from a source located, in a Euclidean space, at the angular diameter distance $D_A$ derived from the Planck cosmological parameters \citep{planck_collaboration_2020}.}
\end{table*}

We illustrate the application by using the emission from a single shock with velocity $V_s=50$~km~s$^{-1}$ propagating into gas with density $n_{\rm H}=10^4$~cm$^{-3}$, taken from the grid of models with magnetic parameter $b=1$ and no external radiation field. The tophat shock specific intensity of the CH$^+$(1-0) line $I_{\nu,1}\sim 7.2\times 10^{-14}$~erg~s$^{-1}$~cm$^{-2}$~Hz$^{-1}$~sr$^{-1}$ and optical depth $\tau_1\sim 0.5$. The emission from successive shock layers overlaps in velocities on the order of the width of their emission line profiles. In Sec.~\ref{sec:RT_chp_profiles} we found that the line broadening is  dominated by the microturbulent velocity dispersion, so we choose an overlap velocity bin $\Delta V=5$~km~s$^{-1}$ for all shocks.

With the observational parameters chosen and shock radiative transfer computed, we can then constrain the total shock surface area. Setting $N(V)=\bar{N}(V)$ in Eq.~\eqref{eq:emission1}, the background subracted ensemble flux density at the peak ($V=0$) averaged over many instances is then
\begin{align}
F_\nu(0) -F_{\nu,c}^{\rm obs} = \left[ \frac{F_{\nu,c}^{\rm obs}}{\Omega_{\rm obs}} \left( \exp \left( - \tau_1\right)   - 1 \right)+ I_{\nu,1}  \right]  \Omega_1 \mathcal{N} \Delta V f(0,\sigma_{\rm obs}).
\end{align}
Equalizing the right hand side of this expression to the observed value $F_{\nu,p}^{\rm obs}$ gives the total shock solid angle
\begin{align}\label{eq:total_shock_surface}
\Omega_1 \mathcal{N} = \frac{F_{\nu,p}^{\rm obs} \sigma_{\rm obs}\sqrt{2\pi}}{ \Delta V} \left[ \frac{F_{\nu,c}^{\rm obs}}{\Omega_{\rm obs}} \left( \exp \left( - \tau_1\right)   - 1 \right)+ I_{\nu,1} \right]^{-1}.
\end{align}
Note that if a shock is a net absorber of the background continuum then $\Omega_1 \mathcal{N}$ will be negative. In this case the observation cannot be reproduced no matter the value of $\mathcal{N}$ and we can rule out the ensemble with that particular shock model. For this shock the total shock solid angle, $\Omega_1 \mathcal{N} =2.1 \times 10^{-10}$ sr, can then be used in Eq.~\eqref{eq:overlap_vel_space}, giving a maximum average overlap of shocks in space and velocity $\bar{N}_{S}(V=0)\sim 2.3\times 10^{-2}$. Hence Eq.~\eqref{eq:emission1} can be used at all velocities over the entire emission line.

The total shock surface gives the total rate of energy dissipated by the shocks in this ensemble
\begin{align}
\mathcal{L}_K = \frac{1}{2} \rho_0 V_s^3 \mathcal{A},
\end{align}
where $\rho_0$ is the preshock mass density, and $\mathcal{A}$ is the total shock surface area. The area is
\begin{align}
\mathcal{A} = \Omega_1 \mathcal{N} D_A^2,
\end{align}
where the angular diameter distance $D_A=1734$~Mpc is calculated assuming the Planck cosmological parameters \citep{planck_collaboration_2020} at redshift $z=2.3$. For this $50$~km~s$^{-1}$ shock and distance to this galaxy, the area $\mathcal{A} \sim 640$ kpc$^2$ and the dissipation rate $\mathcal{L}_K \sim 8.9 \times 10^{45}$~erg~s$^{-1}\sim 2.3 \times 10^{12}$~L$_\odot$.

After using the \textit{average} ensemble flux density we can use the constrained $\Omega_1 \mathcal{N}$ to build a \textit{particular} ensemble by replacing $N$ in Eq.~\eqref{eq:emission1} with a different integer drawn from the Poisson distribution for each velocity bin. Two particular ensembles are shown in Fig.~\ref{fig:distr_shocks} to illustrate the effect of different total number of shocks. The ensemble shock emission has been further binned to $\Delta W = 50$~km~s$^{-1}$ to match the observational resolution. The obvious difference between the scenarios is the relative amplitude of the shot noise, given by
\begin{align}\label{eq:shotnoise}
\frac{\sigma_{\Delta W}(V)}{\bar{N}(V)} = \frac{1}{\sqrt{\bar{N}(V) } },
\end{align}
where $\bar{N}(V)$ is computed with the observational resolution $\Delta W$. Since this noise is a real feature of the discrete velocity sampling of shocks, it could in principle be measured if the noise in the line centre signal were observed to be larger than in the continuum far enough from the line. In \cite{falgarone_2017}, after removing Gaussian fits to the emission and absorption lines, the amplitude of the residuals was found to be constant over the band, and therefore the noise was dominated by instrumental noise. This gives an upper limit on the shot noise, and hence constrains the lower limit on the number of shocks in the galaxy $\mathcal{N}$. The minimum number of shocks is then found by comparing the relative shot noise, Eq.~\eqref{eq:shotnoise}, to the signal-to-noise ratio, leading to the inequality
\begin{align}
\mathcal{N} > \frac{\sigma_{\rm obs} \sqrt{2\pi}}{\Delta W} {\rm SNR}^2.
\end{align}
Note that this is independent of the radiative transfer calculation for any particular shock model, and is just a result of discrete sampling from a Gaussian distribution. For this observation, the signal-to-noise ratio, SNR$\sim 6$, giving a minimum number of shocks of $\mathcal{N}_{\rm min} \sim 1000$. This value can be combined with the fit for the total shock solid angle (Eq.~\eqref{eq:total_shock_surface}) to give a maximum surface area of each shock in the beam, equivalent to a circle of diameter $d_{\rm max} \sim 900$~pc. It is important that this number is much larger than the thickness of the shock layer $L_{\rm sh}\sim 250$~au so that we indeed have a sheet geometry that can be accurately modeled with a 1-dimensional shock code. For any particular observed system if further arguments can be made to constrain this maximum size of shock sheets, for example with regards to inhomogeneities of the medium, then the minimum number of shocks can be further constrained. Reversing the argument, if we demand shock surfaces to have a minimum equivalent diameter of $d_{\rm min} = 100 L_{\rm sh} \sim 10^{-1}$~pc, then we constrain the upper limit of the number of shocks at $\mathcal{N}_{\rm max} \sim 10^{11}$.

\subsection{Application with a grid of shock models}\label{sec:ensemble_app_grid}
We now repeat the treatment of the previous section over the wide range of shock models discussed in Sect.~\ref{sec:model_grids} on the same observational source, the Eyelash starburst galaxy. Here we compare the grids of shocks with magnetic parameter $b=1$ and external radiation parameters $G_0=0$ or $G_0=10^3$.

\begin{figure}
\centering
  \includegraphics[width=\columnwidth]{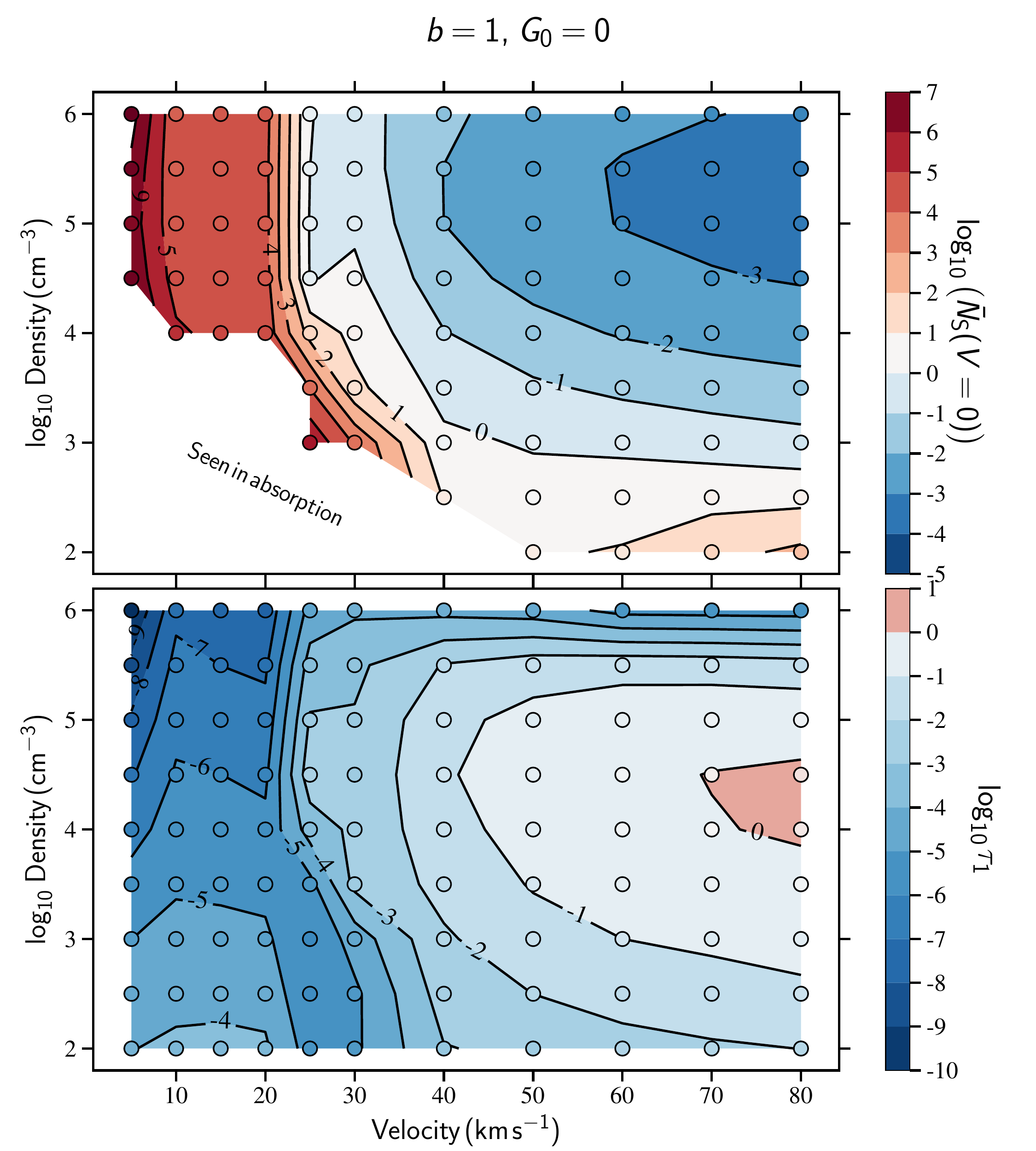}
  \caption{(Top) Average number of shocks in the velocity bin $\pm 2.5$~km/s around $V=0$ overlapping in space according to Eq.~\eqref{eq:overlap_vel_space}, and (bottom) peak optical depth $\tau_1$ for each shock for the grid with $G_0=0$ and $b=1$, applied to the Eyelash galaxy observation.}
  \label{fig:overlap_tau_g0}
\end{figure}

\begin{figure}
\centering
  \includegraphics[width=\columnwidth]{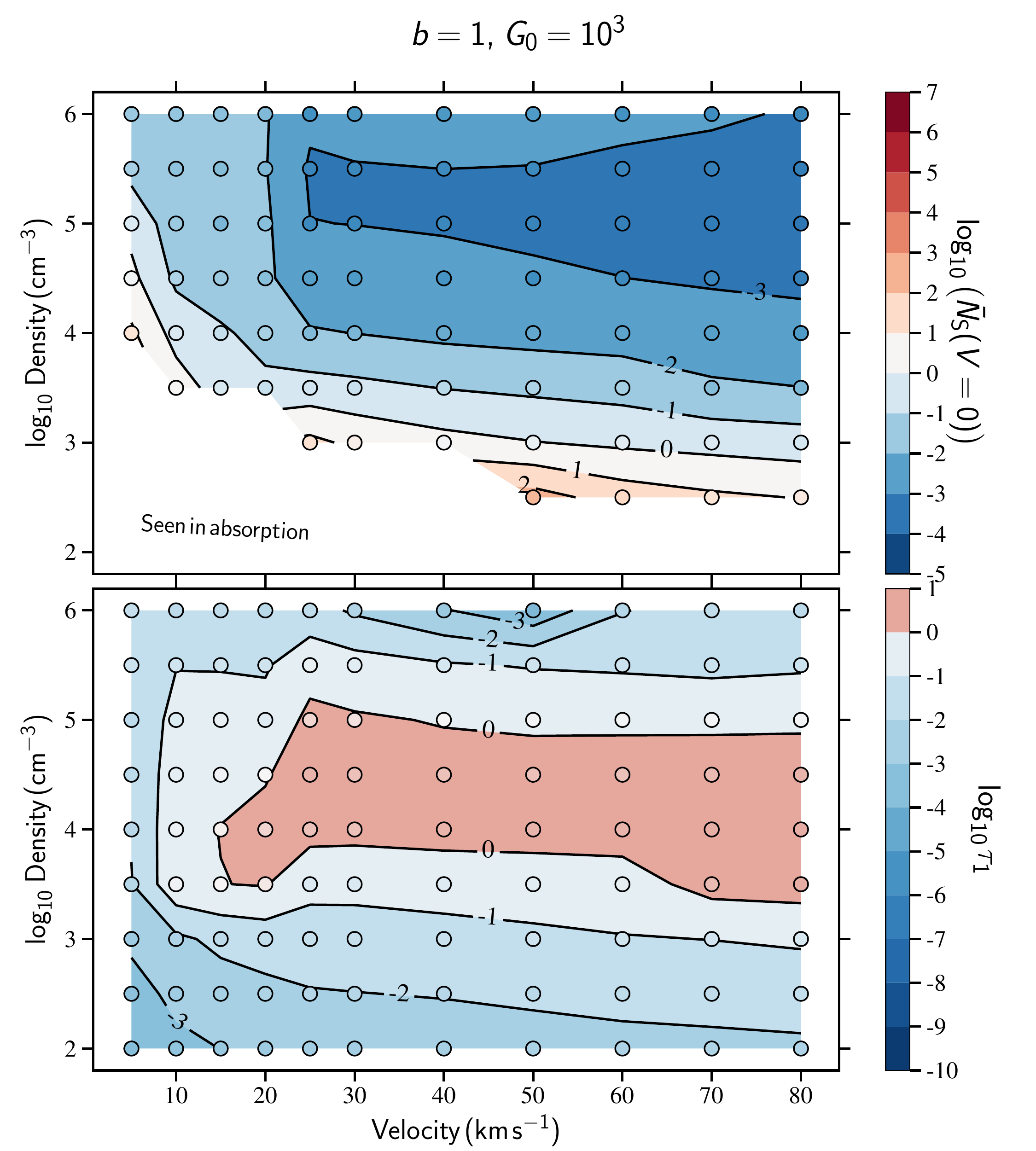}
  \caption{(Top) Average number of shocks in the velocity range $\pm 2.5$~km/s overlapping in space according to Eq.~\eqref{eq:overlap_vel_space}, and (bottom) peak optical depth $\tau_1$ for each shock for the grid with $G_0=10^3$ and $b=1$, applied to the Eyelash galaxy observation.}
  \label{fig:overlap_tau_g3}
\end{figure}

First, we consider the choice between Eqs.~\eqref{eq:emission1} and \eqref{eq:emission2}. As previously, we assume Eq.~\eqref{eq:emission1} in order to derive the total shock solid angle using Eq.~\eqref{eq:total_shock_surface}. In Figs.~\ref{fig:overlap_tau_g0} and \ref{fig:overlap_tau_g3} we show the maximum number of shock surfaces that overlap in velocity and space (top panels) and the peak optical depth through each shock (bottom panels). Note that the blank space in the top panels represents the shock models seen in absorption and which therefore cannot reproduce the observation. The top panels show that Eq.~\eqref{eq:emission1} is immediately verified ($\bar{N}_{S} < 1$) for the shocks at large velocities and densities for the $G_0=0$ grid, and for almost all the shocks in the externally irradiated grid. For the models requiring many overlapping shocks, if we consider also the peak optical depth there is no model in the two grids with a cumulative optical depth $\bar{N}_{S}\tau_1 > 1$. Hence we consider Eq.~\eqref{eq:emission1} accurate.

\begin{figure}
\centering
  \includegraphics[width=\columnwidth]{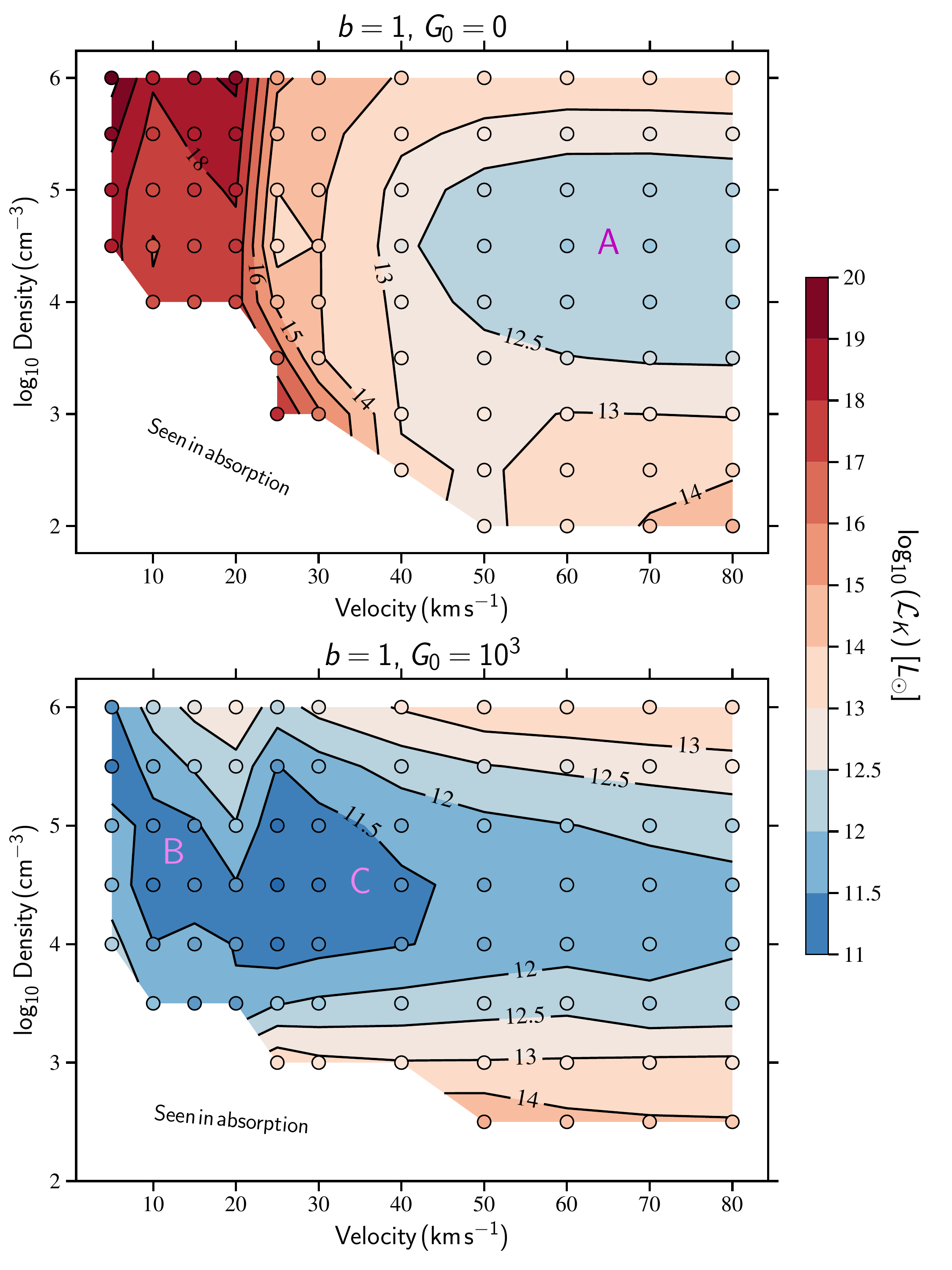}
  \caption{Total shock dissipation rate in the Eyelash galaxy for ensembles of shocks with no external radiation field $G_0=0$ (top) and for shocks with a strong external radiation field $G_0=10^3$ (bottom). Note that one solar luminosity $L_\odot=3.8\times 10^{33}$~erg~s$^{-1}$. The violet labels ``A'', ``B'', and ``C'' mark the regions of most efficient models discussed in Sect.~\ref{sec:ensemble_predictions}.}
  \label{fig:distr_energy}
\end{figure}

In Fig.~\ref{fig:distr_energy} we show the total kinetic energy dissipation rate required to reproduce the observation with the ensemble model for shocks with external radiation parameters $G_0=0$ (top) or $G_0=10^3$ (bottom). We find that no matter the shock type, the presence of a UV radiation field is key to providing the C$^+$, via photoionization of C, required for the shock heated gas to produce significant levels of CH$^+$. With no external radiation field only the higher velocity shocks ($V_s > 30$~km~s$^{-1}$) can produce enough UV, in Ly$\alpha$ and Ly$\beta$ lines, to sufficiently photoionize C (see Paper 1). In contrast, the strong external radiation field allows lower velocity shocks to efficiently produce CH$^+$ emission over a wide range of densities. These figures can be understood as representing the efficiency of the production of CH$^+$ emission with respect to the available shock mechanical energy. In the parameter domain, there are two distinct regions in which externally irradiated shocks are one to two orders of magnitude more efficient at producing CH$^+$(1-0) emission than the high velocity shocks with no external radiation (region A in Fig.~\ref{fig:distr_energy}): the C$^*$- and CJ-type shocks with velocities $V_s=$5-15~km~s$^{-1}$ and proton densities $n_{\rm H} = 10^{4}$--$10^5$~cm$^{-3}$, and J-type shocks with velocities $V_s=25$-40~km~s$^{-1}$ and proton densities $n_{\rm H} = 10^4$--$10^5$~cm$^{-3}$ (regions B and C in Fig.~\ref{fig:distr_energy}, respectively). Both these regions require a mechanical energy injection rate $\mathcal{L}_K \sim 10^{11}$~$L_\odot$.

The bolometric luminosity of the Eyelash galaxy $\mathcal{L}_{\rm bol} = 2.3 \times 10^{12}$~$L_\odot$ \citep{Ivison2010}. This value gives a strong upper limit of the mechanical energy budget.
The lowest density medium possibly responsible for shocked CH$^+$ emission is $n_{\rm H}\sim 10^{3.5}$~cm$^{-3}$. If the shocks occur in an environment with no external radiation field, only high velocity J-type shocks ($V_s > 50$~km~s$^{-1}$) can explain the CH$^+$ emission (region A). It should be noted, however, that these shocks would require a mechanical energy injection rate comparable to the bolometric luminosity. If a strong UV field is present, then the full range of velocities could reproduce the line. The lowest velocity shocks ($V_s < 40$~km~s$^{-1}$, regions B and C) would require the lowest injection rate of energy, with an injection rate $\sim 20$ times smaller than $\mathcal{L}_{\rm bol}$. Such low velocities could be expected theoretically, as simulations of MHD turbulence in circumgalactic and interstellar medium conditions indicate that the dissipation of turbulence by shocks is dominated by the lowest velocities with a sharp drop-off towards higher velocities \citep{lehmann_shockfind_2016,park_2019}. 

While we have gained a lot of information from the analysis of just one line, the emission from these shocks is not limited to CH$^+$. They produce other tracers, such as strong emission from rovibrational lines of H$_2$ that could be observed with the JWST. Therefore observations of these additional tracers can be used to disentangle these two irradiation scenarios and give a definite value of the mechanical energy injection rate. In the next section we provide predictions of such tracers.

\section{Predictions}\label{sec:predictions}
In this section we consider shock tracers other than CH$^+$, for example Ly$\alpha$ and H$_2$. First, we look at emission coming from a single shock layer. Then we consider these other tracers in the framework of our shock ensemble toy model applied to the Eyelash galaxy.

\subsection{Ly$\alpha$ and H$_2$ from a single shock}\label{sec:predictions_lya_H2}

\begin{figure}
\centering
  \includegraphics[width=\columnwidth]{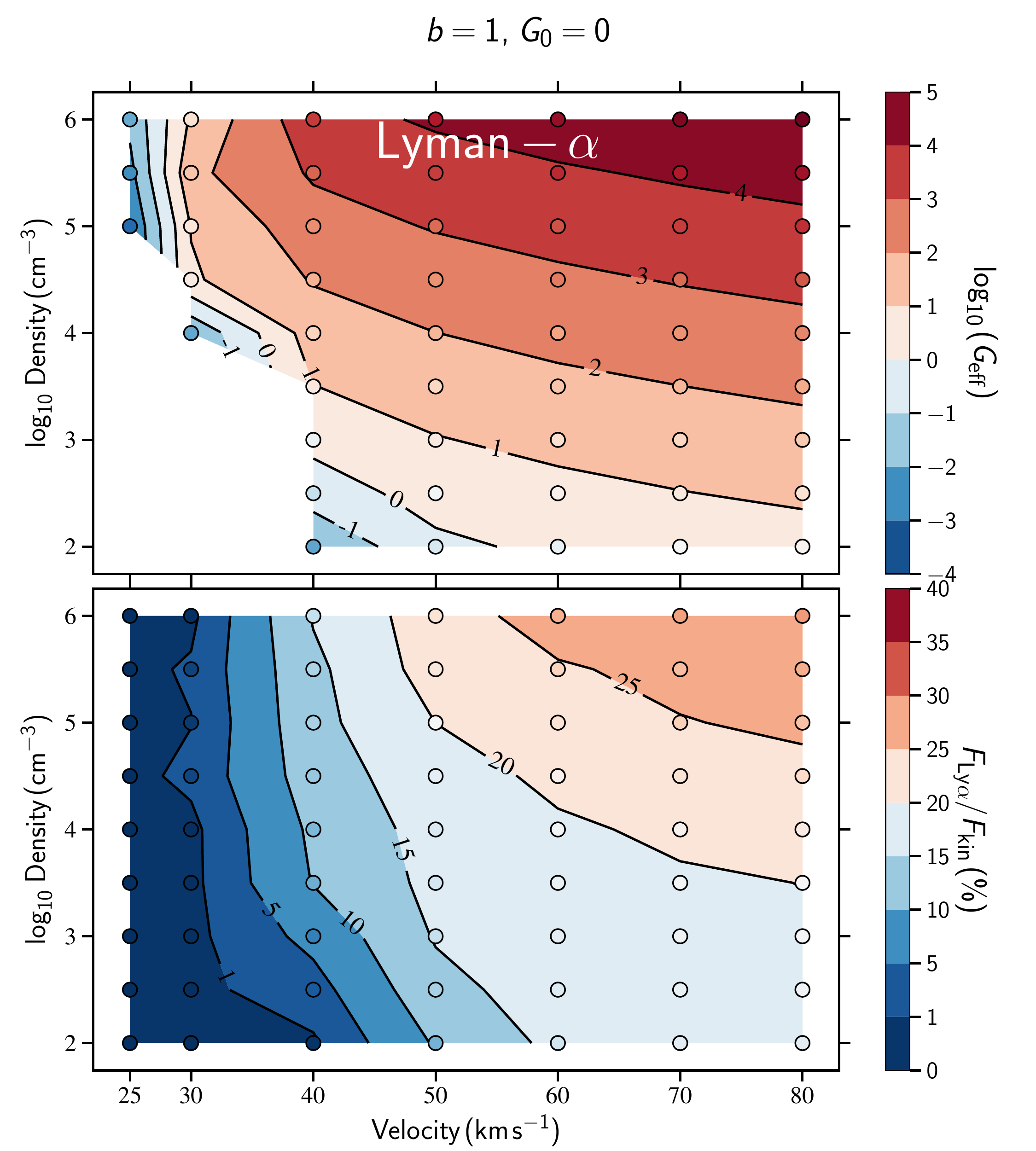}
  \caption{Ly$\alpha$ contribution to the UV flux escaping the shock front for shocks with magnetic parameter $b=1$ and no external radiation field $G_0=0$. (Top) flux relative to the Mathis ISRF, with flux equal to the standard ISRF at the 0 contour. (Bottom) flux relative to the kinetic flux of the shock $1/2 \rho_0 V_s^3$ as a percentage.}
  \label{fig:lyman_alpha}
\end{figure}

In Paper 1 we defined a photon flux normalised to the standard \cite{mathis_interstellar_1983} ISRF given by
\begin{align}
G_{\rm{eff}} = \frac{\int d\nu \, c u_\nu / h\nu }{1.55\times 10^8 \, \rm{ph/s/cm}^2},
\end{align}
where $u_\nu$ is the energy density of the radiation field at frequency $\nu$, and $h$ is the Planck constant. For the shock models considered in that work (preshock proton density of $10^4$~cm$^{-3}$), we found a critical shock velocity $\sim 32$~km~s$^{-1}$ above which this photon flux integrated over just the Ly$\alpha$ line escaping into the precursor region is greater than the entire photon flux of the standard Mathis interstellar radiation field, that is $G_{\rm{eff}} > 1$. The top panel of Fig.~\ref{fig:lyman_alpha} shows the escaping Ly$\alpha$ photon flux for shocks with $b=1$ and no external radiation field. This confirms the previous result for the critical velocity for shocks with preshock density of $10^4$~cm$^{-3}$. This critical velocity depends inversely on density, and once this threshold is crossed the Ly$\alpha$ flux shows a linear proportionality with density. However, no matter how large the density, the critical velocity is never below $\sim 30$~km~s$^{-1}$, which corresponds to the minimum velocity at which the temperature in the viscous jump is large enough to generate enough atomic H from the collisional dissociation of H$_2$.

The bottom panel of Fig~\ref{fig:lyman_alpha} shows the escaping Ly$\alpha$ photon flux relative to the kinetic flux of the shock $1/2 \rho_0 V_s^3$. This shows that once shocks become strong enough to destroy H$_2$ in the initial viscous jump (e.g. $V_s > 30$~km~s$^{-1}$), atomic H becomes a strong coolant. At these intermediate velocities, up to 30\% of the shock mechanical energy is reprocessed and escapes ahead of the shock in the form of Ly$\alpha$. For a broad range of densities and velocities, this fraction is between 10-20\%. Note that initially about twice these percentages are converted to Ly$\alpha$ in the shocks, but that the half of the photons that escape further into the shock are quickly reabsorbed by dust. The fractions at 80~km~s$^{-1}$ are in line with the Ly$\alpha$ fluxes of the $V_s=100$~km~s$^{-1}$ shock models in the database of \citep{alarie_2019} using shocks computed with the \textsc{MAPPINGS V} code \citep{sutherland_effects_2017}.

\begin{figure*}

\subfloat{ \includegraphics[clip,width=\textwidth]{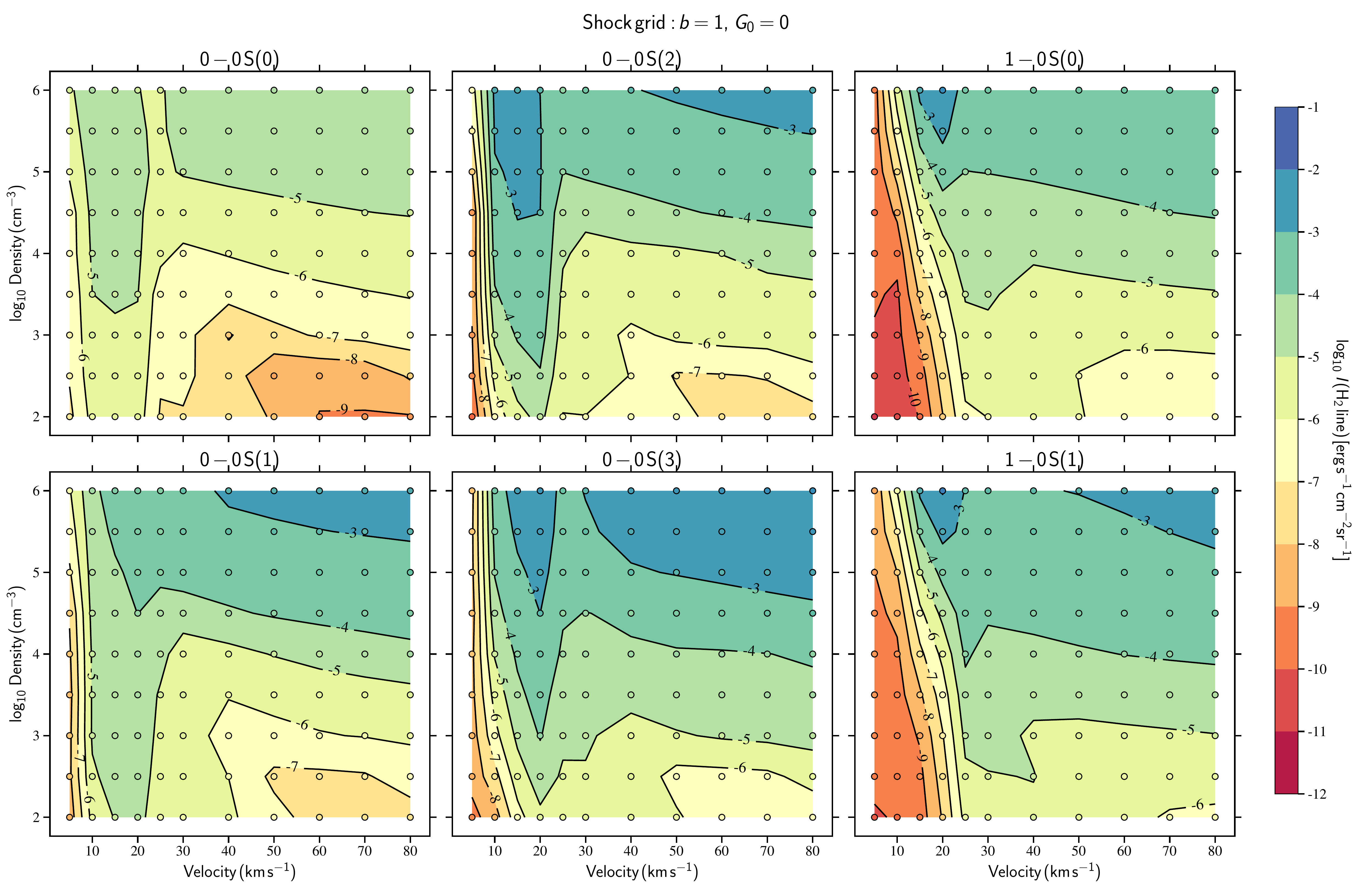} }

\subfloat{ \includegraphics[clip,width=\textwidth]{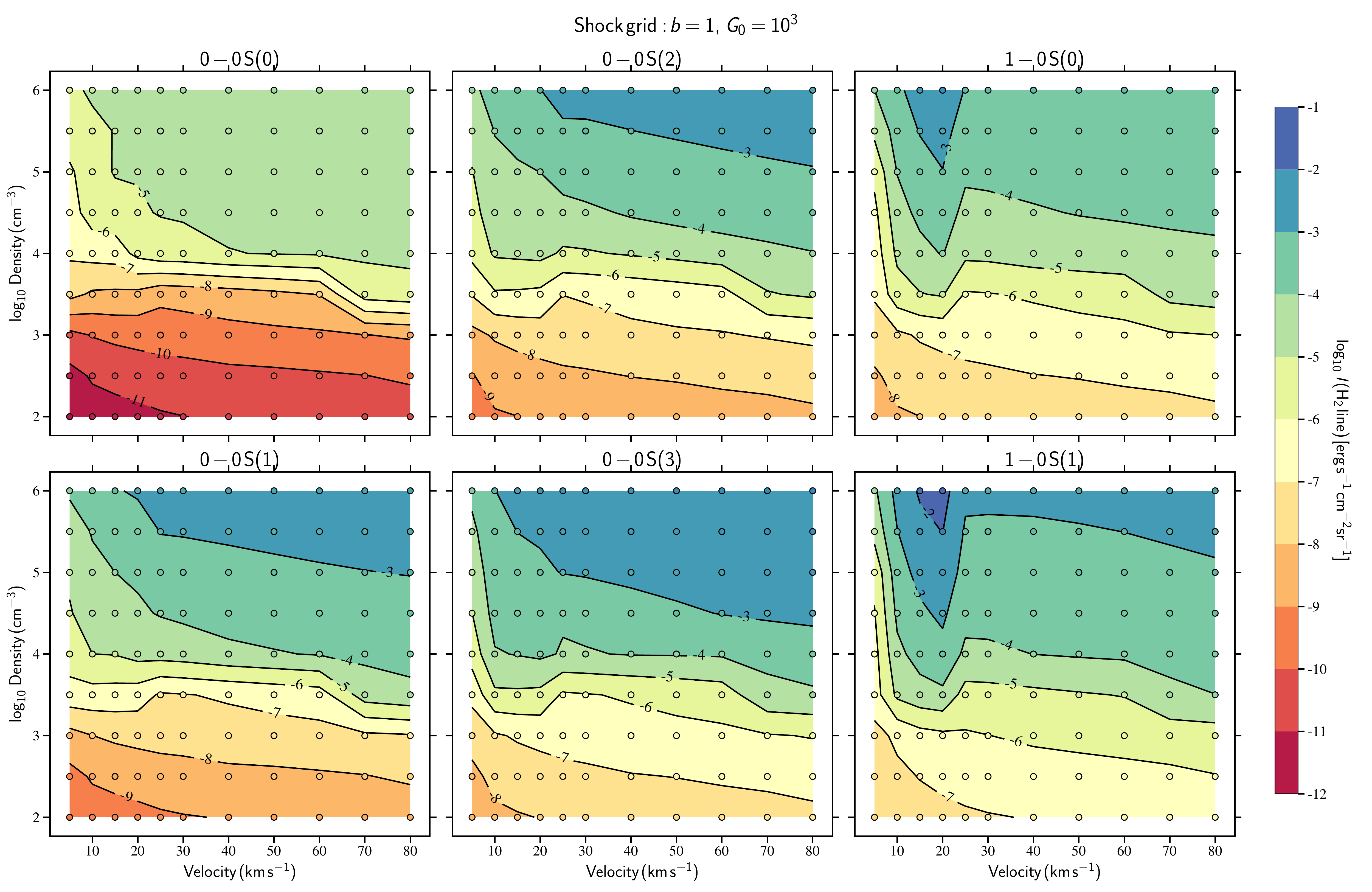} }

\caption{Line intensities of a selection of H$_2$ rovibrational lines, for shocks with no external radiation field (top) and for radiation field $G_0=10^3$ (bottom).}

\label{fig:predictions_h2}
\end{figure*}

In the panels of Fig.~\ref{fig:predictions_h2} we show the line intensities of a selection of H$_2$ rovibrational lines for the grid of shocks with $b=1$ and no external radiation field (upper panels) or external radiation field $G_0=10^3$ (lower panels). These figures show that in dissociative shocks at high velocities, despite the strong destruction of H$_2$ in the initial temperature jump, there is strong rotational and rovibrational H$_2$ emission. This is due to the reformation of H$_2$ in the postshock cooling layer, which pumps even the highest H$_2$ levels. This process only weakly depends on the gas temperature, and so the emission of H$_2$ in its rovibrational lines is weakly dependent on the velocity at high shock velocity. In the absence of an external UV field there is a sharp transition between low- and high-velocity shocks, near $V_s \sim 20$~km~s$^{-1}$ depending on density, in the vibrational lines of H$_2$. Indeed, low-velocity shocks do not reach temperatures high enough to either dissociate H$_2$ or collisionally excite its high energy levels. With an external UV field, the vibrational lines of H$_2$ are excited even at low velocities. This is because the electronic levels of H$_2$ are pumped by UV photons and induce a fluorescent cascade in the ground electronic state. Finally, these figures show perhaps surprisingly that strong emission from rovibrational lines of H$_2$ and strong Ly$\alpha$ emission can emerge from the same shocks over a broad range of parameters.

In Appendix~\ref{app:otherfluxes}, we tabulate the integrated line intensities for a selection of H$_2$ and CH$^+$ rovibrational lines, CO rotational lines, atomic H lines and 2-photon continuum, and atomic fine structure and metastable lines. The tables are given for shock models in all three grids, at velocities at every 10~km~s$^{-1}$ and integer $\log n_{\rm H}$. These lines are a subset of those output by the code.

\subsection{Application to the Eyelash Galaxy}\label{sec:ensemble_predictions}

For the shocks with no external irradiation that give ensemble total dissipation rates below the bolometric luminosity of the Eyelash galaxy ($V_s \geq 50$~km~s$^{-1}$ and $n_{\rm H}=10^4$-$10^5$~cm$^{-3}$), $\sim$15-25\% of the shock kinetic energy is re-emitted in Ly$\alpha$ photons that escape into the preshock medium, as shown in Fig.~\ref{fig:lyman_alpha}. This means there is up to $\sim 4 \times 10^{11}$~L$_{\odot}$ in Ly$\alpha$ photons permeating this region, bathing the surrounding dense, fully molecular medium. This leads to widespread heating of both dust and gas.

In Sec.~\ref{sec:ensemble_app1} and \ref{sec:ensemble_app_grid}, we pushed the limits of what information could be extracted using just the CH$^+$(1-0) rotational line. We were able to constrain regions of allowable shock models within the grids with or without external irradiation. Here we consider how to further observationally constrain the mechanical energy and physical conditions of extragalactic systems. We consider 3 regions of parameters for the most efficient ensemble shock models, denoted ``A'', ``B'', and ``C'' in Fig.~\ref{fig:distr_energy}. Region A contains high-velocity ($V_s\geq 50$~km~s$^{-1}$) J-type shocks, propagating in dense gas ($n_{\rm H}=10^{4-5}$~cm$^{-3}$) with no external irradiation driven by a mechanical energy injection rate $\sim 10^{12}$~$L_{\odot}$, region B contains low-velocity ($V_s\sim 10$~km~s$^{-1}$) C$^*$-type shocks with an external UV field $G_0=10^3$ driven by a mechanical energy injection rate $\sim 10^{11}$~$L_{\odot}$, and region C contains low-velocity ($V_s = 25$-40~km~s$^{-1}$) J-type shocks with an external UV field $G_0=10^3$ driven by a mechanical energy injection rate $\sim 10^{11}$~$L_{\odot}$. 

For each shock we use the outcome of Sect.~\ref{sec:ensemble_app1} in which the ensemble CH$^+$(1-0) emission is fit to the observation. We then predict the ensemble emission counterpart of rovibrational lines of CH$^+$, rovibrational lines of H$_2$, pure rotational lines of CO, and atomic metastable and fine structure lines. The intensities of these lines are multiplied by the total shock solid angle given by Eq.~\eqref{eq:total_shock_surface} to give the total line fluxes for the Eyelash galaxy. Note that the rovibrational lines of H$_2$ and atomic lines are computed by the shock code with an optically thin assumption while the CO lines are computed with the LVG approximation like CH$^+$. 

\begin{figure*}
\centering
  \includegraphics[width=\textwidth,trim = 0cm 5.5cm 0cm 3cm, clip,angle=0]{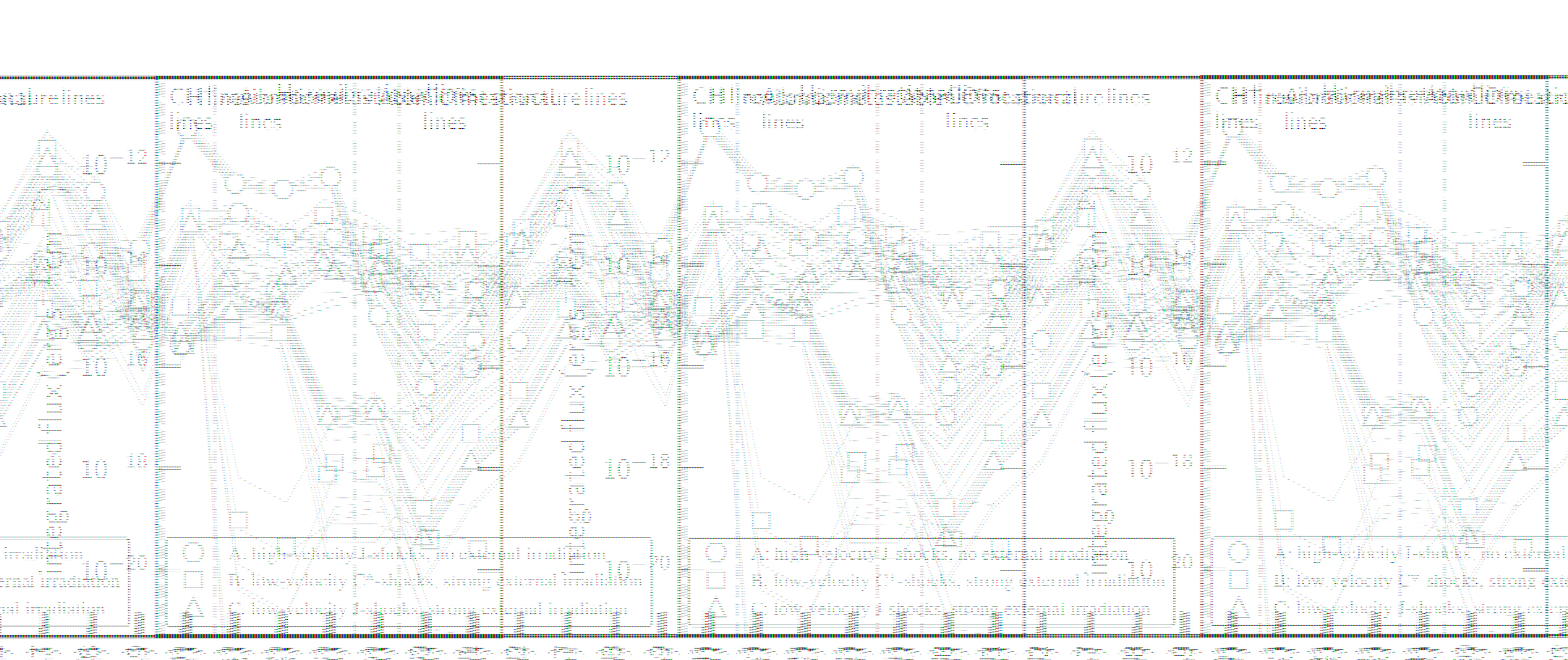}
  \caption{Line fluxes of a selection of H$_2$ and CH$^+$ rovibrational lines, CO rotational lines, and atomic fine structure and metastable lines emitted from ensembles of shocks around the Eyelash galaxy. Ensemble models are grouped by regions in Fig.~\ref{fig:distr_energy}, denoted ``A'' (blue) for high-velocity J-type shocks with no external irradiation driven by a mechanical energy injection rate $\mathcal{L}_K \sim 10^{12}$~$L_{\odot}$, ``B'' (purple) for low-velocity C$^*$-type shocks with an external UV field $G_0=10^3$  driven by a mechanical energy injection rate $\mathcal{L}_K \sim 10^{11}$~$L_{\odot}$, and ``C'' (green) for low-velocity J-type shocks with an external UV field $G_0=10^3$ driven by a mechanical energy injection rate $\mathcal{L}_K \sim 10^{11}$~$L_{\odot}$. Markers indicate fluxes averaged over all the models in the defined region.}
  \label{fig:predictions_comparison}
\end{figure*}

In Fig.~\ref{fig:predictions_comparison} we show a selection of total line fluxes for the efficient ensemble shock models. There are many differences between the models, suggesting that even just one more observation is able to provide strong constraints on galactic environments. Taking note that the observed wavelength, $\lambda$, is redshifted by $\lambda = \lambda_0 \left( 1 + z \right)$ where $\lambda_0$ is the rest-frame wavelength and $z\sim 2.3$ is the cosmic redshift of the Eyelash galaxy, many of the lines from this system fall in observable wavebands of current and near-future state-of-the-art observatories, for example JWST and ALMA.

In molecular emission, H$_2$ rovibrational lines are good at distinguishing between shock types. In J-type shocks (regions A and C) these lines are typically an order of magnitude weaker than in the C$^*$-type shocks (region B). Stronger differences are found in CO lines, with 2-3 orders of magnitude stronger fluxes in the high-velocity ensemble in both low- and high-J rotational lines. This is due to the efficient photodissociation of CO by the broad-spectrum external radiation field in the externally irradiated models. By contrast the high-velocity model does not efficiently photodissociate CO, despite the similarly strong UV flux, due to a lack of overlap between CO absorption lines and the Ly$\alpha$ line. The high temperatures reached in the J-type shocks allow for stronger emission in atomic metastable lines. The low-velocity shocks only weakly excite these lines, and so they are typically around 4 orders of magnitude weaker than from the high-velocity ensemble. So, H$_2$ and atomic metastable lines probe shock type, CO lines probe the external radiation conditions, and H lines distinguish high and low velocity shocks.

In the end, a multiwavelength study is able to probe the physical conditions and energetics of galaxy scale molecular outflows. An extremely broad CH$^+$(1-0) emission line may uncover the presence of shocks in a turbulent cascade. In this framework, we are able to disentangle whether the ensemble of shocks is composed of high-velocity self-irradiated shocks or low-velocity externally irradiated shocks. Even just one extra line to complement the CH$^+$(1-0) observation would strongly tighten constraints on the density and radiation conditions of the turbulent gas as well as give a minimum mechanical energy input from the outflow. These results open new perspectives for observational programs using state-of-the-art facilities such as the JWST and ALMA.

\section{Conclusion}
We have presented models of magnetised, molecular shocks with the broadest published range of physical parameters, with shock velocities $V_s=5$--$80$~km~s$^{-1}$, preshock proton densities $n_{\rm H}=10^2$--$10^6$~cm$^{-3}$, weak to moderate magnetic field strengths $B=1$--$1000$~$\mu$G, and with or without a strong external radiation field $G_0$=0 or $10^3$. Post-processed molecular emission lines from these shocks were used in a simple model of an ensemble of identical low-velocity shocks transported by high-velocity bulk flows, representing the dissipative structures in a turbulent cascade driven by a large-scale injection of mechanical energy. As an illustrative test case, we successfully used the ensemble model to interpret the observation of broad ($\sigma_{\rm obs} > 500$~km~s$^{-1}$) CH$^+$(1-0) emission from the high-$z$ starburst galaxy The Cosmic Eyelash. With just this one emission line, we were able to constrain the medium responsible for a shock origin of this line, and suggested key atomic and molecular lines that could further constrain these conditions. We summarise the key results as follows:

\begin{enumerate}
\item For the physical conditions explored, the full diversity of stationary MHD shock types arise (C-, C$^*$-, CJ- and J-type) emphasising the requirement for a multi-fluid treatment and molecular chemical network to model molecular shocks in these conditions.\\

\item A significant fraction of the shock mechanical energy is reprocessed in the excitation of H. Up to $\sim$30~\% of the shock kinetic flux escapes as Ly$\alpha$ photons.\\

\item Strong emission of H$_2$ rovibrational lines is found over a broad range of parameters, and is the brightest in the same density and velocity regimes that show the strongest Ly$\alpha$ emission. At low velocities ($V_s \leq 40$~km~s$^{-1}$), vibrational lines are strong tracers of the presence of an external UV field.\\

\item CH$^+$ can be a significant pathway for reprocessing the mechanical energy, radiating up to one percent of the kinetic flux in shocks irradiated by a strong external UV radiation field. The UV field, whether self-generated or already present in the shocked environment, is a key component enhancing the production and excitation of CH$^+$ in shocks. With respect to the required mechanical energy input, low-velocity C* and J-type shocks ($V_s \leq 40$~km~s$^{-1}$) with a strong external radiation field ($G_0=10^3$) are at least 10 times more efficient than any of the shocks with no external radiation field.\\

\item We have demonstrated how to use these shock models to interpret extragalactic observations of molecular emission. Our model of emission from an ensemble of shocks is able to extract the mechanical energy and physical conditions of the turbulent cascade triggered by high velocity flows around external galaxies.\\

\item In the Eyelash galaxy, we find that at least $10^{11}$~$L_\odot$ of mechanical energy must be transferred from the large-scale, high-velocity flows to $10^3$--$10^{11}$ dissipative structures at lower velocity via a turbulent cascade. This minimum amount of mechanical energy is 20 times smaller than the total bolometric luminosity and is comparable to estimates of the galaxy scale input of mechanical energy due to, for example, supernovae driven galactic outflows \citep{Veilleux2020}. Non irradiated shocks are barely allowed while irradiated shocks could be a viable scenario for explaining the CH$^+$ emission observed in the Eyelash galaxy. The stringent constraint given by the bolometric luminosity shows, however, that this scenario should be extended to shocks irradiated by stronger radiation field and improved by including the contribution of photodissociation regions to the CH$^+$ emission. In any case, a strong UV radiation field is mandatory to explain the CH$^+$ emission observed in the Eyelash galaxy. The presence of this UV field is supported by the analysis of the spectral energy distribution of the Eyelash \citep{Ivison2010} that is well fitted by two dust components at temperatures of 30 and 60 K.\\

\item We have estimated the H$_2$, CH$^+$, CO, and atomic line counterpart from an ensemble of low-velocity C$^*$-type ($V_s \sim 10$~km~s$^{-1}$) or J-type ($V_s \sim 25$~km~s$^{-1}$) externally irradiated shocks, or high-velocity ($V_s = 50$~km~s$^{-1}$) J-type shocks with no external irradiation. CO lines are particularly good at distinguishing the two irradiation scenarios, as strong photodissociation by the external UV field leads to CO emission more than 2 orders of magnitude weaker across the whole rotational ladder for the low-velocity ensemble. In addition, the high-velocity ensemble produces shock-heated gas temperatures high enough to excite atomic lines, so that Ly$\alpha$, Ly$\beta$, and all atomic metastable lines are at least 4 orders of magnitude brighter than from the low-velocity ensemble.

\end{enumerate}

We have shown how accurate shock modeling over a wide range of physical conditions allows for the interpretation of molecular and atomic line emission from galaxies. The tool developed in this paper has been applied to observations from the Eyelash galaxy. However, this new method opens up a promising avenue for probing the physical conditions and energetics of galaxies through observations of molecular outflows.

\section*{Acknowledgments}
We are very grateful to the referee for their thorough reading and their clever comments which triggered important discussions and helped us correct an error in the interpretation of observations from high-$z$ galaxies. The research leading to these results has received fundings from the European Research Council, under the European Community’s Seventh framework Programme, through the Advanced Grant MIST (FP7/2017-2022, No 787813). We would also like to acknowledge the support from the Programme National “Physique et Chimie du Milieu Interstellaire” (PCMI) of CNRS/INSU with INC/INP co-funded by CEA and CNES.

\bibliographystyle{aa}
\bibliography{selfirrad}

\appendix

\section{Grid of weakly magnetised shocks}\label{app:g0b0}
Here we consider shocks with magnetic parameter $b=0.1$ and no external radiation $G_0=0$.

\begin{figure*}
\centering
  \includegraphics[width=0.95\textwidth]{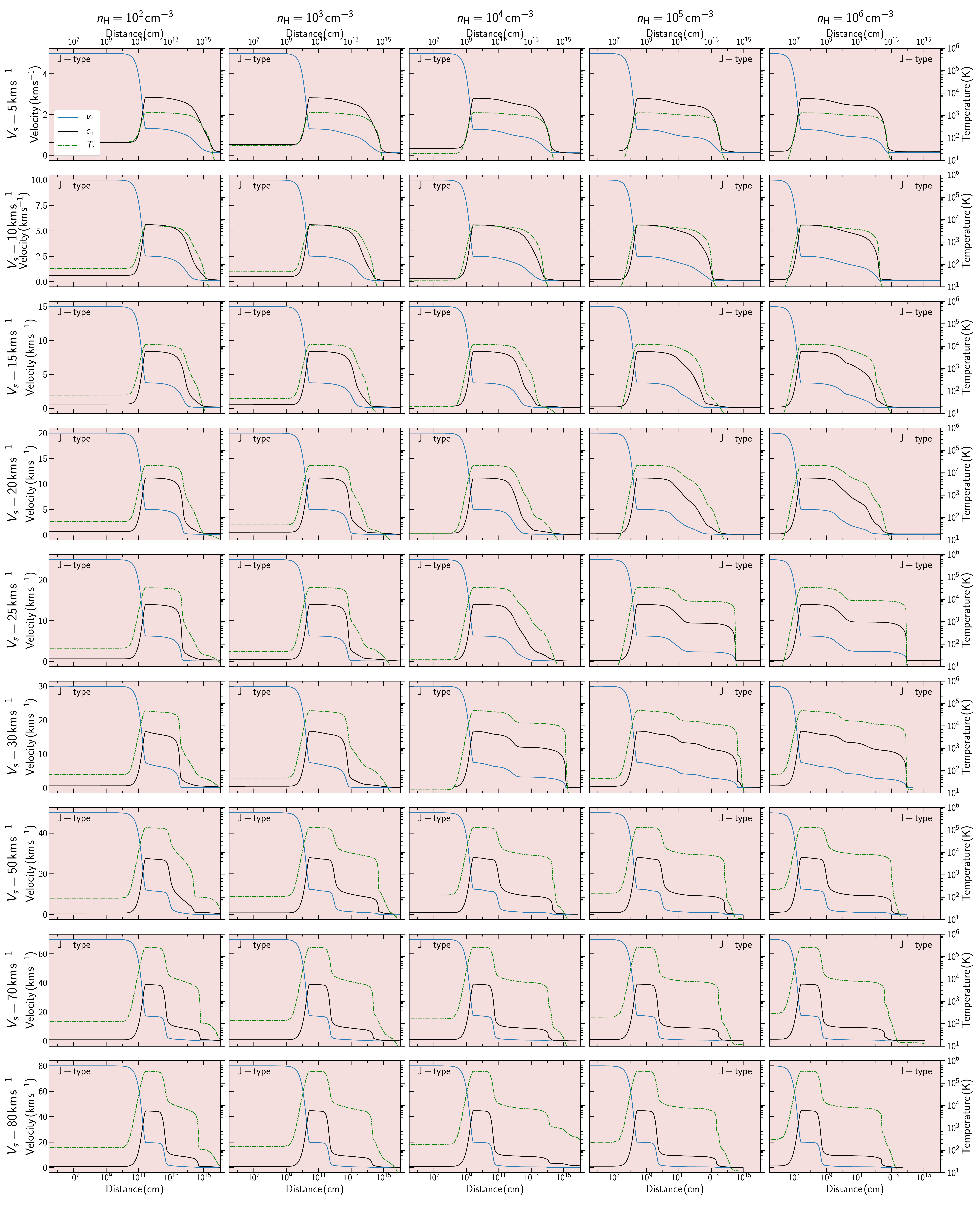}
  \caption{Shock profiles of ion (red solid) and neutral (blue solid) fluid velocities, local sound speed (black solid), and neutral (green dash-dotted) temperature. The shocks propagate from right to left, and the velocities are given in the reference frame in which the shock is stationary. These shocks were computed with magnetic parameter $b=0.1$ and no external radiation field $G_0=0$. We have excluded profiles for shocks with velocities $V_s=40$ and 60~km~s$^{-1}$ for brevity.}
  \label{fig:temps_g0b0}
\end{figure*}

In Fig.~\ref{fig:temps_g0b0} we show profiles of the neutral fluid temperatures and velocities, and local sound speed. The initial ion magnetosonic speed is less than 5~km~s$^{-1}$ for these conditions, and so all the models are J-type shocks. For shock velocities $V_S \geq 25$~km~s$^{-1}$, these shocks are thinner by an order of magnitude than the equivalent shocks with stronger magnetic field, shown in Fig.~\ref{fig:temps_g0}. This decrease in size is due to the weaker magnetic pressure support in the postshock. The lengthscales for the shocks to cool down to 100~K are in range of $10^{12}$-$10^{15}$~cm, or in timescales 1-$10^5$~yrs.

\begin{figure}
\centering
  \includegraphics[width=\columnwidth]{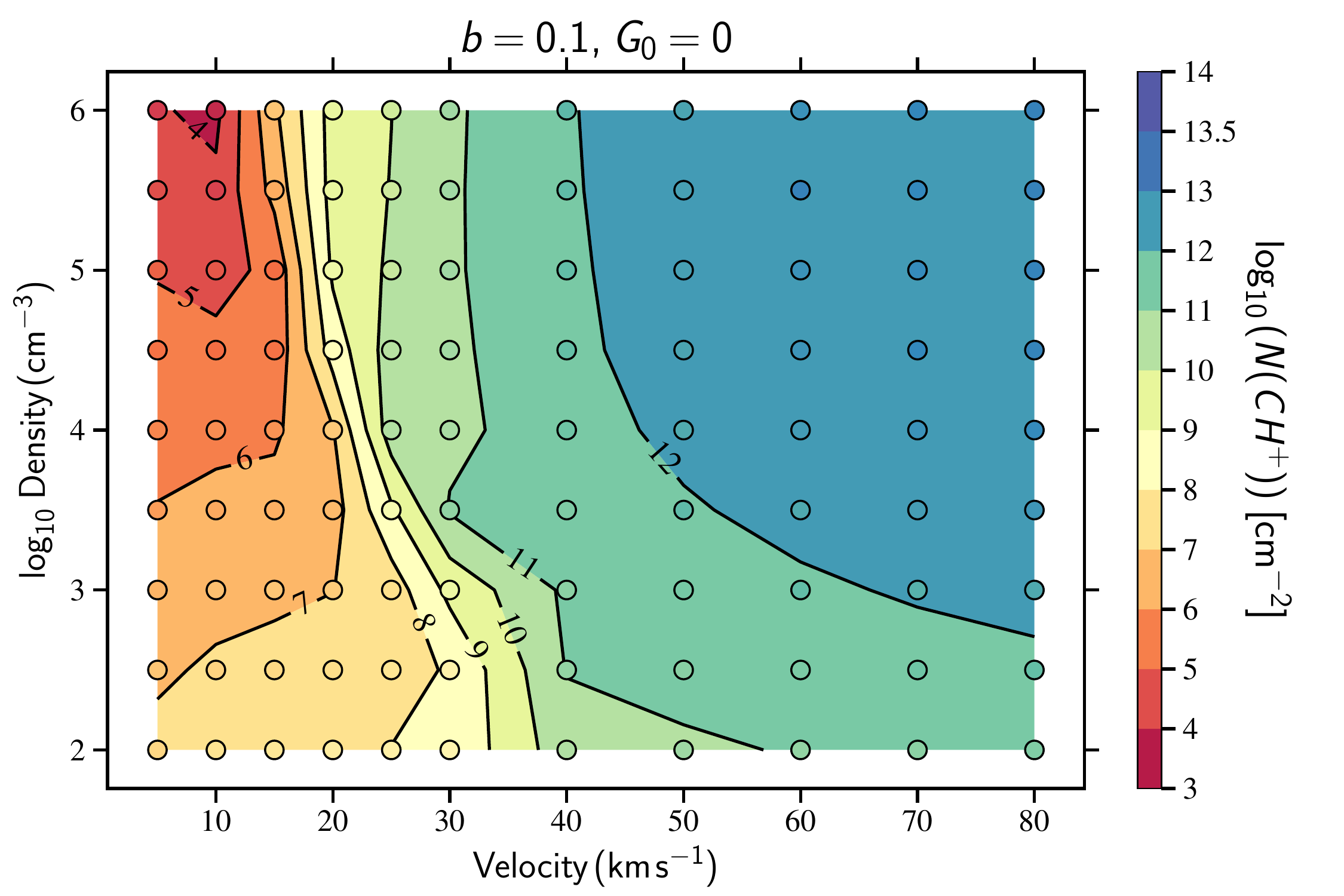}
  \caption{Column density of CH$^+$ for shocks with magnetic parameter $b=0.1$, and no external radiation field $G_0=0$.}
  \label{fig:chp_column_g0b0}
\end{figure}

\begin{figure}
\centering
  \includegraphics[width=\columnwidth]{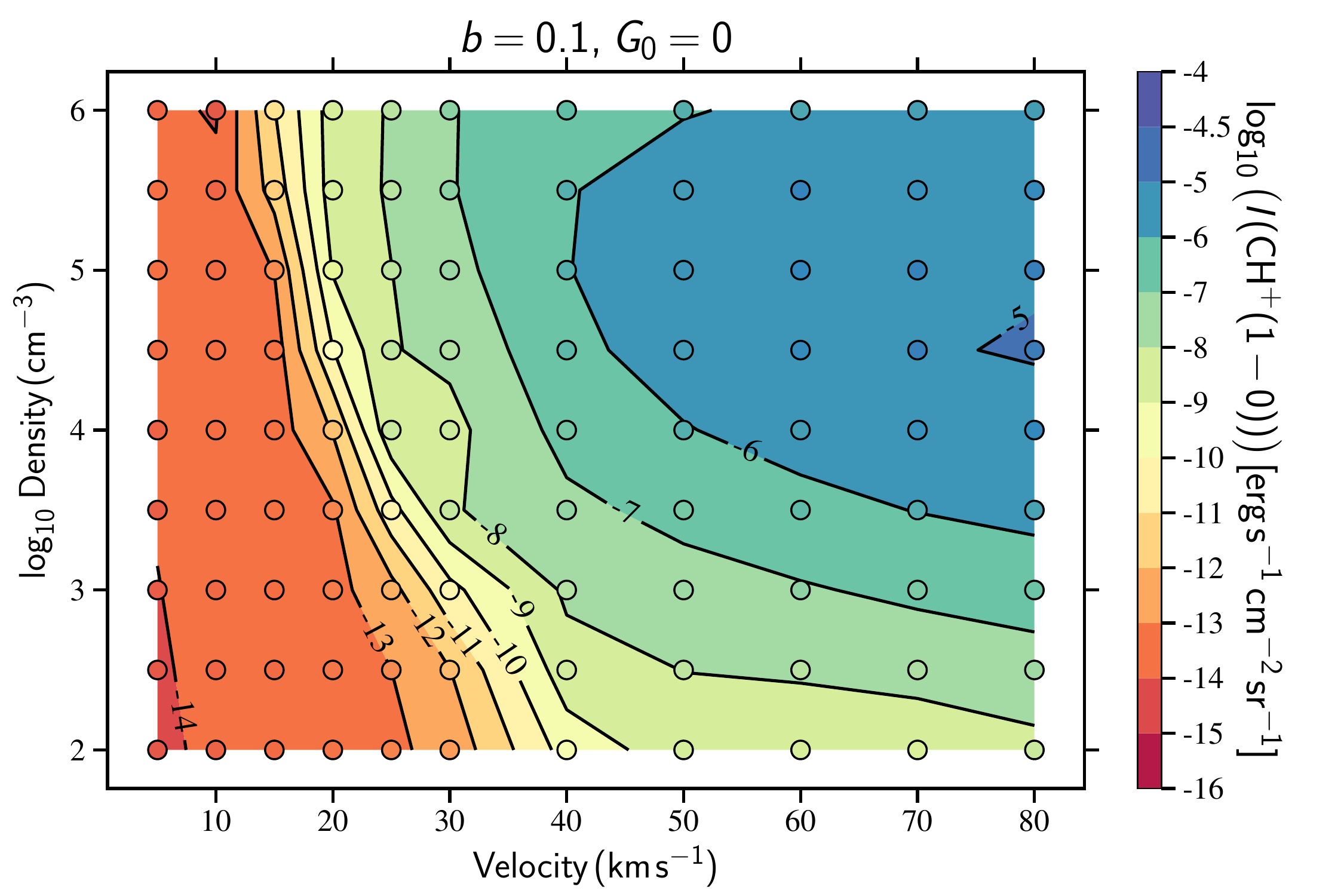}
  \caption{Line intensities of CH$^+$(1-0) for shocks with magnetic parameter $b=0.1$, and no external radiation field $G_0=0$.}
  \label{fig:chp_1to0_g0b0}
\end{figure}

\begin{figure}
\centering
  \includegraphics[width=\columnwidth]{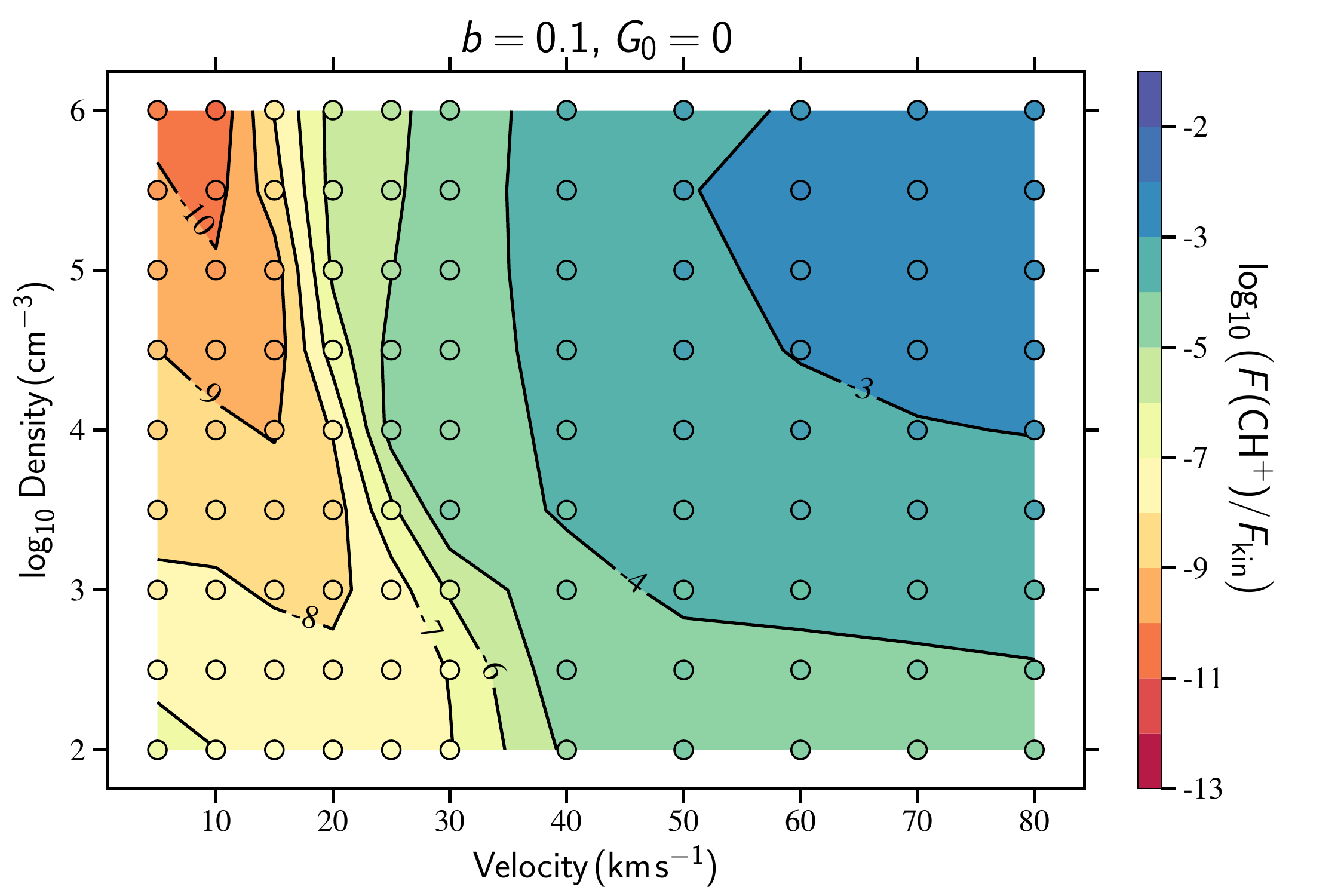}
  \caption{Total CH$^+$ flux as a fraction of shock kinetic flux for shocks with magnetic parameter $b=0.1$, and no external radiation field $G_0=0$.}
  \label{fig:chp_energy_fraction_g0b0}
\end{figure}

In Figs.~\ref{fig:chp_column_g0b0}, \ref{fig:chp_1to0_g0b0}, \ref{fig:chp_energy_fraction_g0b0} we give the CH$^+$ column density, integrated intensity of the 1-0 pure rotational line, and fraction of kinetic flux reprocessed in any rovibrational line, respectively. The column densities show the same general trends as the shocks with stronger magnetic field strength shown in the upper panel of Fig.~\ref{fig:chp_column}. At high velocities ($V_s > 40$~km~s$^{-1}$), the column densities are reduced by an order of magnitude compared to the stronger B field shocks, with a maximum column of $10^{13}$~cm$^{-2}$. This is simply due to the shorter length scales. At low velocities ($V_s < 30$~km~s$^{-1}$) the J-type shocks are much less efficient at producing CH$^+$ than the C-type shocks at the same velocities and densities in the $b=1$ grid of shocks.

\begin{figure}
\centering
  \includegraphics[width=\columnwidth]{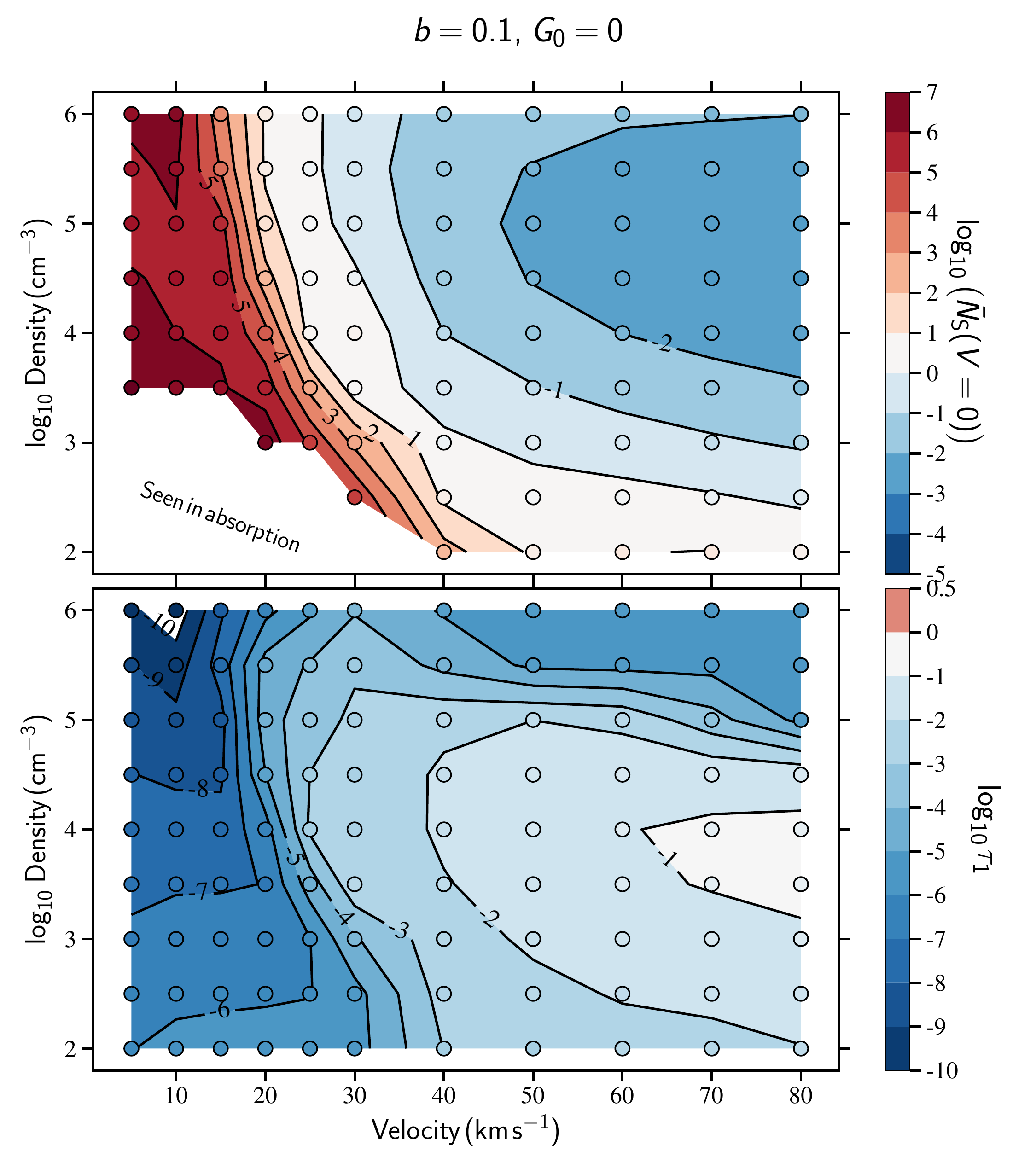}
  \caption{(Top) Average number of shocks in the velocity bin $\pm 2.5$~km/s around $V=0$ overlapping in space according to Eq.~\eqref{eq:overlap_vel_space}, and (bottom) peak optical depth $\tau_1$ for each shock for the grid with $G_0=0$ and $b=0.1$, applied to the Eyelash galaxy observation.}
  \label{fig:overlap_tau_g0b0}
\end{figure}

\begin{figure}
\centering
  \includegraphics[width=\columnwidth]{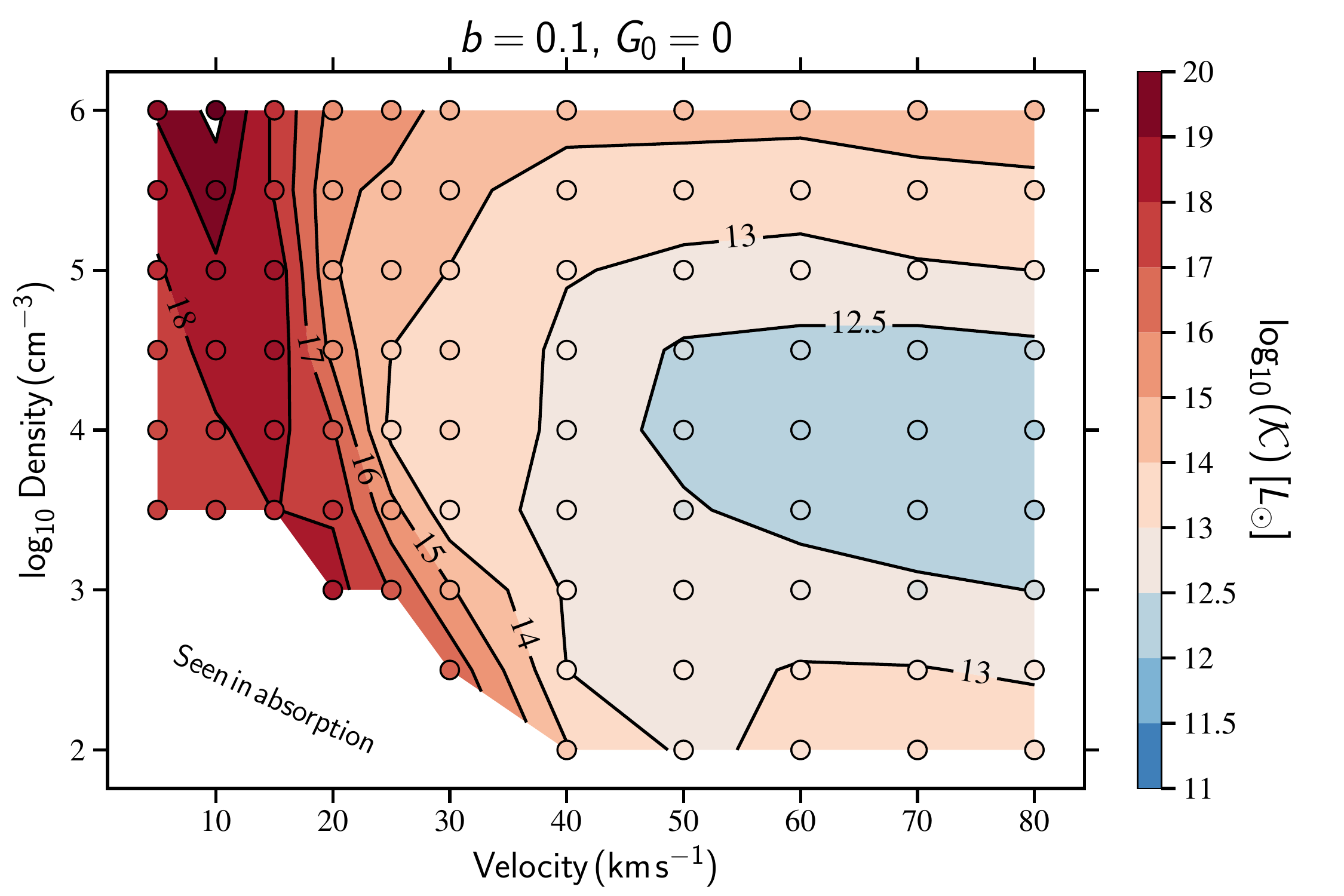}
  \caption{Total shock dissipation rate in the Eyelash galaxy for ensembles of shocks with magnetic parameter $b=0.1$ and no external radiation field $G_0=0$. Note that one solar luminosity $L_\odot=3.8\times 10^{33}$~erg~s$^{-1}$.}
  \label{fig:distr_energy_g0b0}
\end{figure}

To apply this grid to the Eyelash galaxy described in Sec.~\ref{sec:ensemble_app_grid}, we show in Fig.~\ref{fig:overlap_tau_g0b0} the maximum number of shocks overlapping in velocity and space and space in the observed beam (upper panel), and the peak optical depth for the CH$^+$(1-0) line for each shock in the grid. The cumulative optical depth $\bar{N}_{S}\tau_1 < 1$ for all ensemble models, so Eq.~\eqref{eq:emission1} is appropriate over the whole grid. In Fig.~\ref{fig:distr_energy_g0b0} we show the mechanical energy injection rate required to drive the ensembles of shocks. This figure does not significantly differ from Fig.~\ref{fig:distr_energy} for the $b=1$ grid.

\begin{figure}
\centering
  \includegraphics[width=\columnwidth]{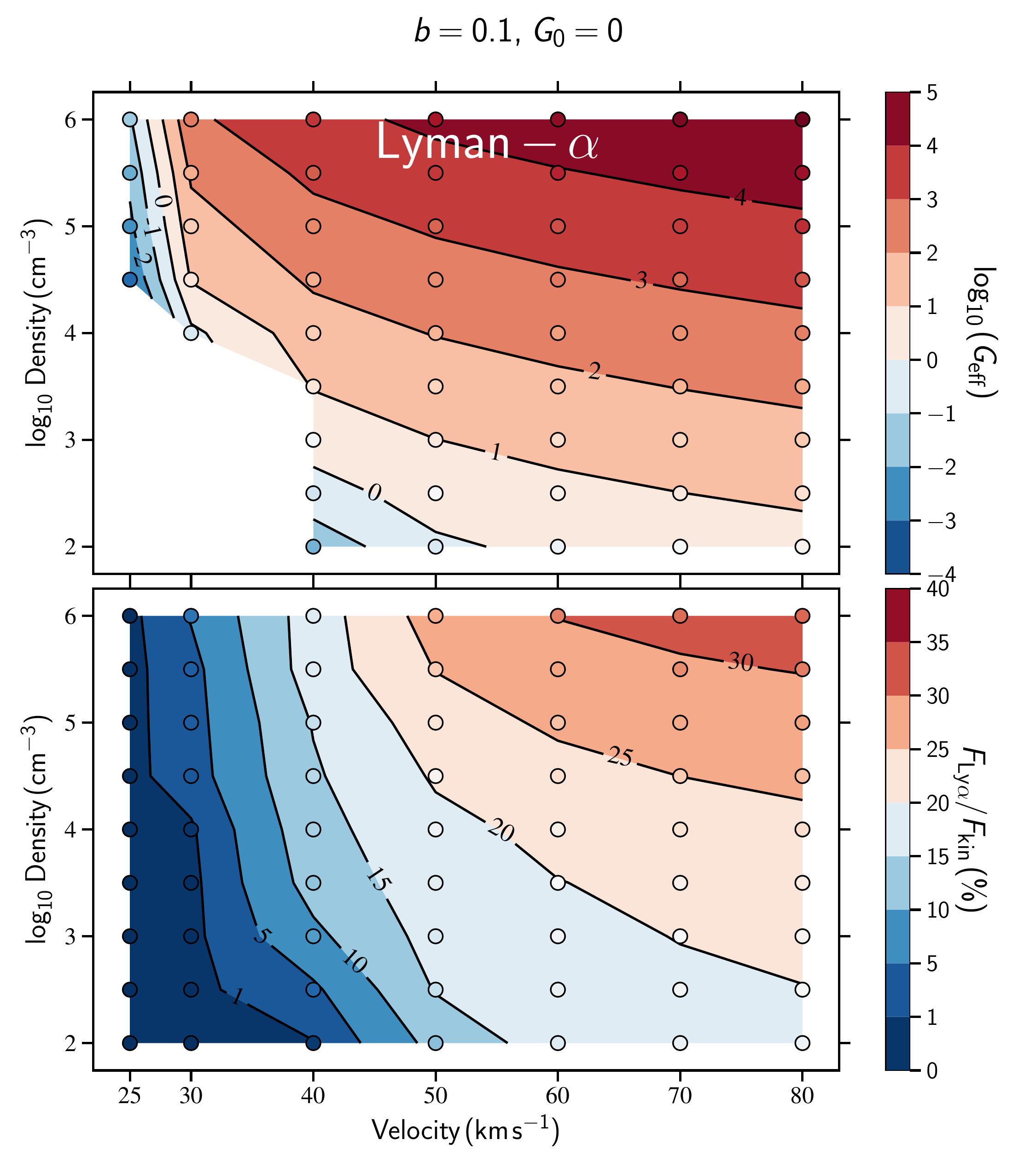}
  \caption{Ly$\alpha$ contribution to the UV flux escaping the shock front for shocks with magnetic parameter $b=0.1$ and no external radiation field $G_0=0$. (Top) flux relative to the Mathis ISRF, with flux equal to the standard ISRF at the 0 contour. (Bottom) flux relative to the kinetic flux of the shock $1/2 \rho V_s^3$ as a percentage.}
  \label{fig:lyman_alpha_g0b0}
\end{figure}

\begin{figure*}
\centering
  \includegraphics[width=\textwidth]{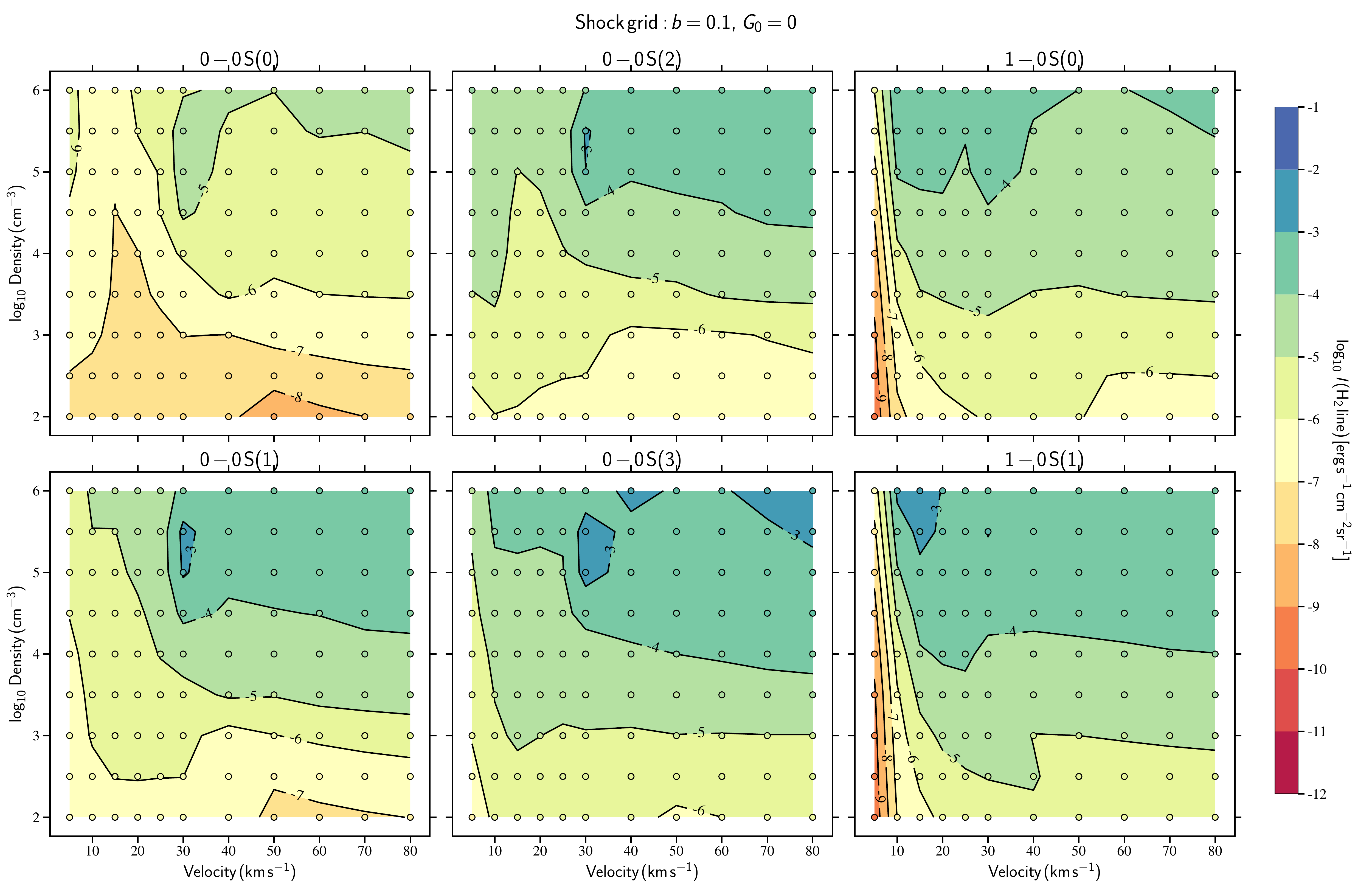}
  \caption{Intensities of H$_2$ rovibrational lines for shocks with magnetic parameter $b=0.1$ and no external radiation field $G_0=0$.}
  \label{fig:predictions_h2_g0b0}
\end{figure*}

For shock tracers other than CH$^+$, we show in Fig.~\ref{fig:lyman_alpha_g0b0} the Ly$\alpha$ photon flux relative to the Mathis ISRF (upper panel) and energy flux relative to the shock kinetic flux (lower panel). In addition, in Fig.~\ref{fig:predictions_h2_g0b0} we show a selection of intensities of H$_2$ rovibrational lines.

\clearpage

\section{Intensities from single shocks}\label{app:otherfluxes}

In Tables B.1-B.15 we tabulate the integrated line intensities for a selection of H$_2$ and CH$^+$ rovibrational lines, CO pure rotational lines, atomic H lines and 2-photon continuum, and atomic fine structure and metastable lines. The tables are given for shock models in all three grids, at velocities at every 10~km~s$^{-1}$ and integer $\log n_{\rm H}$. Note that these tabulated lines are a subset of those output by the code which contains
\begin{itemize}
\item 200 rovibrational lines of H$_2$,
\item 100 rovibrational lines of CH$^+$,
\item 14 pure rotational lines of CO,
\item Ly$\alpha$, Ly$\beta$, H$\alpha$, and 2-photon continuum (<2400~$\AA$),
\item 81 atomic infrared and optical lines including
	\begin{itemize}
	\item C      [230.41 $\mu$m], [370.37 $\mu$m], [609.75 $\mu$m], [4621.57 $\AA$], [4627.35 $\AA$], [8727.13 $\AA$], [9808.32 $\AA$], [9824.13 $\AA$], and [9850.26 $\AA$],
	\item C$^+$  [157.68 $\mu$m], [2323.50 $\AA$], [2324.69 $\AA$], [2325.40 $\AA$], [2326.93 $\AA$], and [2328.12 $\AA$],
	\item N      [1.04006 $\mu$m], [1.04010 $\mu$m], [1.04100 $\mu$m], [1.04104 $\mu$m], [1.147 mm], [6527.23 $\AA$], [6548.05 $\AA$], and [6583.45 $\AA$], 
	\item N$^+$  [76.45 $\mu$m], [121.80 $\mu$m], [205.34 $\mu$m], [3062.83 $\AA$], [3070.55 $\AA$], [6527.23 $\AA$], [6548.05 $\AA$], and [6583.45 $\AA$],
	\item O      [44.06 $\mu$m], [63.19 $\mu$m], [145.53 $\mu$m], [2958.37 $\AA$], [2972.29 $\AA$], [5577.34 $\AA$], [6300.30 $\AA$], [6363.78 $\AA$], and [6391.73 $\AA$],
	\item O$^+$  [0.4995 mm],[2470.219 $\AA$], [2470.341 $\AA$], [3726.032 $\AA$], [3728.815 $\AA$], [7318.92 $\AA$], [7319.99 $\AA$], [7329.67 $\AA$], and [7330.73 $\AA$],
	\item S      [1.08241 $\mu$m], [1.13089 $\mu$m], [1.15407 $\mu$m], [17.43 $\mu$m], [25.25 $\mu$m], [56.31 $\mu$m], [4507.31 $\AA$], [4589.26 $\AA$], and [7725.05 $\AA$],
	\item S$^+$  [1.0290 $\mu$m], [1.0323 $\mu$m], [1.0339 $\mu$m], [1.0373 $\mu$m], [0.31456 mm], [4068.60 $\AA$], [4076.35 $\AA$], [6716.44 $\AA$], and [6730.82 $\AA$],
	\item Si     [1.5876 $\mu$m], [1.6073 $\mu$m], [1.6459 $\mu$m], [44.812 $\mu$m], [68.472 $\mu$m], [129.682 $\mu$m], [6526.78 $\AA$] and [6589.61 $\AA$],
	\item Si$^+$ [34.814 $\mu$m], [2328.52 $\AA$], [2334.40 $\AA$], [2334.60 $\AA$], [2344.20 $\AA$], and [2350.17 $\AA$].
	\end{itemize}
\end{itemize}


\begin{table*}
  \centering
  \caption{Integrated line and two-photon continuum intensities (erg s$^{-1}$ cm$^{-2}$ sr$^{-1}$) escaping the shock front as a function of shock velocity (km s$^{-1}$) for shocks with preshock density $n_{\rm H}=10^2$~cm$^{-3}$, magnetic parameter $b=0.1$, and no external radiation field. Emission is integrated only in the postshock region (i.e. without radiative precursor) until the temperature drops to 100~K. These lines are a subset of those output by the code.}
\begin{tabular}{lcccccccc} \hline \hline
\multicolumn{3}{l}{$n_{\rm H}=10^2$ cm$^{-3}$, $b=0.1$, $G_0=0$} &  & & & &\\
& \multicolumn{8}{c}{Velocities (km s$^{-1}$)} \\ \cline{2-9} 
\multicolumn{9}{c}{\vspace{-0.3cm}} \\ 
Line & 10 & 20 & 30 & 40 & 50 & 60 & 70 & 80 \\ \cline{1-9}
\multicolumn{9}{c}{\vspace{-0.3cm}} \\ 
\cline{1-9} \multicolumn{9}{c}{H$_2$ rovibrational lines} \\
0-0 S(0) (28 $\mu$m) & $3.3(-08)$ & $2.3(-08)$ & $2.6(-08)$ & $1.8(-08)$ & $1.5(-09)$ & $5.6(-09)$ & $1.0(-08)$ & $1.7(-08)$ \\
0-0 S(1) (17 $\mu$m) & $5.2(-07)$ & $6.4(-07)$ & $6.7(-07)$ & $5.0(-07)$ & $4.6(-08)$ & $7.0(-08)$ & $8.6(-08)$ & $1.1(-07)$ \\
\end{tabular}

\tablefoot{Numbers in parentheses denote powers of ten.}
\end{table*}


\begin{table*}
  \centering
  \caption{Integrated line and two-photon continuum intensities (erg s$^{-1}$ cm$^{-2}$ sr$^{-1}$) escaping the shock front as a function of shock velocity (km s$^{-1}$) for shocks with preshock density $n_{\rm H}=10^3$~cm$^{-3}$, magnetic parameter $b=0.1$, and no external radiation field. Emission is integrated only in the postshock region (i.e. without radiative precursor) until the temperature drops to 100~K. These lines are a subset of those output by the code.}
\begin{tabular}{lcccccccc} \hline \hline
\multicolumn{3}{l}{$n_{\rm H}=10^3$ cm$^{-3}$, $b=0.1$, $G_0=0$} &  & & & &\\
 & \multicolumn{8}{c}{Velocities (km s$^{-1}$)} \\ \cline{2-9} 
\multicolumn{9}{c}{\vspace{-0.3cm}} \\ 
Line & 10 & 20 & 30 & 40 & 50 & 60 & 70 & 80 \\ \cline{1-9}
\multicolumn{9}{c}{\vspace{-0.3cm}} \\ 
\cline{1-9} \multicolumn{9}{c}{H$_2$ rovibrational lines} \\
0-0 S(0) (28 $\mu$m) & $1.4(-07)$ & $4.6(-08)$ & $1.0(-07)$ & $9.7(-08)$ & $1.8(-07)$ & $2.4(-07)$ & $3.2(-07)$ & $3.9(-07)$ \\
0-0 S(1) (17 $\mu$m) & $1.1(-06)$ & $1.6(-06)$ & $1.8(-06)$ & $4.3(-07)$ & $9.4(-07)$ & $1.6(-06)$ & $2.5(-06)$ & $3.6(-06)$ \\
\end{tabular}

\tablefoot{Numbers in parentheses denote powers of ten.}
\end{table*}


\begin{table*}
  \centering
  \caption{Integrated line and two-photon continuum intensities (erg s$^{-1}$ cm$^{-2}$ sr$^{-1}$) escaping the shock front as a function of shock velocity (km s$^{-1}$) for shocks with preshock density $n_{\rm H}=10^4$~cm$^{-3}$, magnetic parameter $b=0.1$, and no external radiation field. Emission is integrated only in the postshock region (i.e. without radiative precursor) until the temperature drops to 100~K. These lines are a subset of those output by the code.}
\begin{tabular}{lcccccccc} \hline \hline
\multicolumn{3}{l}{$n_{\rm H}=10^4$ cm$^{-3}$, $b=0.1$, $G_0=0$} &  & & & &\\
 & \multicolumn{8}{c}{Velocities (km s$^{-1}$)} \\ \cline{2-9} 
\multicolumn{9}{c}{\vspace{-0.3cm}} \\ 
Line & 10 & 20 & 30 & 40 & 50 & 60 & 70 & 80 \\ \cline{1-9}
\multicolumn{9}{c}{\vspace{-0.3cm}} \\ 
\cline{1-9} \multicolumn{9}{c}{H$_2$ rovibrational lines} \\
0-0 S(0) (28 $\mu$m) & $3.7(-07)$ & $9.0(-08)$ & $1.2(-06)$ & $1.5(-06)$ & $1.7(-06)$ & $1.9(-06)$ & $2.1(-06)$ & $5.5(-04)$ \\
0-0 S(1) (17 $\mu$m) & $1.9(-06)$ & $2.8(-06)$ & $2.7(-05)$ & $3.7(-05)$ & $4.5(-05)$ & $5.1(-05)$ & $5.8(-05)$ & $1.3(-02)$ \\
\end{tabular}

\tablefoot{Numbers in parentheses denote powers of ten.}
\end{table*}


\begin{table*}
  \centering
  \caption{Integrated line and two-photon continuum intensities (erg s$^{-1}$ cm$^{-2}$ sr$^{-1}$) escaping the shock front as a function of shock velocity (km s$^{-1}$) for shocks with preshock density $n_{\rm H}=10^5$~cm$^{-3}$, magnetic parameter $b=0.1$, and no external radiation field. Emission is integrated only in the postshock region (i.e. without radiative precursor) until the temperature drops to 100~K. These lines are a subset of those output by the code.}
\begin{tabular}{lcccccccc} \hline \hline
\multicolumn{3}{l}{$n_{\rm H}=10^5$ cm$^{-3}$, $b=0.1$, $G_0=0$} &  & & & &\\
 & \multicolumn{8}{c}{Velocities (km s$^{-1}$)} \\ \cline{2-9} 
\multicolumn{9}{c}{\vspace{-0.3cm}} \\ 
Line & 10 & 20 & 30 & 40 & 50 & 60 & 70 & 80 \\ \cline{1-9}
\multicolumn{9}{c}{\vspace{-0.3cm}} \\ 
\cline{1-9} \multicolumn{9}{c}{H$_2$ rovibrational lines} \\
0-0 S(0) (28 $\mu$m) & $5.2(-07)$ & $7.6(-07)$ & $4.3(-05)$ & $4.5(-06)$ & $5.0(-06)$ & $5.4(-06)$ & $6.4(-06)$ & $8.1(-06)$ \\
0-0 S(1) (17 $\mu$m) & $5.8(-06)$ & $1.6(-05)$ & $1.3(-03)$ & $1.4(-04)$ & $1.6(-04)$ & $1.8(-04)$ & $2.0(-04)$ & $2.2(-04)$ \\
\end{tabular}

\tablefoot{Numbers in parentheses denote powers of ten.}
\end{table*}


\begin{table*}
  \centering
  \caption{Integrated line and two-photon continuum intensities (erg s$^{-1}$ cm$^{-2}$ sr$^{-1}$) escaping the shock front as a function of shock velocity (km s$^{-1}$) for shocks with preshock density $n_{\rm H}=10^6$~cm$^{-3}$, magnetic parameter $b=0.1$, and no external radiation field. Emission is integrated only in the postshock region (i.e. without radiative precursor) until the temperature drops to 100~K. These lines are a subset of those output by the code.}
\begin{tabular}{lcccccccc} \hline \hline
\multicolumn{3}{l}{$n_{\rm H}=10^6$ cm$^{-3}$, $b=0.1$, $G_0=0$} &  & & & &\\
 & \multicolumn{8}{c}{Velocities (km s$^{-1}$)} \\ \cline{2-9} 
\multicolumn{9}{c}{\vspace{-0.3cm}} \\ 
Line & 10 & 20 & 30 & 40 & 50 & 60 & 70 & 80 \\ \cline{1-9}
\multicolumn{9}{c}{\vspace{-0.3cm}} \\ 
\cline{1-9} \multicolumn{9}{c}{H$_2$ rovibrational lines} \\
0-0 S(0) (28 $\mu$m) & $3.3(-07)$ & $1.3(-06)$ & $7.1(-06)$ & $1.7(-05)$ & $1.0(-05)$ & $1.2(-05)$ & $1.7(-05)$ & $1.7(-05)$ \\
0-0 S(1) (17 $\mu$m) & $1.4(-05)$ & $3.1(-05)$ & $1.7(-04)$ & $6.1(-04)$ & $3.0(-04)$ & $3.3(-04)$ & $4.4(-04)$ & $4.1(-04)$ \\
\end{tabular}

\tablefoot{Numbers in parentheses denote powers of ten.}
\end{table*}

\begin{table*}
  \centering
  \caption{Integrated line and two-photon continuum intensities (erg s$^{-1}$ cm$^{-2}$ sr$^{-1}$) escaping the shock front as a function of shock velocity (km s$^{-1}$) for shocks with preshock density $n_{\rm H}=10^2$~cm$^{-3}$, magnetic parameter $b=1$, and no external radiation field. Emission is integrated only in the postshock region (i.e. without radiative precursor) until the temperature drops to 100~K. These lines are a subset of those output by the code.}
\begin{tabular}{lcccccccc} \hline \hline
\multicolumn{3}{l}{$n_{\rm H}=10^2$ cm$^{-3}$, $b=1$, $G_0=0$} &  & & & &\\
 & \multicolumn{8}{c}{Velocities (km s$^{-1}$)} \\ \cline{2-9} 
\multicolumn{9}{c}{\vspace{-0.3cm}} \\ 
Line & 10 & 20 & 30 & 40 & 50 & 60 & 70 & 80 \\ \cline{1-9}
\multicolumn{9}{c}{\vspace{-0.3cm}} \\ 
\cline{1-9} \multicolumn{9}{c}{H$_2$ rovibrational lines} \\
0-0 S(0) (28 $\mu$m) & $2.0(-06)$ & $2.8(-06)$ & $8.7(-08)$ & $6.6(-08)$ & $7.0(-09)$ & $9.4(-10)$ & $8.9(-10)$ & $8.7(-10)$ \\
0-0 S(1) (17 $\mu$m) & $2.1(-06)$ & $2.7(-05)$ & $1.6(-06)$ & $1.4(-06)$ & $2.3(-07)$ & $3.3(-08)$ & $3.3(-08)$ & $3.3(-08)$ \\
\end{tabular}

\tablefoot{Numbers in parentheses denote powers of ten.}
\end{table*}


\begin{table*}
  \centering
  \caption{Integrated line and two-photon continuum intensities (erg s$^{-1}$ cm$^{-2}$ sr$^{-1}$) escaping the shock front as a function of shock velocity (km s$^{-1}$) for shocks with preshock density $n_{\rm H}=10^3$~cm$^{-3}$, magnetic parameter $b=1$, and no external radiation field. Emission is integrated only in the postshock region (i.e. without radiative precursor) until the temperature drops to 100~K. These lines are a subset of those output by the code.}
\begin{tabular}{lcccccccc} \hline \hline
\multicolumn{3}{l}{$n_{\rm H}=10^3$ cm$^{-3}$, $b=1$, $G_0=0$} &  & & & &\\
 & \multicolumn{8}{c}{Velocities (km s$^{-1}$)} \\ \cline{2-9} 
\multicolumn{9}{c}{\vspace{-0.3cm}} \\ 
Line & 10 & 20 & 30 & 40 & 50 & 60 & 70 & 80 \\ \cline{1-9}
\multicolumn{9}{c}{\vspace{-0.3cm}} \\ 
\cline{1-9} \multicolumn{9}{c}{H$_2$ rovibrational lines} \\
0-0 S(0) (28 $\mu$m) & $6.4(-06)$ & $7.3(-06)$ & $2.5(-07)$ & $8.3(-09)$ & $5.4(-08)$ & $1.2(-07)$ & $2.1(-07)$ & $3.2(-07)$ \\
0-0 S(1) (17 $\mu$m) & $1.3(-05)$ & $5.3(-05)$ & $5.3(-06)$ & $2.7(-07)$ & $6.4(-07)$ & $8.8(-07)$ & $1.1(-06)$ & $1.4(-06)$ \\
\end{tabular}

\tablefoot{Numbers in parentheses denote powers of ten.}
\end{table*}


\begin{table*}
  \centering
  \caption{Integrated line and two-photon continuum intensities (erg s$^{-1}$ cm$^{-2}$ sr$^{-1}$) escaping the shock front as a function of shock velocity (km s$^{-1}$) for shocks with preshock density $n_{\rm H}=10^4$~cm$^{-3}$, magnetic parameter $b=1$, and no external radiation field. Emission is integrated only in the postshock region (i.e. without radiative precursor) until the temperature drops to 100~K. These lines are a subset of those output by the code.}
\begin{tabular}{lcccccccc} \hline \hline
\multicolumn{3}{l}{$n_{\rm H}=10^4$ cm$^{-3}$, $b=1$, $G_0=0$} &  & & & &\\
 & \multicolumn{8}{c}{Velocities (km s$^{-1}$)} \\ \cline{2-9} 
\multicolumn{9}{c}{\vspace{-0.3cm}} \\ 
Line & 10 & 20 & 30 & 40 & 50 & 60 & 70 & 80 \\ \cline{1-9}
\multicolumn{9}{c}{\vspace{-0.3cm}} \\ 
\cline{1-9} \multicolumn{9}{c}{H$_2$ rovibrational lines} \\
0-0 S(0) (28 $\mu$m) & $1.4(-05)$ & $1.4(-05)$ & $6.5(-07)$ & $1.1(-06)$ & $1.9(-06)$ & $2.8(-06)$ & $3.6(-06)$ & $4.3(-06)$ \\
0-0 S(1) (17 $\mu$m) & $1.4(-05)$ & $6.1(-05)$ & $2.7(-06)$ & $5.0(-06)$ & $1.1(-05)$ & $2.3(-05)$ & $3.8(-05)$ & $5.8(-05)$ \\
\end{tabular}

\tablefoot{Numbers in parentheses denote powers of ten.}
\end{table*}


\begin{table*}
  \centering
  \caption{Integrated line and two-photon continuum intensities (erg s$^{-1}$ cm$^{-2}$ sr$^{-1}$) escaping the shock front as a function of shock velocity (km s$^{-1}$) for shocks with preshock density $n_{\rm H}=10^5$~cm$^{-3}$, magnetic parameter $b=1$, and no external radiation field. Emission is integrated only in the postshock region (i.e. without radiative precursor) until the temperature drops to 100~K. These lines are a subset of those output by the code.}
\begin{tabular}{lcccccccc} \hline \hline
\multicolumn{3}{l}{$n_{\rm H}=10^5$ cm$^{-3}$, $b=1$, $G_0=0$} &  & & & &\\
 & \multicolumn{8}{c}{Velocities (km s$^{-1}$)} \\ \cline{2-9} 
\multicolumn{9}{c}{\vspace{-0.3cm}} \\ 
Line & 10 & 20 & 30 & 40 & 50 & 60 & 70 & 80 \\ \cline{1-9}
\multicolumn{9}{c}{\vspace{-0.3cm}} \\ 
\cline{1-9} \multicolumn{9}{c}{H$_2$ rovibrational lines} \\
0-0 S(0) (28 $\mu$m) & $2.6(-05)$ & $1.3(-05)$ & $1.1(-05)$ & $1.4(-05)$ & $1.6(-05)$ & $1.8(-05)$ & $1.9(-05)$ & $2.1(-05)$ \\
0-0 S(1) (17 $\mu$m) & $2.8(-05)$ & $1.7(-04)$ & $2.5(-04)$ & $3.6(-04)$ & $4.4(-04)$ & $5.1(-04)$ & $5.7(-04)$ & $6.2(-04)$ \\
\end{tabular}

\tablefoot{Numbers in parentheses denote powers of ten.}
\end{table*}


\begin{table*}
  \centering
  \caption{Integrated line and two-photon continuum intensities (erg s$^{-1}$ cm$^{-2}$ sr$^{-1}$) escaping the shock front as a function of shock velocity (km s$^{-1}$) for shocks with preshock density $n_{\rm H}=10^6$~cm$^{-3}$, magnetic parameter $b=1$, and no external radiation field. Emission is integrated only in the postshock region (i.e. without radiative precursor) until the temperature drops to 100~K. These lines are a subset of those output by the code.}
\begin{tabular}{lcccccccc} \hline \hline
\multicolumn{3}{l}{$n_{\rm H}=10^6$ cm$^{-3}$, $b=1$, $G_0=0$} &  & & & &\\
 & \multicolumn{8}{c}{Velocities (km s$^{-1}$)} \\ \cline{2-9} 
\multicolumn{9}{c}{\vspace{-0.3cm}} \\ 
Line & 10 & 20 & 30 & 40 & 50 & 60 & 70 & 80 \\ \cline{1-9}
\multicolumn{9}{c}{\vspace{-0.3cm}} \\ 
\cline{1-9} \multicolumn{9}{c}{H$_2$ rovibrational lines} \\
0-0 S(0) (28 $\mu$m) & $4.4(-05)$ & $1.0(-05)$ & $2.8(-05)$ & $3.9(-05)$ & $4.3(-05)$ & $4.6(-05)$ & $4.9(-05)$ & $5.1(-05)$ \\
0-0 S(1) (17 $\mu$m) & $6.6(-05)$ & $3.1(-04)$ & $6.6(-04)$ & $1.2(-03)$ & $1.4(-03)$ & $1.5(-03)$ & $1.6(-03)$ & $1.8(-03)$ \\
\end{tabular}

\tablefoot{Numbers in parentheses denote powers of ten.}
\end{table*}

\begin{table*}
  \centering
  \caption{Integrated line and two-photon continuum intensities (erg s$^{-1}$ cm$^{-2}$ sr$^{-1}$) escaping the shock front as a function of shock velocity (km s$^{-1}$) for shocks with preshock density $n_{\rm H}=10^2$~cm$^{-3}$, magnetic parameter $b=1$, and external radiation field $G_0=10^3$. Emission is integrated only in the postshock region (i.e. without radiative precursor) until the temperature drops to 100~K. These lines are a subset of those output by the code.}
\begin{tabular}{lcccccccc} \hline \hline
\multicolumn{3}{l}{$n_{\rm H}=10^2$ cm$^{-3}$, $b=1$, $G_0=10^3$} &  & & & &\\
 & \multicolumn{8}{c}{Velocities (km s$^{-1}$)} \\ \cline{2-9} 
\multicolumn{9}{c}{\vspace{-0.3cm}} \\ 
Line & 10 & 20 & 30 & 40 & 50 & 60 & 70 & 80 \\ \cline{1-9}
\multicolumn{9}{c}{\vspace{-0.3cm}} \\ 
\cline{1-9} \multicolumn{9}{c}{H$_2$ rovibrational lines} \\
0-0 S(0) (28 $\mu$m) & $4.2(-12)$ & $5.9(-12)$ & $9.7(-12)$ & $1.2(-11)$ & $1.6(-11)$ & $2.1(-11)$ & $2.3(-11)$ & $2.9(-11)$ \\
0-0 S(1) (17 $\mu$m) & $3.9(-10)$ & $5.5(-10)$ & $9.0(-10)$ & $1.1(-09)$ & $1.5(-09)$ & $1.9(-09)$ & $2.1(-09)$ & $2.7(-09)$ \\
0-0 S(2) (12 $\mu$m) & $8.2(-10)$ & $1.2(-09)$ & $2.0(-09)$ & $2.4(-09)$ & $3.3(-09)$ & $4.2(-09)$ & $4.7(-09)$ & $6.0(-09)$ \\
0-0 S(3) (10 $\mu$m) & $6.8(-09)$ & $1.1(-08)$ & $1.7(-08)$ & $2.2(-08)$ & $2.9(-08)$ & $3.8(-08)$ & $4.3(-08)$ & $5.5(-08)$ \\
0-0 S(4) (8.0 $\mu$m) & $3.5(-09)$ & $6.0(-09)$ & $9.9(-09)$ & $1.3(-08)$ & $1.7(-08)$ & $2.2(-08)$ & $2.5(-08)$ & $3.2(-08)$ \\
0-0 S(5) (6.9 $\mu$m) & $1.2(-08)$ & $2.3(-08)$ & $3.8(-08)$ & $4.9(-08)$ & $6.6(-08)$ & $8.5(-08)$ & $9.7(-08)$ & $1.2(-07)$ \\
0-0 S(6) (6.1 $\mu$m) & $3.9(-09)$ & $8.5(-09)$ & $1.4(-08)$ & $1.8(-08)$ & $2.4(-08)$ & $3.0(-08)$ & $3.5(-08)$ & $4.4(-08)$ \\
0-0 S(7) (5.5 $\mu$m) & $1.1(-08)$ & $2.5(-08)$ & $4.2(-08)$ & $5.5(-08)$ & $7.3(-08)$ & $9.2(-08)$ & $1.1(-07)$ & $1.3(-07)$ \\
0-0 S(8) (5.1 $\mu$m) & $3.1(-09)$ & $8.1(-09)$ & $1.4(-08)$ & $1.8(-08)$ & $2.3(-08)$ & $2.9(-08)$ & $3.3(-08)$ & $4.0(-08)$ \\
0-0 S(9) (4.7 $\mu$m) & $8.7(-09)$ & $2.1(-08)$ & $3.7(-08)$ & $4.8(-08)$ & $6.3(-08)$ & $7.8(-08)$ & $8.9(-08)$ & $1.1(-07)$ \\
1-0 S(0) (2.22 $\mu$m) & $8.0(-09)$ & $1.3(-08)$ & $2.1(-08)$ & $2.5(-08)$ & $3.3(-08)$ & $4.2(-08)$ & $4.7(-08)$ & $5.9(-08)$ \\
1-0 S(1) (2.12 $\mu$m) & $3.3(-08)$ & $5.4(-08)$ & $8.8(-08)$ & $1.1(-07)$ & $1.4(-07)$ & $1.8(-07)$ & $2.0(-07)$ & $2.6(-07)$ \\
1-0 S(2) (2.03 $\mu$m) & $1.2(-08)$ & $2.1(-08)$ & $3.4(-08)$ & $4.3(-08)$ & $5.6(-08)$ & $7.1(-08)$ & $7.9(-08)$ & $9.8(-08)$ \\
1-0 S(3) (1.96 $\mu$m) & $2.7(-08)$ & $5.2(-08)$ & $8.7(-08)$ & $1.1(-07)$ & $1.4(-07)$ & $1.8(-07)$ & $2.0(-07)$ & $2.5(-07)$ \\
2-1 S(0) (2.36 $\mu$m) & $4.0(-09)$ & $5.8(-09)$ & $9.4(-09)$ & $1.1(-08)$ & $1.4(-08)$ & $1.8(-08)$ & $1.9(-08)$ & $2.4(-08)$ \\
2-1 S(1) (2.25 $\mu$m) & $1.6(-08)$ & $2.4(-08)$ & $3.9(-08)$ & $4.7(-08)$ & $6.1(-08)$ & $7.6(-08)$ & $8.3(-08)$ & $1.0(-07)$ \\
2-1 S(2) (2.15 $\mu$m) & $5.8(-09)$ & $9.1(-09)$ & $1.5(-08)$ & $1.8(-08)$ & $2.3(-08)$ & $2.9(-08)$ & $3.2(-08)$ & $3.9(-08)$ \\
2-1 S(3) (2.07 $\mu$m) & $1.2(-08)$ & $2.2(-08)$ & $3.5(-08)$ & $4.4(-08)$ & $5.6(-08)$ & $6.9(-08)$ & $7.6(-08)$ & $9.3(-08)$ \\
\cline{1-9} \multicolumn{9}{c}{CH$^+$ rovibrational lines} \\
0-0 (J=1-0) (359 $\mu$m) & $4.9(-12)$ & $2.5(-11)$ & $2.2(-11)$ & $1.6(-11)$ & $7.4(-11)$ & $1.5(-10)$ & $2.4(-10)$ & $3.4(-10)$ \\
0-0 (J=2-1) (180 $\mu$m) & $5.3(-12)$ & $3.2(-11)$ & $8.3(-11)$ & $1.5(-10)$ & $2.6(-10)$ & $3.7(-10)$ & $5.0(-10)$ & $6.4(-10)$ \\
0-0 (J=3-2) (120 $\mu$m) & $4.8(-12)$ & $3.4(-11)$ & $8.6(-11)$ & $1.6(-10)$ & $2.5(-10)$ & $3.5(-10)$ & $4.6(-10)$ & $6.0(-10)$ \\
1-0 (J=0-1) (3.69 $\mu$m) & $3.2(-14)$ & $4.7(-13)$ & $9.8(-13)$ & $1.5(-12)$ & $2.1(-12)$ & $2.8(-12)$ & $3.6(-12)$ & $4.7(-12)$ \\
1-0 (J=1-0) (3.61 $\mu$m) & $3.6(-14)$ & $5.8(-13)$ & $8.2(-13)$ & $9.3(-13)$ & $1.3(-12)$ & $1.8(-12)$ & $2.6(-12)$ & $3.8(-12)$ \\
\cline{1-9} \multicolumn{9}{c}{CO pure rotational lines} \\
J=1-0 (2601 $\mu$m) & $7.8(-19)$ & $3.9(-17)$ & $8.1(-17)$ & $1.2(-16)$ & $1.7(-16)$ & $2.2(-16)$ & $2.7(-16)$ & $3.4(-16)$ \\
J=2-1 (1300 $\mu$m) & $1.5(-17)$ & $1.1(-15)$ & $2.8(-15)$ & $4.5(-15)$ & $6.3(-15)$ & $8.0(-15)$ & $9.9(-15)$ & $1.2(-14)$ \\
J=3-2 (867 $\mu$m) & $3.7(-17)$ & $3.8(-15)$ & $1.4(-14)$ & $2.7(-14)$ & $4.2(-14)$ & $5.7(-14)$ & $7.3(-14)$ & $9.2(-14)$ \\
J=4-3 (650 $\mu$m) & $4.5(-17)$ & $6.0(-15)$ & $2.8(-14)$ & $6.7(-14)$ & $1.2(-13)$ & $1.8(-13)$ & $2.4(-13)$ & $3.2(-13)$ \\
J=8-7 (434 $\mu$m) & $2.9(-17)$ & $4.3(-15)$ & $2.6(-14)$ & $8.0(-14)$ & $1.8(-13)$ & $3.4(-13)$ & $5.7(-13)$ & $9.0(-13)$ \\
J=9-8 (372 $\mu$m) & $2.4(-17)$ & $3.5(-15)$ & $2.1(-14)$ & $6.4(-14)$ & $1.5(-13)$ & $2.8(-13)$ & $4.7(-13)$ & $7.5(-13)$ \\
J=10-9 (325 $\mu$m) & $2.0(-17)$ & $2.9(-15)$ & $1.7(-14)$ & $5.0(-14)$ & $1.2(-13)$ & $2.2(-13)$ & $3.7(-13)$ & $6.0(-13)$ \\
\cline{1-9} \multicolumn{9}{c}{Atomic H lines} \\
Ly$\alpha$ (1215.7 \AA) & 0 & 0 & 0 & $1.4(-05)$ & $6.6(-05)$ & $2.0(-04)$ & $3.5(-04)$ & $5.6(-04)$ \\
Ly$\beta$ (1025.7 \AA) & 0 & 0 & 0 & $3.6(-07)$ & $3.9(-06)$ & $1.6(-05)$ & $4.0(-05)$ & $7.5(-05)$ \\
H$\alpha$ (6564.6 \AA) & 0 & 0 & 0 & $1.1(-05)$ & $2.3(-05)$ & $4.0(-05)$ & $5.9(-05)$ & $8.2(-05)$ \\
2ph (<2400 \AA) & 0 & 0 & 0 & $3.2(-05)$ & $5.8(-05)$ & $9.3(-05)$ & $1.4(-04)$ & $1.9(-04)$ \\
\cline{1-9} \multicolumn{9}{c}{Atomic fine structure lines} \\
C$^+$ (158 $\mu$m) & $2.2(-06)$ & $7.5(-06)$ & $7.8(-06)$ & $7.1(-06)$ & $6.6(-06)$ & $6.4(-06)$ & $6.3(-06)$ & $6.6(-06)$ \\
C (609.8 $\mu$m) & $4.1(-12)$ & $1.0(-11)$ & $3.5(-11)$ & $6.2(-11)$ & $1.0(-10)$ & $1.5(-10)$ & $2.0(-10)$ & $2.6(-10)$ \\
C (370.4 $\mu$m) & $1.8(-11)$ & $6.4(-11)$ & $2.5(-10)$ & $4.8(-10)$ & $8.1(-10)$ & $1.2(-09)$ & $1.7(-09)$ & $2.2(-09)$ \\
O (145.5 $\mu$m) & $3.9(-07)$ & $3.9(-06)$ & $4.9(-06)$ & $4.9(-06)$ & $4.9(-06)$ & $5.0(-06)$ & $5.2(-06)$ & $5.6(-06)$ \\
O (63.2 $\mu$m) & $4.8(-06)$ & $4.7(-05)$ & $6.1(-05)$ & $6.6(-05)$ & $6.8(-05)$ & $7.3(-05)$ & $7.9(-05)$ & $8.8(-05)$ \\
S (25.3 $\mu$m) & $2.1(-12)$ & $5.1(-11)$ & $5.8(-10)$ & $2.0(-09)$ & $5.0(-09)$ & $1.1(-08)$ & $2.4(-08)$ & $4.5(-08)$ \\
Si (68.5 $\mu$m) & $8.7(-14)$ & $8.6(-13)$ & $6.4(-12)$ & $2.1(-11)$ & $5.3(-11)$ & $1.0(-10)$ & $1.6(-10)$ & $2.2(-10)$ \\
Si$^+$ (34.8 $\mu$m) & $1.4(-07)$ & $8.4(-07)$ & $1.4(-06)$ & $2.1(-06)$ & $3.3(-06)$ & $4.7(-06)$ & $5.8(-06)$ & $6.8(-06)$ \\
\cline{1-9} \multicolumn{9}{c}{Atomic metastable lines} \\
O (6300 \AA) & $5.9(-12)$ & $1.5(-07)$ & $9.1(-07)$ & $2.2(-06)$ & $4.0(-06)$ & $6.1(-06)$ & $8.1(-06)$ & $1.0(-05)$ \\
O (6364 \AA) & $1.9(-12)$ & $4.9(-08)$ & $2.9(-07)$ & $7.0(-07)$ & $1.3(-06)$ & $1.9(-06)$ & $2.6(-06)$ & $3.3(-06)$ \\
S$^+$ (6731 \AA) & $5.4(-11)$ & $1.8(-07)$ & $1.8(-06)$ & $5.4(-06)$ & $1.2(-05)$ & $2.3(-05)$ & $3.6(-05)$ & $5.4(-05)$ \\
S$^+$ (6716 \AA) & $8.0(-11)$ & $2.7(-07)$ & $2.6(-06)$ & $7.6(-06)$ & $1.6(-05)$ & $2.4(-05)$ & $3.2(-05)$ & $4.2(-05)$ \\
N (5200 \AA) & $3.7(-13)$ & $4.0(-08)$ & $5.3(-07)$ & $1.2(-06)$ & $1.9(-06)$ & $2.2(-06)$ & $2.2(-06)$ & $2.2(-06)$ \\
N (5198 \AA) & $2.5(-13)$ & $2.7(-08)$ & $3.6(-07)$ & $8.7(-07)$ & $1.5(-06)$ & $2.0(-06)$ & $2.4(-06)$ & $2.7(-06)$ \\
\cline{1-9}
\end{tabular}

\tablefoot{Numbers in parentheses denote powers of ten.}
\end{table*}

\begin{table*}
  \centering
  \caption{Integrated line and two-photon continuum intensities (erg s$^{-1}$ cm$^{-2}$ sr$^{-1}$) escaping the shock front as a function of shock velocity (km s$^{-1}$) for shocks with preshock density $n_{\rm H}=10^3$~cm$^{-3}$, magnetic parameter $b=1$, and external radiation field $G_0=10^3$. Emission is integrated only in the postshock region (i.e. without radiative precursor) until the temperature drops to 100~K. These lines are a subset of those output by the code.}
\begin{tabular}{lcccccccc} \hline \hline
\multicolumn{3}{l}{$n_{\rm H}=10^3$ cm$^{-3}$, $b=1$, $G_0=10^3$} &  & & & &\\
 & \multicolumn{8}{c}{Velocities (km s$^{-1}$)} \\ \cline{2-9} 
\multicolumn{9}{c}{\vspace{-0.3cm}} \\ 
Line & 10 & 20 & 30 & 40 & 50 & 60 & 70 & 80 \\ \cline{1-9}
\multicolumn{9}{c}{\vspace{-0.3cm}} \\ 
\cline{1-9} \multicolumn{9}{c}{H$_2$ rovibrational lines} \\
0-0 S(0) (28 $\mu$m) & $1.1(-10)$ & $2.2(-10)$ & $3.2(-10)$ & $5.0(-10)$ & $5.9(-10)$ & $7.0(-10)$ & $9.6(-10)$ & $1.3(-09)$ \\
0-0 S(1) (17 $\mu$m) & $9.8(-09)$ & $1.9(-08)$ & $2.6(-08)$ & $3.9(-08)$ & $4.4(-08)$ & $4.8(-08)$ & $6.4(-08)$ & $8.2(-08)$ \\
0-0 S(2) (12 $\mu$m) & $1.3(-08)$ & $2.8(-08)$ & $4.2(-08)$ & $6.7(-08)$ & $7.7(-08)$ & $8.6(-08)$ & $1.2(-07)$ & $1.5(-07)$ \\
0-0 S(3) (10 $\mu$m) & $7.5(-08)$ & $1.8(-07)$ & $2.8(-07)$ & $4.5(-07)$ & $5.4(-07)$ & $6.2(-07)$ & $8.5(-07)$ & $1.1(-06)$ \\
0-0 S(4) (8.0 $\mu$m) & $4.2(-08)$ & $1.0(-07)$ & $1.7(-07)$ & $2.7(-07)$ & $3.2(-07)$ & $3.8(-07)$ & $5.2(-07)$ & $6.9(-07)$ \\
0-0 S(5) (6.9 $\mu$m) & $1.3(-07)$ & $3.7(-07)$ & $6.0(-07)$ & $9.5(-07)$ & $1.2(-06)$ & $1.4(-06)$ & $1.9(-06)$ & $2.5(-06)$ \\
0-0 S(6) (6.1 $\mu$m) & $4.8(-08)$ & $1.5(-07)$ & $2.4(-07)$ & $3.8(-07)$ & $4.8(-07)$ & $5.8(-07)$ & $7.7(-07)$ & $1.0(-06)$ \\
0-0 S(7) (5.5 $\mu$m) & $1.2(-07)$ & $4.5(-07)$ & $7.3(-07)$ & $1.1(-06)$ & $1.4(-06)$ & $1.7(-06)$ & $2.3(-06)$ & $2.9(-06)$ \\
0-0 S(8) (5.1 $\mu$m) & $3.5(-08)$ & $1.5(-07)$ & $2.5(-07)$ & $3.7(-07)$ & $4.7(-07)$ & $5.6(-07)$ & $7.4(-07)$ & $9.5(-07)$ \\
0-0 S(9) (4.7 $\mu$m) & $8.4(-08)$ & $4.4(-07)$ & $6.9(-07)$ & $1.0(-06)$ & $1.3(-06)$ & $1.6(-06)$ & $2.0(-06)$ & $2.6(-06)$ \\
1-0 S(0) (2.22 $\mu$m) & $6.3(-08)$ & $1.8(-07)$ & $2.7(-07)$ & $4.1(-07)$ & $5.1(-07)$ & $6.0(-07)$ & $7.8(-07)$ & $9.9(-07)$ \\
1-0 S(1) (2.12 $\mu$m) & $1.6(-07)$ & $6.1(-07)$ & $9.7(-07)$ & $1.4(-06)$ & $1.8(-06)$ & $2.2(-06)$ & $2.8(-06)$ & $3.5(-06)$ \\
1-0 S(2) (2.03 $\mu$m) & $9.3(-08)$ & $3.0(-07)$ & $4.6(-07)$ & $6.9(-07)$ & $8.4(-07)$ & $9.8(-07)$ & $1.3(-06)$ & $1.6(-06)$ \\
1-0 S(3) (1.96 $\mu$m) & $1.6(-07)$ & $7.2(-07)$ & $1.1(-06)$ & $1.6(-06)$ & $2.1(-06)$ & $2.5(-06)$ & $3.1(-06)$ & $3.8(-06)$ \\
2-1 S(0) (2.36 $\mu$m) & $2.5(-08)$ & $6.0(-08)$ & $8.3(-08)$ & $1.2(-07)$ & $1.3(-07)$ & $1.4(-07)$ & $1.9(-07)$ & $2.4(-07)$ \\
2-1 S(1) (2.25 $\mu$m) & $5.6(-08)$ & $1.9(-07)$ & $2.7(-07)$ & $3.9(-07)$ & $4.4(-07)$ & $4.9(-07)$ & $6.2(-07)$ & $7.6(-07)$ \\
2-1 S(2) (2.15 $\mu$m) & $3.6(-08)$ & $1.0(-07)$ & $1.4(-07)$ & $2.0(-07)$ & $2.3(-07)$ & $2.5(-07)$ & $3.2(-07)$ & $3.9(-07)$ \\
2-1 S(3) (2.07 $\mu$m) & $5.4(-08)$ & $2.3(-07)$ & $3.3(-07)$ & $4.6(-07)$ & $5.4(-07)$ & $5.9(-07)$ & $7.3(-07)$ & $8.8(-07)$ \\
\cline{1-9} \multicolumn{9}{c}{CH$^+$ rovibrational lines} \\
0-0 (J=1-0) (359 $\mu$m) & $3.5(-10)$ & $3.7(-09)$ & $9.5(-09)$ & $1.8(-08)$ & $3.0(-08)$ & $4.6(-08)$ & $6.3(-08)$ & $8.2(-08)$ \\
0-0 (J=2-1) (180 $\mu$m) & $4.3(-10)$ & $5.0(-09)$ & $1.3(-08)$ & $2.5(-08)$ & $4.1(-08)$ & $6.2(-08)$ & $8.6(-08)$ & $1.2(-07)$ \\
0-0 (J=3-2) (120 $\mu$m) & $4.2(-10)$ & $5.5(-09)$ & $1.5(-08)$ & $2.7(-08)$ & $4.4(-08)$ & $6.6(-08)$ & $9.2(-08)$ & $1.3(-07)$ \\
1-0 (J=0-1) (3.69 $\mu$m) & $2.6(-12)$ & $5.2(-11)$ & $1.4(-10)$ & $2.4(-10)$ & $4.0(-10)$ & $5.8(-10)$ & $7.9(-10)$ & $1.1(-09)$ \\
1-0 (J=1-0) (3.61 $\mu$m) & $3.1(-12)$ & $7.0(-11)$ & $1.8(-10)$ & $3.2(-10)$ & $5.3(-10)$ & $7.7(-10)$ & $1.0(-09)$ & $1.4(-09)$ \\
\cline{1-9} \multicolumn{9}{c}{CO pure rotational lines} \\
J=1-0 (2601 $\mu$m) & $3.7(-16)$ & $3.9(-15)$ & $9.2(-15)$ & $1.6(-14)$ & $2.5(-14)$ & $3.5(-14)$ & $4.6(-14)$ & $6.1(-14)$ \\
J=2-1 (1300 $\mu$m) & $1.4(-14)$ & $1.4(-13)$ & $3.1(-13)$ & $5.2(-13)$ & $8.0(-13)$ & $1.1(-12)$ & $1.5(-12)$ & $2.0(-12)$ \\
J=3-2 (867 $\mu$m) & $9.8(-14)$ & $1.1(-12)$ & $2.4(-12)$ & $4.0(-12)$ & $6.2(-12)$ & $8.7(-12)$ & $1.1(-11)$ & $1.5(-11)$ \\
J=4-3 (650 $\mu$m) & $3.1(-13)$ & $4.3(-12)$ & $1.0(-11)$ & $1.7(-11)$ & $2.6(-11)$ & $3.7(-11)$ & $4.8(-11)$ & $6.3(-11)$ \\
J=8-7 (434 $\mu$m) & $6.0(-13)$ & $2.5(-11)$ & $1.1(-10)$ & $2.7(-10)$ & $4.9(-10)$ & $7.7(-10)$ & $1.1(-09)$ & $1.5(-09)$ \\
J=9-8 (372 $\mu$m) & $4.8(-13)$ & $2.3(-11)$ & $1.2(-10)$ & $3.1(-10)$ & $6.1(-10)$ & $1.0(-09)$ & $1.5(-09)$ & $2.1(-09)$ \\
J=10-9 (325 $\mu$m) & $3.7(-13)$ & $2.0(-11)$ & $1.1(-10)$ & $3.1(-10)$ & $6.7(-10)$ & $1.2(-09)$ & $1.8(-09)$ & $2.6(-09)$ \\
\cline{1-9} \multicolumn{9}{c}{Atomic H lines} \\
Ly$\alpha$ (1215.7 \AA) & 0 & 0 & 0 & $6.1(-04)$ & $1.2(-03)$ & $2.2(-03)$ & $3.5(-03)$ & $4.5(-03)$ \\
Ly$\beta$ (1025.7 \AA) & 0 & 0 & 0 & $3.2(-06)$ & $3.6(-05)$ & $1.5(-04)$ & $3.8(-04)$ & $7.3(-04)$ \\
H$\alpha$ (6564.6 \AA) & 0 & 0 & 0 & $1.1(-04)$ & $2.4(-04)$ & $4.0(-04)$ & $6.0(-04)$ & $8.6(-04)$ \\
2ph (<2400 \AA) & 0 & 0 & 0 & $3.2(-04)$ & $5.9(-04)$ & $9.3(-04)$ & $1.3(-03)$ & $1.9(-03)$ \\
\cline{1-9} \multicolumn{9}{c}{Atomic fine structure lines} \\
C$^+$ (158 $\mu$m) & $5.3(-06)$ & $2.1(-05)$ & $1.8(-05)$ & $1.6(-05)$ & $1.4(-05)$ & $1.4(-05)$ & $1.5(-05)$ & $1.6(-05)$ \\
C (609.8 $\mu$m) & $8.2(-11)$ & $4.8(-10)$ & $1.2(-09)$ & $2.2(-09)$ & $3.1(-09)$ & $4.0(-09)$ & $5.3(-09)$ & $6.7(-09)$ \\
C (370.4 $\mu$m) & $6.2(-10)$ & $3.9(-09)$ & $1.0(-08)$ & $1.9(-08)$ & $2.7(-08)$ & $3.4(-08)$ & $4.5(-08)$ & $5.7(-08)$ \\
O (145.5 $\mu$m) & $3.3(-06)$ & $3.0(-05)$ & $3.1(-05)$ & $2.8(-05)$ & $2.6(-05)$ & $2.6(-05)$ & $2.9(-05)$ & $3.2(-05)$ \\
O (63.2 $\mu$m) & $5.0(-05)$ & $5.1(-04)$ & $6.4(-04)$ & $6.5(-04)$ & $6.6(-04)$ & $7.1(-04)$ & $8.0(-04)$ & $9.3(-04)$ \\
S (25.3 $\mu$m) & $7.7(-10)$ & $2.7(-08)$ & $1.6(-07)$ & $4.0(-07)$ & $8.1(-07)$ & $1.5(-06)$ & $2.6(-06)$ & $4.1(-06)$ \\
Si (68.5 $\mu$m) & $2.5(-11)$ & $3.5(-10)$ & $1.4(-09)$ & $2.7(-09)$ & $4.1(-09)$ & $5.2(-09)$ & $6.6(-09)$ & $8.3(-09)$ \\
Si$^+$ (34.8 $\mu$m) & $1.5(-06)$ & $1.1(-05)$ & $1.8(-05)$ & $2.4(-05)$ & $2.9(-05)$ & $3.4(-05)$ & $4.1(-05)$ & $4.8(-05)$ \\
\cline{1-9} \multicolumn{9}{c}{Atomic metastable lines} \\
O (6300 \AA) & $5.3(-11)$ & $1.9(-06)$ & $8.6(-06)$ & $2.2(-05)$ & $4.4(-05)$ & $7.1(-05)$ & $1.0(-04)$ & $1.4(-04)$ \\
O (6364 \AA) & $1.7(-11)$ & $6.2(-07)$ & $2.7(-06)$ & $7.0(-06)$ & $1.4(-05)$ & $2.3(-05)$ & $3.3(-05)$ & $4.5(-05)$ \\
S$^+$ (6731 \AA) & $4.1(-10)$ & $1.4(-06)$ & $1.3(-05)$ & $5.2(-05)$ & $1.3(-04)$ & $2.0(-04)$ & $2.6(-04)$ & $3.1(-04)$ \\
S$^+$ (6716 \AA) & $6.0(-10)$ & $2.1(-06)$ & $1.8(-05)$ & $6.0(-05)$ & $1.0(-04)$ & $1.3(-04)$ & $1.5(-04)$ & $1.7(-04)$ \\
N (5200 \AA) & $3.2(-12)$ & $3.1(-07)$ & $4.2(-06)$ & $9.4(-06)$ & $1.1(-05)$ & $1.0(-05)$ & $9.2(-06)$ & $9.6(-06)$ \\
N (5198 \AA) & $2.1(-12)$ & $2.1(-07)$ & $2.9(-06)$ & $8.5(-06)$ & $1.3(-05)$ & $1.4(-05)$ & $1.4(-05)$ & $1.5(-05)$ \\
\cline{1-9}
\end{tabular}

\tablefoot{Numbers in parentheses denote powers of ten.}
\end{table*}

\begin{table*}
  \centering
  \caption{Integrated line and two-photon continuum intensities (erg s$^{-1}$ cm$^{-2}$ sr$^{-1}$) escaping the shock front as a function of shock velocity (km s$^{-1}$) for shocks with preshock density $n_{\rm H}=10^4$~cm$^{-3}$, magnetic parameter $b=1$, and external radiation field $G_0=10^3$. Emission is integrated only in the postshock region (i.e. without radiative precursor) until the temperature drops to 100~K. These lines are a subset of those output by the code.}
\begin{tabular}{lcccccccc} \hline \hline
\multicolumn{3}{l}{$n_{\rm H}=10^4$ cm$^{-3}$, $b=1$, $G_0=10^3$} &  & & & &\\
 & \multicolumn{8}{c}{Velocities (km s$^{-1}$)} \\ \cline{2-9} 
\multicolumn{9}{c}{\vspace{-0.3cm}} \\ 
Line & 10 & 20 & 30 & 40 & 50 & 60 & 70 & 80 \\ \cline{1-9}
\multicolumn{9}{c}{\vspace{-0.3cm}} \\ 
\cline{1-9} \multicolumn{9}{c}{H$_2$ rovibrational lines} \\
0-0 S(0) (28 $\mu$m) & $2.2(-07)$ & $2.0(-06)$ & $4.8(-06)$ & $9.2(-06)$ & $1.1(-05)$ & $1.3(-05)$ & $1.4(-05)$ & $1.5(-05)$ \\
0-0 S(1) (17 $\mu$m) & $9.9(-06)$ & $2.2(-05)$ & $3.0(-05)$ & $6.0(-05)$ & $8.5(-05)$ & $1.2(-04)$ & $1.6(-04)$ & $2.1(-04)$ \\
0-0 S(2) (12 $\mu$m) & $1.3(-05)$ & $1.9(-05)$ & $8.1(-06)$ & $1.2(-05)$ & $2.0(-05)$ & $3.4(-05)$ & $5.7(-05)$ & $9.2(-05)$ \\
0-0 S(3) (10 $\mu$m) & $8.4(-05)$ & $1.5(-04)$ & $8.8(-05)$ & $1.1(-04)$ & $1.2(-04)$ & $1.3(-04)$ & $1.7(-04)$ & $2.7(-04)$ \\
0-0 S(4) (8.0 $\mu$m) & $3.6(-05)$ & $8.4(-05)$ & $4.4(-05)$ & $6.5(-05)$ & $7.4(-05)$ & $8.0(-05)$ & $8.9(-05)$ & $1.1(-04)$ \\
0-0 S(5) (6.9 $\mu$m) & $1.0(-04)$ & $3.5(-04)$ & $1.7(-04)$ & $2.5(-04)$ & $2.9(-04)$ & $3.1(-04)$ & $3.4(-04)$ & $3.7(-04)$ \\
0-0 S(6) (6.1 $\mu$m) & $2.5(-05)$ & $1.3(-04)$ & $6.2(-05)$ & $9.4(-05)$ & $1.1(-04)$ & $1.3(-04)$ & $1.4(-04)$ & $1.4(-04)$ \\
0-0 S(7) (5.5 $\mu$m) & $3.8(-05)$ & $3.7(-04)$ & $1.8(-04)$ & $2.7(-04)$ & $3.3(-04)$ & $3.7(-04)$ & $4.1(-04)$ & $4.3(-04)$ \\
0-0 S(8) (5.1 $\mu$m) & $7.5(-06)$ & $1.1(-04)$ & $5.9(-05)$ & $9.0(-05)$ & $1.1(-04)$ & $1.2(-04)$ & $1.4(-04)$ & $1.4(-04)$ \\
0-0 S(9) (4.7 $\mu$m) & $9.3(-06)$ & $2.2(-04)$ & $1.6(-04)$ & $2.4(-04)$ & $2.9(-04)$ & $3.3(-04)$ & $3.7(-04)$ & $3.9(-04)$ \\
1-0 S(0) (2.22 $\mu$m) & $1.6(-05)$ & $1.1(-04)$ & $1.8(-05)$ & $3.0(-05)$ & $3.7(-05)$ & $4.5(-05)$ & $5.4(-05)$ & $6.3(-05)$ \\
1-0 S(1) (2.12 $\mu$m) & $5.9(-05)$ & $5.0(-04)$ & $6.7(-05)$ & $1.0(-04)$ & $1.2(-04)$ & $1.5(-04)$ & $1.7(-04)$ & $2.0(-04)$ \\
1-0 S(2) (2.03 $\mu$m) & $2.2(-05)$ & $2.0(-04)$ & $3.5(-05)$ & $5.2(-05)$ & $6.3(-05)$ & $7.3(-05)$ & $8.3(-05)$ & $9.3(-05)$ \\
1-0 S(3) (1.96 $\mu$m) & $4.4(-05)$ & $5.2(-04)$ & $9.6(-05)$ & $1.4(-04)$ & $1.7(-04)$ & $1.9(-04)$ & $2.1(-04)$ & $2.3(-04)$ \\
2-1 S(0) (2.36 $\mu$m) & $8.2(-07)$ & $1.3(-05)$ & $5.4(-06)$ & $8.6(-06)$ & $9.8(-06)$ & $1.1(-05)$ & $1.1(-05)$ & $1.2(-05)$ \\
2-1 S(1) (2.25 $\mu$m) & $2.9(-06)$ & $5.5(-05)$ & $2.0(-05)$ & $2.8(-05)$ & $3.1(-05)$ & $3.3(-05)$ & $3.6(-05)$ & $3.7(-05)$ \\
2-1 S(2) (2.15 $\mu$m) & $1.9(-06)$ & $2.4(-05)$ & $1.1(-05)$ & $1.6(-05)$ & $1.8(-05)$ & $1.9(-05)$ & $2.1(-05)$ & $2.2(-05)$ \\
2-1 S(3) (2.07 $\mu$m) & $3.9(-06)$ & $6.3(-05)$ & $3.0(-05)$ & $3.9(-05)$ & $4.3(-05)$ & $4.5(-05)$ & $4.9(-05)$ & $5.1(-05)$ \\
\cline{1-9} \multicolumn{9}{c}{CH$^+$ rovibrational lines} \\
0-0 (J=1-0) (359 $\mu$m) & $1.1(-07)$ & $5.7(-07)$ & $2.4(-06)$ & $3.9(-06)$ & $5.1(-06)$ & $6.3(-06)$ & $7.9(-06)$ & $9.8(-06)$ \\
0-0 (J=2-1) (180 $\mu$m) & $2.0(-07)$ & $1.5(-06)$ & $4.9(-06)$ & $8.9(-06)$ & $1.3(-05)$ & $1.7(-05)$ & $2.3(-05)$ & $3.2(-05)$ \\
0-0 (J=3-2) (120 $\mu$m) & $2.0(-07)$ & $1.6(-06)$ & $4.8(-06)$ & $8.5(-06)$ & $1.3(-05)$ & $1.8(-05)$ & $2.5(-05)$ & $3.5(-05)$ \\
1-0 (J=0-1) (3.69 $\mu$m) & $3.6(-10)$ & $1.9(-08)$ & $3.3(-08)$ & $5.1(-08)$ & $7.9(-08)$ & $1.2(-07)$ & $1.6(-07)$ & $2.3(-07)$ \\
1-0 (J=1-0) (3.61 $\mu$m) & $5.2(-10)$ & $2.4(-08)$ & $4.5(-08)$ & $6.8(-08)$ & $1.1(-07)$ & $1.5(-07)$ & $2.2(-07)$ & $3.0(-07)$ \\
\cline{1-9} \multicolumn{9}{c}{CO pure rotational lines} \\
J=1-0 (2601 $\mu$m) & $2.8(-12)$ & $4.0(-11)$ & $2.1(-10)$ & $7.6(-10)$ & $1.1(-09)$ & $1.3(-09)$ & $1.5(-09)$ & $1.7(-09)$ \\
J=2-1 (1300 $\mu$m) & $9.2(-11)$ & $1.2(-09)$ & $6.5(-09)$ & $2.3(-08)$ & $3.3(-08)$ & $3.9(-08)$ & $4.5(-08)$ & $5.1(-08)$ \\
J=3-2 (867 $\mu$m) & $7.1(-10)$ & $8.6(-09)$ & $4.5(-08)$ & $1.6(-07)$ & $2.3(-07)$ & $2.7(-07)$ & $3.1(-07)$ & $3.6(-07)$ \\
J=4-3 (650 $\mu$m) & $3.0(-09)$ & $3.2(-08)$ & $1.7(-07)$ & $5.9(-07)$ & $8.5(-07)$ & $1.0(-06)$ & $1.2(-06)$ & $1.3(-06)$ \\
J=8-7 (434 $\mu$m) & $4.4(-08)$ & $3.5(-07)$ & $2.0(-06)$ & $6.6(-06)$ & $1.0(-05)$ & $1.3(-05)$ & $1.5(-05)$ & $1.7(-05)$ \\
J=9-8 (372 $\mu$m) & $5.2(-08)$ & $4.0(-07)$ & $2.4(-06)$ & $7.8(-06)$ & $1.2(-05)$ & $1.6(-05)$ & $1.9(-05)$ & $2.2(-05)$ \\
J=10-9 (325 $\mu$m) & $5.3(-08)$ & $4.1(-07)$ & $2.4(-06)$ & $8.0(-06)$ & $1.3(-05)$ & $1.7(-05)$ & $2.2(-05)$ & $2.6(-05)$ \\
\cline{1-9} \multicolumn{9}{c}{Atomic H lines} \\
Ly$\alpha$ (1215.7 \AA) & 0 & 0 & 0 & $5.3(-03)$ & $1.3(-02)$ & $2.7(-02)$ & $4.7(-02)$ & $7.4(-02)$ \\
Ly$\beta$ (1025.7 \AA) & 0 & 0 & 0 & $8.9(-05)$ & $8.3(-04)$ & $2.6(-03)$ & $5.5(-03)$ & $9.6(-03)$ \\
H$\alpha$ (6564.6 \AA) & 0 & 0 & 0 & $7.6(-04)$ & $1.9(-03)$ & $3.3(-03)$ & $4.8(-03)$ & $7.4(-03)$ \\
2ph (<2400 \AA) & 0 & 0 & 0 & $2.3(-03)$ & $4.6(-03)$ & $7.2(-03)$ & $9.8(-03)$ & $1.3(-02)$ \\
\cline{1-9} \multicolumn{9}{c}{Atomic fine structure lines} \\
C$^+$ (158 $\mu$m) & $5.0(-06)$ & $7.9(-05)$ & $2.0(-04)$ & $2.2(-04)$ & $2.0(-04)$ & $1.9(-04)$ & $2.0(-04)$ & $2.0(-04)$ \\
C (609.8 $\mu$m) & $1.9(-08)$ & $4.0(-07)$ & $1.3(-06)$ & $2.3(-06)$ & $2.8(-06)$ & $3.0(-06)$ & $3.4(-06)$ & $3.6(-06)$ \\
C (370.4 $\mu$m) & $1.6(-07)$ & $2.9(-06)$ & $9.5(-06)$ & $1.7(-05)$ & $2.1(-05)$ & $2.3(-05)$ & $2.6(-05)$ & $2.7(-05)$ \\
O (145.5 $\mu$m) & $9.2(-06)$ & $7.7(-05)$ & $3.9(-04)$ & $5.1(-04)$ & $5.9(-04)$ & $6.8(-04)$ & $8.0(-04)$ & $9.0(-04)$ \\
O (63.2 $\mu$m) & $2.2(-04)$ & $2.8(-03)$ & $1.5(-02)$ & $2.0(-02)$ & $2.4(-02)$ & $2.8(-02)$ & $3.3(-02)$ & $3.7(-02)$ \\
S (25.3 $\mu$m) & $3.6(-07)$ & $7.0(-06)$ & $1.1(-04)$ & $2.2(-04)$ & $3.6(-04)$ & $5.3(-04)$ & $7.7(-04)$ & $1.1(-03)$ \\
Si (68.5 $\mu$m) & $7.0(-09)$ & $3.6(-07)$ & $1.8(-06)$ & $3.4(-06)$ & $4.5(-06)$ & $5.2(-06)$ & $6.3(-06)$ & $6.9(-06)$ \\
Si$^+$ (34.8 $\mu$m) & $1.2(-05)$ & $1.3(-04)$ & $7.0(-04)$ & $1.0(-03)$ & $1.3(-03)$ & $1.6(-03)$ & $1.9(-03)$ & $2.3(-03)$ \\
\cline{1-9} \multicolumn{9}{c}{Atomic metastable lines} \\
O (6300 \AA) & $5.5(-12)$ & $1.1(-08)$ & $1.7(-04)$ & $2.7(-04)$ & $5.2(-04)$ & $8.9(-04)$ & $1.3(-03)$ & $1.8(-03)$ \\
O (6364 \AA) & $1.8(-12)$ & $3.6(-09)$ & $5.4(-05)$ & $8.5(-05)$ & $1.7(-04)$ & $2.8(-04)$ & $4.2(-04)$ & $5.8(-04)$ \\
S$^+$ (6731 \AA) & $3.1(-10)$ & $6.6(-08)$ & $1.1(-04)$ & $4.0(-04)$ & $6.4(-04)$ & $7.2(-04)$ & $7.8(-04)$ & $8.7(-04)$ \\
S$^+$ (6716 \AA) & $4.5(-10)$ & $9.7(-08)$ & $1.5(-04)$ & $3.2(-04)$ & $3.8(-04)$ & $4.0(-04)$ & $4.3(-04)$ & $4.9(-04)$ \\
N (5200 \AA) & $7.1(-13)$ & $1.0(-08)$ & $3.1(-05)$ & $4.6(-05)$ & $3.9(-05)$ & $3.0(-05)$ & $2.8(-05)$ & $3.3(-05)$ \\
N (5198 \AA) & $4.7(-13)$ & $6.8(-09)$ & $2.2(-05)$ & $5.8(-05)$ & $5.8(-05)$ & $4.6(-05)$ & $4.3(-05)$ & $4.9(-05)$ \\
\cline{1-9}
\end{tabular}

\tablefoot{Numbers in parentheses denote powers of ten.}
\end{table*}

\begin{table*}
  \centering
  \caption{Integrated line and two-photon continuum intensities (erg s$^{-1}$ cm$^{-2}$ sr$^{-1}$) escaping the shock front as a function of shock velocity (km s$^{-1}$) for shocks with preshock density $n_{\rm H}=10^5$~cm$^{-3}$, magnetic parameter $b=1$, and external radiation field $G_0=10^3$. Emission is integrated only in the postshock region (i.e. without radiative precursor) until the temperature drops to 100~K. These lines are a subset of those output by the code.}
\begin{tabular}{lcccccccc} \hline \hline
\multicolumn{3}{l}{$n_{\rm H}=10^5$ cm$^{-3}$, $b=1$, $G_0=10^3$} &  & & & &\\
 & \multicolumn{8}{c}{Velocities (km s$^{-1}$)} \\ \cline{2-9} 
\multicolumn{9}{c}{\vspace{-0.3cm}} \\ 
Line & 10 & 20 & 30 & 40 & 50 & 60 & 70 & 80 \\ \cline{1-9}
\multicolumn{9}{c}{\vspace{-0.3cm}} \\ 
\cline{1-9} \multicolumn{9}{c}{H$_2$ rovibrational lines} \\
0-0 S(0) (28 $\mu$m) & $4.6(-06)$ & $1.1(-05)$ & $2.2(-05)$ & $2.6(-05)$ & $2.8(-05)$ & $3.0(-05)$ & $3.2(-05)$ & $3.3(-05)$ \\
0-0 S(1) (17 $\mu$m) & $6.5(-05)$ & $1.3(-04)$ & $5.4(-04)$ & $6.8(-04)$ & $7.8(-04)$ & $8.9(-04)$ & $9.7(-04)$ & $1.1(-03)$ \\
0-0 S(2) (12 $\mu$m) & $6.5(-05)$ & $1.0(-04)$ & $4.0(-04)$ & $5.3(-04)$ & $6.4(-04)$ & $7.6(-04)$ & $8.5(-04)$ & $9.4(-04)$ \\
0-0 S(3) (10 $\mu$m) & $4.1(-04)$ & $7.3(-04)$ & $1.3(-03)$ & $1.8(-03)$ & $2.2(-03)$ & $2.7(-03)$ & $3.1(-03)$ & $3.5(-03)$ \\
0-0 S(4) (8.0 $\mu$m) & $2.0(-04)$ & $4.9(-04)$ & $3.1(-04)$ & $4.3(-04)$ & $5.4(-04)$ & $6.7(-04)$ & $7.9(-04)$ & $9.0(-04)$ \\
0-0 S(5) (6.9 $\mu$m) & $6.5(-04)$ & $2.3(-03)$ & $6.8(-04)$ & $8.5(-04)$ & $1.0(-03)$ & $1.2(-03)$ & $1.4(-03)$ & $1.6(-03)$ \\
0-0 S(6) (6.1 $\mu$m) & $1.8(-04)$ & $9.6(-04)$ & $2.3(-04)$ & $2.7(-04)$ & $2.9(-04)$ & $3.2(-04)$ & $3.5(-04)$ & $3.8(-04)$ \\
0-0 S(7) (5.5 $\mu$m) & $3.5(-04)$ & $3.0(-03)$ & $7.0(-04)$ & $7.7(-04)$ & $8.1(-04)$ & $8.5(-04)$ & $9.0(-04)$ & $9.2(-04)$ \\
0-0 S(8) (5.1 $\mu$m) & $7.6(-05)$ & $9.3(-04)$ & $2.6(-04)$ & $2.9(-04)$ & $3.1(-04)$ & $3.2(-04)$ & $3.3(-04)$ & $3.3(-04)$ \\
0-0 S(9) (4.7 $\mu$m) & $1.2(-04)$ & $2.1(-03)$ & $7.5(-04)$ & $8.3(-04)$ & $8.7(-04)$ & $9.1(-04)$ & $9.4(-04)$ & $9.6(-04)$ \\
1-0 S(0) (2.22 $\mu$m) & $1.1(-04)$ & $9.4(-04)$ & $1.6(-04)$ & $1.8(-04)$ & $2.1(-04)$ & $2.4(-04)$ & $2.7(-04)$ & $3.0(-04)$ \\
1-0 S(1) (2.12 $\mu$m) & $4.0(-04)$ & $4.3(-03)$ & $4.7(-04)$ & $5.4(-04)$ & $6.1(-04)$ & $7.1(-04)$ & $8.0(-04)$ & $9.0(-04)$ \\
1-0 S(2) (2.03 $\mu$m) & $1.4(-04)$ & $1.7(-03)$ & $2.0(-04)$ & $2.2(-04)$ & $2.3(-04)$ & $2.6(-04)$ & $2.8(-04)$ & $3.0(-04)$ \\
1-0 S(3) (1.96 $\mu$m) & $3.0(-04)$ & $4.6(-03)$ & $4.9(-04)$ & $5.3(-04)$ & $5.7(-04)$ & $6.2(-04)$ & $6.6(-04)$ & $7.1(-04)$ \\
2-1 S(0) (2.36 $\mu$m) & $6.6(-06)$ & $1.0(-04)$ & $1.9(-05)$ & $2.0(-05)$ & $2.1(-05)$ & $2.1(-05)$ & $2.2(-05)$ & $2.2(-05)$ \\
2-1 S(1) (2.25 $\mu$m) & $2.0(-05)$ & $4.7(-04)$ & $6.9(-05)$ & $7.1(-05)$ & $7.3(-05)$ & $7.5(-05)$ & $7.8(-05)$ & $8.1(-05)$ \\
2-1 S(2) (2.15 $\mu$m) & $1.1(-05)$ & $1.9(-04)$ & $3.9(-05)$ & $4.0(-05)$ & $4.2(-05)$ & $4.3(-05)$ & $4.5(-05)$ & $4.6(-05)$ \\
2-1 S(3) (2.07 $\mu$m) & $2.2(-05)$ & $5.2(-04)$ & $9.8(-05)$ & $9.9(-05)$ & $1.0(-04)$ & $1.0(-04)$ & $1.1(-04)$ & $1.1(-04)$ \\
\cline{1-9} \multicolumn{9}{c}{CH$^+$ rovibrational lines} \\
0-0 (J=1-0) (359 $\mu$m) & $7.8(-07)$ & $1.5(-06)$ & $2.2(-05)$ & $2.5(-05)$ & $3.0(-05)$ & $3.9(-05)$ & $4.8(-05)$ & $6.2(-05)$ \\
0-0 (J=2-1) (180 $\mu$m) & $1.1(-06)$ & $2.5(-06)$ & $7.8(-05)$ & $9.6(-05)$ & $1.3(-04)$ & $1.9(-04)$ & $3.2(-04)$ & $4.0(-04)$ \\
0-0 (J=3-2) (120 $\mu$m) & $8.2(-07)$ & $2.1(-06)$ & $9.6(-05)$ & $1.2(-04)$ & $1.6(-04)$ & $2.5(-04)$ & $5.7(-04)$ & $9.7(-04)$ \\
1-0 (J=0-1) (3.69 $\mu$m) & $1.4(-09)$ & $3.2(-09)$ & $7.5(-07)$ & $8.7(-07)$ & $1.2(-06)$ & $1.8(-06)$ & $5.3(-06)$ & $1.1(-05)$ \\
1-0 (J=1-0) (3.61 $\mu$m) & $1.7(-09)$ & $4.6(-09)$ & $1.0(-06)$ & $1.2(-06)$ & $1.6(-06)$ & $2.4(-06)$ & $6.4(-06)$ & $1.3(-05)$ \\
\cline{1-9} \multicolumn{9}{c}{CO pure rotational lines} \\
J=1-0 (2601 $\mu$m) & $6.5(-10)$ & $1.8(-09)$ & $5.0(-09)$ & $7.4(-09)$ & $8.2(-09)$ & $8.8(-09)$ & $1.1(-08)$ & $1.1(-08)$ \\
J=2-1 (1300 $\mu$m) & $1.9(-08)$ & $5.6(-08)$ & $1.5(-07)$ & $2.2(-07)$ & $2.5(-07)$ & $2.6(-07)$ & $3.3(-07)$ & $3.3(-07)$ \\
J=3-2 (867 $\mu$m) & $1.3(-07)$ & $3.9(-07)$ & $1.0(-06)$ & $1.5(-06)$ & $1.7(-06)$ & $1.8(-06)$ & $2.3(-06)$ & $2.3(-06)$ \\
J=4-3 (650 $\mu$m) & $4.8(-07)$ & $1.5(-06)$ & $3.9(-06)$ & $5.6(-06)$ & $6.2(-06)$ & $6.7(-06)$ & $9.0(-06)$ & $9.0(-06)$ \\
J=8-7 (434 $\mu$m) & $5.5(-06)$ & $2.1(-05)$ & $6.1(-05)$ & $8.0(-05)$ & $9.1(-05)$ & $1.0(-04)$ & $1.5(-04)$ & $1.7(-04)$ \\
J=9-8 (372 $\mu$m) & $6.7(-06)$ & $2.9(-05)$ & $8.6(-05)$ & $1.1(-04)$ & $1.3(-04)$ & $1.5(-04)$ & $2.1(-04)$ & $2.5(-04)$ \\
J=10-9 (325 $\mu$m) & $7.3(-06)$ & $3.5(-05)$ & $1.1(-04)$ & $1.5(-04)$ & $1.7(-04)$ & $1.9(-04)$ & $2.8(-04)$ & $3.3(-04)$ \\
\cline{1-9} \multicolumn{9}{c}{Atomic H lines} \\
Ly$\alpha$ (1215.7 \AA) & 0 & 0 & 0 & $7.6(-02)$ & $2.1(-01)$ & $4.4(-01)$ & $7.6(-01)$ & $1.2(+00)$ \\
Ly$\beta$ (1025.7 \AA) & 0 & 0 & 0 & $2.5(-03)$ & $1.3(-02)$ & $3.3(-02)$ & $6.5(-02)$ & $1.1(-01)$ \\
H$\alpha$ (6564.6 \AA) & 0 & 0 & 0 & $6.8(-03)$ & $1.6(-02)$ & $3.0(-02)$ & $4.2(-02)$ & $6.4(-02)$ \\
2ph (<2400 \AA) & 0 & 0 & 0 & $1.7(-02)$ & $2.8(-02)$ & $3.7(-02)$ & $4.4(-02)$ & $5.3(-02)$ \\
\cline{1-9} \multicolumn{9}{c}{Atomic fine structure lines} \\
C$^+$ (158 $\mu$m) & $4.3(-05)$ & $4.6(-05)$ & $1.0(-04)$ & $6.7(-05)$ & $5.8(-05)$ & $6.4(-05)$ & $7.8(-05)$ & $9.0(-05)$ \\
C (609.8 $\mu$m) & $1.4(-06)$ & $3.6(-06)$ & $5.0(-06)$ & $5.9(-06)$ & $6.1(-06)$ & $6.1(-06)$ & $6.2(-06)$ & $5.8(-06)$ \\
C (370.4 $\mu$m) & $1.0(-05)$ & $2.5(-05)$ & $3.9(-05)$ & $4.6(-05)$ & $4.7(-05)$ & $4.8(-05)$ & $4.8(-05)$ & $4.6(-05)$ \\
O (145.5 $\mu$m) & $1.5(-04)$ & $3.8(-04)$ & $1.7(-03)$ & $1.7(-03)$ & $1.7(-03)$ & $1.8(-03)$ & $2.0(-03)$ & $2.1(-03)$ \\
O (63.2 $\mu$m) & $6.4(-03)$ & $1.9(-02)$ & $6.9(-02)$ & $7.2(-02)$ & $7.6(-02)$ & $8.0(-02)$ & $8.4(-02)$ & $8.7(-02)$ \\
S (25.3 $\mu$m) & $1.2(-04)$ & $8.2(-04)$ & $1.8(-02)$ & $2.5(-02)$ & $3.1(-02)$ & $3.6(-02)$ & $4.1(-02)$ & $4.6(-02)$ \\
Si (68.5 $\mu$m) & $4.6(-06)$ & $2.5(-05)$ & $6.6(-05)$ & $7.0(-05)$ & $7.4(-05)$ & $7.4(-05)$ & $7.7(-05)$ & $7.5(-05)$ \\
Si$^+$ (34.8 $\mu$m) & $3.4(-04)$ & $1.0(-03)$ & $4.7(-03)$ & $4.9(-03)$ & $5.2(-03)$ & $5.7(-03)$ & $6.1(-03)$ & $6.5(-03)$ \\
\cline{1-9} \multicolumn{9}{c}{Atomic metastable lines} \\
O (6300 \AA) & $2.3(-11)$ & $1.4(-08)$ & $4.3(-03)$ & $4.6(-03)$ & $7.3(-03)$ & $1.1(-02)$ & $1.6(-02)$ & $2.0(-02)$ \\
O (6364 \AA) & $7.2(-12)$ & $4.4(-09)$ & $1.4(-03)$ & $1.5(-03)$ & $2.3(-03)$ & $3.6(-03)$ & $5.0(-03)$ & $6.3(-03)$ \\
S$^+$ (6731 \AA) & $5.6(-10)$ & $8.2(-09)$ & $1.1(-03)$ & $1.4(-03)$ & $1.6(-03)$ & $2.1(-03)$ & $3.0(-03)$ & $3.9(-03)$ \\
S$^+$ (6716 \AA) & $8.2(-10)$ & $1.2(-08)$ & $1.4(-03)$ & $1.1(-03)$ & $1.2(-03)$ & $1.6(-03)$ & $2.4(-03)$ & $3.1(-03)$ \\
N (5200 \AA) & $3.3(-12)$ & $3.3(-08)$ & $3.2(-04)$ & $1.9(-04)$ & $1.5(-04)$ & $1.4(-04)$ & $1.9(-04)$ & $2.5(-04)$ \\
N (5198 \AA) & $2.2(-12)$ & $2.2(-08)$ & $2.5(-04)$ & $2.2(-04)$ & $1.8(-04)$ & $1.7(-04)$ & $2.1(-04)$ & $2.8(-04)$ \\
\cline{1-9}
\end{tabular}

\tablefoot{Numbers in parentheses denote powers of ten.}
\end{table*}

\begin{table*}
  \centering
  \caption{Integrated line and two-photon continuum intensities (erg s$^{-1}$ cm$^{-2}$ sr$^{-1}$) escaping the shock front as a function of shock velocity (km s$^{-1}$) for shocks with preshock density $n_{\rm H}=10^6$~cm$^{-3}$, magnetic parameter $b=1$, and external radiation field $G_0=10^3$. Emission is integrated only in the postshock region (i.e. without radiative precursor) until the temperature drops to 100~K. These lines are a subset of those output by the code.}
\begin{tabular}{lcccccccc} \hline \hline
\multicolumn{3}{l}{$n_{\rm H}=10^6$ cm$^{-3}$, $b=1$, $G_0=10^3$} &  & & & &\\
 & \multicolumn{8}{c}{Velocities (km s$^{-1}$)} \\ \cline{2-9} 
\multicolumn{9}{c}{\vspace{-0.3cm}} \\ 
Line & 10 & 20 & 30 & 40 & 50 & 60 & 70 & 80 \\ \cline{1-9}
\multicolumn{9}{c}{\vspace{-0.3cm}} \\ 
\cline{1-9} \multicolumn{9}{c}{H$_2$ rovibrational lines} \\
0-0 S(0) (28 $\mu$m) & $1.3(-05)$ & $4.7(-05)$ & $5.9(-05)$ & $6.1(-05)$ & $6.3(-05)$ & $6.5(-05)$ & $6.8(-05)$ & $7.1(-05)$ \\
0-0 S(1) (17 $\mu$m) & $2.4(-04)$ & $1.4(-03)$ & $1.9(-03)$ & $1.9(-03)$ & $2.1(-03)$ & $2.2(-03)$ & $2.3(-03)$ & $2.4(-03)$ \\
0-0 S(2) (12 $\mu$m) & $1.7(-04)$ & $9.8(-04)$ & $1.4(-03)$ & $1.5(-03)$ & $1.6(-03)$ & $1.8(-03)$ & $1.9(-03)$ & $2.0(-03)$ \\
0-0 S(3) (10 $\mu$m) & $9.6(-04)$ & $3.7(-03)$ & $4.0(-03)$ & $4.4(-03)$ & $5.2(-03)$ & $5.7(-03)$ & $6.4(-03)$ & $7.1(-03)$ \\
0-0 S(4) (8.0 $\mu$m) & $5.1(-04)$ & $1.6(-03)$ & $7.4(-04)$ & $8.5(-04)$ & $1.1(-03)$ & $1.2(-03)$ & $1.4(-03)$ & $1.7(-03)$ \\
0-0 S(5) (6.9 $\mu$m) & $2.0(-03)$ & $7.8(-03)$ & $1.1(-03)$ & $1.2(-03)$ & $1.5(-03)$ & $1.8(-03)$ & $2.1(-03)$ & $2.6(-03)$ \\
0-0 S(6) (6.1 $\mu$m) & $6.9(-04)$ & $3.9(-03)$ & $4.0(-04)$ & $3.8(-04)$ & $3.9(-04)$ & $4.1(-04)$ & $4.5(-04)$ & $5.1(-04)$ \\
0-0 S(7) (5.5 $\mu$m) & $1.7(-03)$ & $1.5(-02)$ & $1.3(-03)$ & $1.2(-03)$ & $1.2(-03)$ & $1.1(-03)$ & $1.2(-03)$ & $1.2(-03)$ \\
0-0 S(8) (5.1 $\mu$m) & $4.3(-04)$ & $5.8(-03)$ & $6.7(-04)$ & $6.3(-04)$ & $5.8(-04)$ & $5.6(-04)$ & $5.6(-04)$ & $5.5(-04)$ \\
0-0 S(9) (4.7 $\mu$m) & $8.1(-04)$ & $1.7(-02)$ & $2.1(-03)$ & $2.1(-03)$ & $2.0(-03)$ & $2.0(-03)$ & $2.0(-03)$ & $2.0(-03)$ \\
1-0 S(0) (2.22 $\mu$m) & $6.0(-04)$ & $5.4(-03)$ & $8.3(-04)$ & $8.1(-04)$ & $8.2(-04)$ & $8.5(-04)$ & $8.9(-04)$ & $9.6(-04)$ \\
1-0 S(1) (2.12 $\mu$m) & $2.1(-03)$ & $2.5(-02)$ & $1.4(-03)$ & $1.4(-03)$ & $1.4(-03)$ & $1.5(-03)$ & $1.6(-03)$ & $1.8(-03)$ \\
1-0 S(2) (2.03 $\mu$m) & $7.7(-04)$ & $1.0(-02)$ & $5.2(-04)$ & $4.8(-04)$ & $4.7(-04)$ & $4.8(-04)$ & $5.0(-04)$ & $5.5(-04)$ \\
1-0 S(3) (1.96 $\mu$m) & $1.7(-03)$ & $3.0(-02)$ & $9.9(-04)$ & $9.2(-04)$ & $9.0(-04)$ & $9.1(-04)$ & $9.5(-04)$ & $1.0(-03)$ \\
2-1 S(0) (2.36 $\mu$m) & $3.8(-05)$ & $9.6(-04)$ & $5.8(-05)$ & $5.4(-05)$ & $5.2(-05)$ & $5.3(-05)$ & $5.4(-05)$ & $5.5(-05)$ \\
2-1 S(1) (2.25 $\mu$m) & $9.4(-05)$ & $4.7(-03)$ & $1.4(-04)$ & $1.3(-04)$ & $1.2(-04)$ & $1.2(-04)$ & $1.2(-04)$ & $1.3(-04)$ \\
2-1 S(2) (2.15 $\mu$m) & $6.4(-05)$ & $2.0(-03)$ & $1.1(-04)$ & $9.9(-05)$ & $9.1(-05)$ & $8.9(-05)$ & $8.7(-05)$ & $8.7(-05)$ \\
2-1 S(3) (2.07 $\mu$m) & $9.2(-05)$ & $6.0(-03)$ & $1.9(-04)$ & $1.7(-04)$ & $1.5(-04)$ & $1.5(-04)$ & $1.5(-04)$ & $1.5(-04)$ \\
\cline{1-9} \multicolumn{9}{c}{CH$^+$ rovibrational lines} \\
0-0 (J=1-0) (359 $\mu$m) & $9.9(-07)$ & $1.8(-06)$ & $1.0(-05)$ & $9.9(-06)$ & $9.3(-06)$ & $1.4(-05)$ & $1.7(-05)$ & $2.6(-05)$ \\
0-0 (J=2-1) (180 $\mu$m) & $2.6(-06)$ & $7.4(-06)$ & $1.2(-04)$ & $1.8(-04)$ & $1.7(-04)$ & $3.1(-04)$ & $4.4(-04)$ & $7.3(-04)$ \\
0-0 (J=3-2) (120 $\mu$m) & $2.3(-06)$ & $8.4(-06)$ & $2.1(-04)$ & $5.5(-04)$ & $3.7(-04)$ & $7.3(-04)$ & $1.1(-03)$ & $2.1(-03)$ \\
1-0 (J=0-1) (3.69 $\mu$m) & $3.5(-09)$ & $1.3(-08)$ & $1.7(-06)$ & $2.3(-06)$ & $3.9(-06)$ & $8.8(-06)$ & $1.5(-05)$ & $3.2(-05)$ \\
1-0 (J=1-0) (3.61 $\mu$m) & $4.0(-09)$ & $1.5(-08)$ & $2.3(-06)$ & $3.0(-06)$ & $5.2(-06)$ & $1.2(-05)$ & $2.0(-05)$ & $4.4(-05)$ \\
\cline{1-9} \multicolumn{9}{c}{CO pure rotational lines} \\
J=1-0 (2601 $\mu$m) & $1.1(-08)$ & $1.9(-08)$ & $2.9(-08)$ & $3.5(-08)$ & $3.6(-08)$ & $3.4(-08)$ & $3.9(-08)$ & $4.2(-08)$ \\
J=2-1 (1300 $\mu$m) & $3.3(-07)$ & $5.8(-07)$ & $8.8(-07)$ & $1.1(-06)$ & $1.1(-06)$ & $1.0(-06)$ & $1.2(-06)$ & $1.3(-06)$ \\
J=3-2 (867 $\mu$m) & $2.3(-06)$ & $4.0(-06)$ & $6.3(-06)$ & $7.7(-06)$ & $8.1(-06)$ & $7.4(-06)$ & $8.5(-06)$ & $9.1(-06)$ \\
J=4-3 (650 $\mu$m) & $8.3(-06)$ & $1.5(-05)$ & $2.4(-05)$ & $3.0(-05)$ & $3.2(-05)$ & $2.9(-05)$ & $3.3(-05)$ & $3.5(-05)$ \\
J=8-7 (434 $\mu$m) & $1.2(-04)$ & $2.6(-04)$ & $4.6(-04)$ & $5.9(-04)$ & $7.1(-04)$ & $5.9(-04)$ & $6.5(-04)$ & $7.0(-04)$ \\
J=9-8 (372 $\mu$m) & $1.6(-04)$ & $3.8(-04)$ & $7.0(-04)$ & $9.1(-04)$ & $1.1(-03)$ & $9.2(-04)$ & $1.0(-03)$ & $1.1(-03)$ \\
J=10-9 (325 $\mu$m) & $2.0(-04)$ & $5.3(-04)$ & $1.0(-03)$ & $1.3(-03)$ & $1.7(-03)$ & $1.3(-03)$ & $1.5(-03)$ & $1.6(-03)$ \\
\cline{1-9} \multicolumn{9}{c}{Atomic H lines} \\
Ly$\alpha$ (1215.7 \AA) & 0 & 0 & $3.1(-02)$ & $9.6(-01)$ & $1.8(+00)$ & $3.3(+00)$ & $5.3(+00)$ & $8.5(+00)$ \\
Ly$\beta$ (1025.7 \AA) & 0 & 0 & $7.3(-05)$ & $2.8(-02)$ & $5.0(-02)$ & $1.8(-01)$ & $4.2(-01)$ & $7.8(-01)$ \\
H$\alpha$ (6564.6 \AA) & 0 & 0 & $4.8(-03)$ & $6.3(-02)$ & $2.7(-01)$ & $4.5(-01)$ & $6.7(-01)$ & $9.4(-01)$ \\
2ph (<2400 \AA) & 0 & 0 & $3.3(-02)$ & $8.4(-02)$ & $1.8(-01)$ & $2.0(-01)$ & $2.1(-01)$ & $2.5(-01)$ \\
\cline{1-9} \multicolumn{9}{c}{Atomic fine structure lines} \\
C$^+$ (158 $\mu$m) & $4.6(-06)$ & $4.8(-06)$ & $2.1(-05)$ & $1.3(-05)$ & $2.1(-05)$ & $3.6(-05)$ & $5.3(-05)$ & $1.5(-04)$ \\
C (609.8 $\mu$m) & $2.5(-06)$ & $5.2(-06)$ & $7.8(-06)$ & $7.3(-06)$ & $6.4(-06)$ & $6.3(-06)$ & $6.6(-06)$ & $6.1(-06)$ \\
C (370.4 $\mu$m) & $1.8(-05)$ & $4.1(-05)$ & $6.4(-05)$ & $5.9(-05)$ & $5.3(-05)$ & $5.1(-05)$ & $5.3(-05)$ & $5.0(-05)$ \\
O (145.5 $\mu$m) & $3.7(-04)$ & $1.4(-03)$ & $3.0(-03)$ & $2.8(-03)$ & $2.8(-03)$ & $2.8(-03)$ & $3.0(-03)$ & $3.3(-03)$ \\
O (63.2 $\mu$m) & $2.0(-02)$ & $6.5(-02)$ & $1.3(-01)$ & $1.2(-01)$ & $1.2(-01)$ & $1.2(-01)$ & $1.3(-01)$ & $1.4(-01)$ \\
S (25.3 $\mu$m) & $6.0(-03)$ & $4.4(-02)$ & $1.7(-01)$ & $1.6(-01)$ & $1.6(-01)$ & $1.6(-01)$ & $1.7(-01)$ & $1.7(-01)$ \\
Si (68.5 $\mu$m) & $8.1(-05)$ & $3.1(-04)$ & $5.8(-04)$ & $4.6(-04)$ & $4.0(-04)$ & $4.0(-04)$ & $4.0(-04)$ & $3.8(-04)$ \\
Si$^+$ (34.8 $\mu$m) & $9.5(-04)$ & $2.8(-03)$ & $5.5(-03)$ & $5.9(-03)$ & $6.3(-03)$ & $6.7(-03)$ & $7.1(-03)$ & $8.3(-03)$ \\
\cline{1-9} \multicolumn{9}{c}{Atomic metastable lines} \\
O (6300 \AA) & $9.3(-11)$ & $4.9(-08)$ & $7.6(-02)$ & $6.3(-02)$ & $7.8(-02)$ & $9.8(-02)$ & $1.2(-01)$ & $2.2(-01)$ \\
O (6364 \AA) & $3.0(-11)$ & $1.6(-08)$ & $2.4(-02)$ & $2.0(-02)$ & $2.5(-02)$ & $3.1(-02)$ & $3.8(-02)$ & $7.0(-02)$ \\
S$^+$ (6731 \AA) & $6.8(-10)$ & $1.4(-09)$ & $1.8(-03)$ & $1.9(-03)$ & $4.9(-03)$ & $1.0(-02)$ & $1.7(-02)$ & $5.7(-02)$ \\
S$^+$ (6716 \AA) & $9.5(-10)$ & $1.9(-09)$ & $1.9(-03)$ & $1.5(-03)$ & $3.4(-03)$ & $6.4(-03)$ & $9.8(-03)$ & $3.0(-02)$ \\
N (5200 \AA) & $1.5(-11)$ & $7.4(-08)$ & $1.3(-03)$ & $5.6(-04)$ & $5.8(-04)$ & $8.0(-04)$ & $1.0(-03)$ & $2.8(-03)$ \\
N (5198 \AA) & $1.0(-11)$ & $4.9(-08)$ & $1.3(-03)$ & $6.2(-04)$ & $7.0(-04)$ & $1.1(-03)$ & $1.5(-03)$ & $4.4(-03)$ \\
\cline{1-9}
\end{tabular}

\tablefoot{Numbers in parentheses denote powers of ten.}
\end{table*}

\end{document}